\documentclass[aip,pof,nofootinbib]{revtex4} 

\usepackage{amsmath}
\usepackage{natbib}
\usepackage{amssymb}
\usepackage{graphicx}
\usepackage{dcolumn}
\usepackage{bm}
\usepackage{mathrsfs}
\usepackage{color}
\usepackage{psfrag}
\usepackage{url}

\usepackage[normalem]{ulem}
\usepackage{color}

\newcommand{\rev}[1]{{\color{black}#1}}  

\newcommand{\revCom}[1]{\color{black}#1}  

\newcommand{\revProof}[1]{\color{black}#1}  

\usepackage{booktabs}

\newcommand{\et}{et al.}
\newcommand{\eg}{e.g.}
\newcommand{\fig}{Figure}
\newcommand{\figs}{Figures}

\newcommand{\sect}{Section}
\newcommand{\sects}{Sections}
\newcommand{\app}{Appendix}
\newcommand{\tab}{Table}
\newcommand{\of}{OpenFOAM}
\newcommand{\dak}{Dakota}

\newcommand{\var}{\text{var}}

\newcommand{\BE}{\mathbb{E}}
\newcommand{\BR}{\mathbb{R}}
\newcommand{\BQ}{\mathbb{Q}}
\newcommand{\cR}{\mathcal{R}}
\newcommand{\cY}{\mathcal{Y}}

\newcommand{\grad}{\mathbf{\nabla}}

\newcommand{\dd}{\mbox{d}}
\newcommand{\dash}{\mbox{-}}

\newcommand{\rey}{\mbox{Re}}
\newcommand{\reyt}{\mbox{Re}_\tau}

\newcommand{\bfu}{\bar{\mathbf{u}}}
\newcommand{\sgs}{\mathbf{B}}
\newcommand{\dx}{\Delta x}
\newcommand{\dy}{\Delta y}
\newcommand{\dz}{\Delta z}

\newcommand{\dxp}{\Delta x^+}
\newcommand{\dyp}{\Delta y^+}
\newcommand{\dzp}{\Delta z^+}
\newcommand{\Qx}{\mathbb{Q}_{\Delta x^+}}
\newcommand{\Qz}{\mathbb{Q}_{\Delta z^+}}
\newcommand{\Qy}{\mathbb{Q}_{\Delta y^+_w}}
\newcommand{\Qr}{\mathbb{Q}_{\reyt^\circ}}
\newcommand{\bu}{\bar{u}}
\newcommand{\ut}{u_\tau}
\newcommand{\dudy}{ { {\rm d}\langle u\rangle}/{{\rm d}\eta} }

\newcommand{\U}{\langle u \rangle}
\newcommand{\uv}{\langle u'v'\rangle}
\newcommand{\uu}{\langle u'u'\rangle}

\newcommand{\urms}{u_{\mbox {rms}}}
\newcommand{\vrms}{v_{\mbox {rms}}}
\newcommand{\wrms}{w_{\mbox {rms}}}

\newcommand{\tke}{\mathcal{K}}
\newcommand{\Pk}{\mathcal{P}_{\mathcal{K}}}

\newcommand{\eabs}{\epsilon_{1}}
\newcommand{\el}{\epsilon_{2}}
\newcommand{\einf}{\epsilon_{\infty}}
\newcommand{\epik}{\epsilon_{\rm pp}}
\newcommand{\earea}{\epsilon_{\rm A}}
\newcommand{\lon}{{\rm L}_1}
\newcommand{\ltw}{{\rm L}_2}
\newcommand{\lnf}{{\rm L}_\infty}

\newcommand{\rM}{{\rm M}}

\begin{document}

\title{Effect of grid resolution on large eddy simulation of wall-bounded turbulence}

\author{S. Rezaeiravesh}
\email[]{saleh.rezaeiravesh@it.uu.se}
\affiliation{Uppsala University, Department of Information Technology, Box 337, SE-751 05 Uppsala, Sweden.}
\author{M. Liefvendahl}
\email[]{mattias.liefvendahl@foi.se}
\affiliation{Uppsala University, Department of Information Technology, Box 337, SE-751 05 Uppsala, Sweden.}
\affiliation{FOI, Totalf\"{o}rsvarets forskningsinstitut, SE-164 90 Stockholm, Sweden.}


\date{\today}



\begin{abstract}
  \subsection*{Abstract}
  The effect of grid resolution on large eddy simulation (LES) of wall-bounded turbulent flow
  is investigated. A channel flow simulation campaign involving systematic variation of the
  streamwise ($\Delta x$) and spanwise ($\Delta z$) grid resolution is used for this purpose.
  The main friction-velocity based Reynolds number investigated is~300.
  Near the walls, the grid cell size is determined by the frictional scaling, $\Delta x^+$
  and $\Delta z^+$, and strongly anisotropic cells, with first $\Delta y^+ \sim 1$, thus aiming
  for wall-resolving LES.
  Results are compared to direct numerical simulations (DNS) and several quality measures
  are investigated, including the error in the predicted mean friction velocity and the
  error in cross-channel profiles of flow statistics.
  To reduce the total number of channel flow simulations, techniques from the framework of
  uncertainty quantification (UQ) are employed. In particular, generalized polynomial chaos
  expansion (gPCE) is used to create meta models for the errors over the allowed parameter ranges.
  The differing behavior of the different quality measures is demonstrated and analyzed.
  It is shown that friction velocity, and profiles of velocity and the Reynolds stress tensor,
  are most sensitive to $\Delta z^+$, while the error in the turbulent kinetic energy is
  mostly influenced by $\Delta x^+$.
  Recommendations for grid resolution requirements are given, together with quantification of
  the resulting predictive accuracy.
  The sensitivity of the results to subgrid-scale (SGS) model and varying Reynolds number
  is also investigated. All simulations are carried out with second-order accurate
  finite-volume based solver \revCom{OpenFOAM}. 
  \revCom{It is shown, the choice of numerical scheme for the convective term significantly influences the error portraits.}
  It is emphasized that the proposed methodology, involving gPCE, can be applied to other modeling approaches\revCom{, i.e. other numerical methods and choice of SGS model}.
  
\vspace{0.4cm}
{\footnotesize The following article has been accepted by \emph{Physics of Fluids}. After it is published, it will be found at \url{https://aip.scitation.org/journal/phf}. Copyright 2018 Saleh Rezaeiravesh and Mattias Liefvendahl. This article is distributed under a Creative Commons Attribution (CC-BY-NC-ND 4.0) License.}   
\end{abstract}


\maketitle

\section{Introduction}
Scale-resolving simulation of turbulent flows is very challenging due to the wide range of spatial and temporal scales, as is well known.
Particular difficulties arise for wall-bounded turbulent flows, which are associated with the special structure of the turbulence in boundary layers.
The inner part of the turbulent boundary layers (TBLs) contains energetic strongly-anisotropic flow structures which produce peaks in turbulent quantities such as production, dissipation, kinetic energy, and the Reynolds stress,~\cite{jimenez:pof13,pope}.
Furthermore, these structures scale with the inner (frictional) scales and their ratio to the outer scales decreases with Reynolds ($\rey$-)number.
This implies, among other things, very unfavorable scaling of the grid requirements with the $\rey$-number for wall-resolving large eddy simulations (WRLES),~\cite{chapman79,choi12,saleh:2,LF:3}. 
Thus, it is crucial to construct an appropriate computational grid for such simulations, and to strike the right balance between accuracy and computational cost.

The main aim of the present study is to provide extensive information concerning how the accuracy of the simulation results depends on the grid resolution.
\rev{In general, different sources of error and uncertainty\footnote{According to \cite{oberkampf,smith}, we would refer to a "recognizable" deficiency as error compared to uncertainty which indicates a "potential" discrepancy mainly due to the lack of knowledge or information. Uncertainty may, for instance, be due to the choice of a model to describe a phenomenon or due to influential parameters.} interact, propagate into the simulations, and eventually contaminate the results.}
\rev{Here, epistemic uncertainties are considered which systematically influence the results, however, they can be reduced and even be ideally removed, see~\eg~\cite{smith}.}
Recently, there has been an increasing emphasis in computational fluid dynamics (CFD) on uncertainty quantification (UQ) of the results, see~\eg~\cite{najm,montomoli}.
The focus of the present study is on \rev{employing appropriate techniques developed in the UQ framework to scrutinize} the grid dependence of incompressible WRLES of turbulent channel flow.

First, \rev{to develop the underlying ideas,} some concepts are introduced in the context of direct numerical simulation (DNS). 
For a set of target flow conditions, \rev{(e.g. target $\rey$-number)}, the purpose of the simulation can be to compute a quantity of interest (QoI), $\cY_{\rm DNS} (\varphi_H,\ldots)$, where $\varphi_H$ represents the discrete values of the solution variables.
This QoI can be \eg~the mean wall shear stress in the channel, or the profile across the channel of a component of the Reynolds stress tensor.
The error in $\cY_{\rm DNS}$ depends on both numerical errors and errors associated with the finite simulation time used to calculate statistics.
In the context of the present paper, however, we will use DNS results for channel flow~\cite{iwamoto02,lee15} as a reference \rev{or ``true" data to which the LES results are compared}. 

The governing LES equations are derived by formally filtering the incompressible Navier-Stokes, see~\eg~\cite{sagaut}.
In the case of a fixed filter kernel $G$, and a field $\varphi$, this operation can be expressed by the following spatial convolution integral,
\begin{equation} \label{eq:filter}
\bar{\varphi}(\mathbf{x},t) = \int \varphi(\bm{\xi},t) G(\mathbf{x}-\bm{\xi}) \mbox{d}V_{\xi}.
\end{equation}
Here, bold letters indicate vectors, $t$ denotes time, and the integral is a volume integral.
Applying filtering to the incompressible Navier-Stokes equations, and assuming that it commutes with differentiation, the following equations are obtained.
\begin{equation}\label{eq:LEScont}
\begin{aligned}
\grad \cdot \mathbf{\bfu} &= 0 \,,  \\ 
\frac{\partial \bfu}{\partial t} + (\mathbf{\bfu}\cdot \grad )\mathbf{\bfu} &= -\frac{1}{\rho}\grad \bar{p} + \grad \cdot \bar{\sigma} - \grad \cdot \sgs\,.
\end{aligned}
\end{equation}
Here, $\sgs=\overline{\mathbf{u}\mathbf{u}} - \bfu\bfu$, is the \rev{unresolved stress tensor}, $\bfu$ is the filtered velocity field, $\rho$ is the density, $\bar{p}$ is the filtered pressure, and $\bar{\sigma}$ is the filtered viscous stress.
In the remainder of the paper, the overbar notation is, for convenience, dropped for filtered quantities.

The filtered equations~(\ref{eq:LEScont}) are not closed, i.e.~there are more unknowns than equations. 
In order to close the equations, some form of modeling for $\sgs$ must be applied. 
The closed equations are then discretized by some numerical method, and solved. 
\rev{For the same set of target flow conditions as reference DNS, the QoI of LES can be expressed as $\cY_{\rm LES} (\varphi_h,\cdots)$, where $\varphi_h$ represents discrete solution variables.}
\rev{These QoIs are influenced by both modeling errors and numerical errors.}
\rev{Accounting for the errors in LES is more involved, since} the filtering and the discretization can be dependent and interact.
If the filtering operation is explicit \rev{(explicitly-filtered LES), both the computational grid resolution and the filter width can independently be chosen to control the accuracy of simulations.}
\rev{However,} the computational grid must be finer than the filter width in order to accurately compute the filtering, \rev{see~\eg~\cite{chow03}}.
\rev{In contrast,} it is more common to associate the filtering implicitly with the numerical method used for discretization, see \sect~\ref{sec:LESCFD} below for how this is done in the finite volume framework of the present study.
\rev{In implicitly-filtered LES, the filter width depends on the grid spacing, so consequently numerical and subgrid-scale (SGS) modeling errors (due to modeling of $\sgs$) are intertwined and can be both controlled by grid resolution.}
For more details and discussions of the quantification and the possibility of separation of modeling and numerical errors for LES, reader is referred to \cite{geurts:02,gullbrand:02,pope04,klein:05,celik:05}.

A further complication is that an important component of most models to close \rev{SGS stress tensor $\sgs$} is a dissipative effect, and at the same time most numerical schemes introduce a stabilizing artificial dissipation. 
Inspired by this, the approach of letting the numerical dissipation act as \rev{SGS stress tensor} has been developed, \cite{GMR:1}, which is referred to as implicit LES.

\rev{Due to the mentioned complications, studying the associated effects of each type of errors affecting the LES results is not a trivial task.
Despite this, there have been attempts to develop error assessment techniques, see \cite{lesReli1,lesReli2,geurts:06,celik:06}. 
In particular in the framework of implicitly-filtered LES, Celik \et~\cite{celik:05} derived quality assessment indices in terms of grid cell size, employing Richardson extrapolation technique. 
Klein \cite{klein:05} extended this work to derive error estimates for both numerical and modeling errors. 
However, as it is demonstrated by the systematic study of Meyers and Sagaut, \cite{meyers07}, dealing with the error between LES and DNS of channel flow, as an example of wall-bounded turbulence, is challenging.}

The general approach of the present study \rev{to investigate effect of grid resolution} is similar to that of Meyers and Sagaut, \cite{meyers07}. 
A-posteriori error analysis is carried out by comparing channel flow LES to reference DNS data, and systematic variation of grid resolution in the spanwise and streamwise directions is carried out in the simulation campaign.
\rev{The distance of the first off-wall cell center is chosen such that the requirement of wall-resolving LES is fulfilled.}
This work however significantly extends \cite{meyers07} in a number of directions. 
The UQ framework, described in \sect~\ref{sec:approach}, allows for a significantly smaller number of LES computations for the mapping of the parameter range.
\rev{The error between $\cY_{\rm LES}$ and $\cY_{\rm DNS}$} is investigated for several quantities of interest, including the mean wall shear stress, \rev{$\tau_w$}, and cross-channel profiles of the mean velocity, root-mean-square (rms) velocity fluctuations, and Reynolds stress tensor.
For each quantity, a meta model is constructed which represents the dependence of the error on the grid resolution.
The resulting ``error portraits'' clearly illustrate the challenges that any theoretical error estimation procedure must deal with. 
\rev{As a complement, a variance-based sensitivity analysis is also performed to quantify the influence of the grid resolution on the errors.}

The importance of considering several quantities is demonstrated by the fact that, for a particular grid resolution, the error may be close to zero for \eg~the mean wall shear stress, while certain profiles show clear discrepancies when compared to DNS. 
The differing behavior for different quantities is analyzed.
The bulk of the simulation campaign is carried out at the friction-based Reynolds number,~$\reyt=300$, \rev{where $\reyt=\ut\delta/\nu$, with wall friction velocity defined as $u_\tau=\sqrt{\tau_w/\rho}$, $\delta$ denoting the channel half-height, and $\nu$ representing fluid kinematic viscosity}. In order to check the sensitivity of the results to $\rey$-number, a set of LES is also computed at the higher $\rey_\tau=550$.
Furthermore, the sensitivity of the results to SGS model is also investigated.
Naturally, the results may be affected by the particular SGS modeling and numerical method employed. 
It is however believed that the conclusions are relatively robust since the grid resolution requirements are strongly connected to the length scales of the flow.
It is emphasized that the proposed method, with generalized polynomial chaos expansion (gPCE), \cite{xiu_gPCE,ghanem:91}, and several quantities of interest, should be also applied to other LES methods to complement the results of the present study.

The paper is organized as follows. 
In \sect~\ref{sec:approach}, the CFD method and the UQ techniques, as well as the non-intrusive linking between the two are described.
The simulation campaign and the pre-processing, in particular grid generation, are described in \sect~\ref{sec:simCase}.
All results and analyses are presented in \sect~\ref{sec:results}.
The starting point is constructing the meta models, depending on $\dxp$ and $\dzp$, for the error in the friction velocity at $\reyt=300$.
In subsequent sections, different aspects are developed and the effect of variation in  additional parameters are considered.
Complementary information concerning the convergence of gPCE (with polynomial order), and flow statistics (with the length of the simulation time-averaging interval), are collected in \app~\ref{sec:appendix}.
The most important findings are summarized in \sect~\ref{sec:concl}, where recommendations are also given concerning grid resolution, together with quantification of the resulting errors.


\section{Approach}\label{sec:approach}
\subsection{Non-intrusive parameter study}\label{sec:uqcfdLink}
\begin{figure}[!htbp]
\centering
\includegraphics[scale=.95]{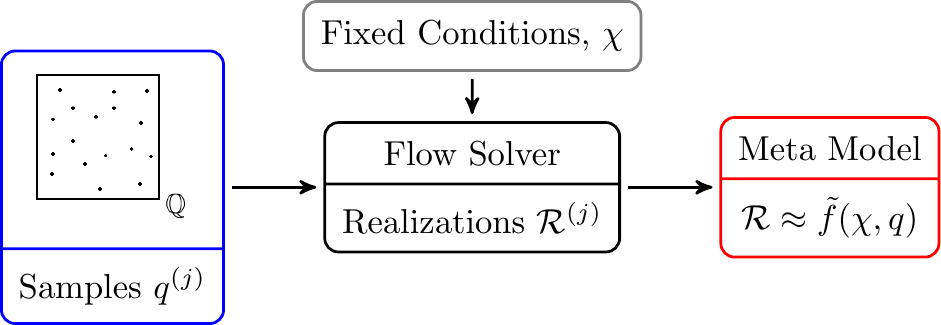}
\caption{Schematic representation of the non-intrusive construction of a meta (surrogate) model from responses of a flow solver, within a UQ forward problem.}\label{fig:uqFWDSchem}
\end{figure}

Within the framework of UQ, the main aim of the forward problem is to study how the uncertainties in parameters propagate into a model response.
To formulate a forward problem in the context of the present study, consider a flow solver is available to simulate a certain type of flow for a given set of fixed conditions $\chi$ \rev{(\eg~target $\rey$-number, domain size, $\cdots$)}, and a set of uncertain inputs and parameters, $q$.
Note that, those parameters whose effects on the responses aimed to be studied are assumed to be uncertain and random, even though they are not actually so, \cite{smith}.
By running the solver for combinations of fixed $\chi$, and sample $q^{(j)}$ taken from the \rev{prescribed} parameters admissible space, $\BQ$, a set of quantities are obtained which are, in turn, used to evaluate associated response, $\cR^{(j)}$, see \fig~\ref{fig:uqFWDSchem}.

\rev{In this study, the (normalized) error between the QoIs of LES and DNS, comprise the responses, as detailed in \sect~\ref{sec:Rdefs}.}
Once enough number of realizations of the responses is obtained, a function can be constructed to parametrize the model responses in terms of the fixed and uncertain parameters as, 
\begin{equation}\label{eq:RModelFunc}
\cR=f(\chi,q) \,.
\end{equation}
Depending on the particular approach adopted to tackle the forward problem, a specific structure for the functional form of $f$ is considered, see~\eg~\cite{smith}. 
Nevertheless, in any case, due to taking limited number of samples for $q$, constructing an exact function for $f$ might not be feasible and instead a surrogate or meta model, $\tilde{f}$, can be obtained. 
The specific approach taken in the current study to construct the meta model is discussed in \sect~\ref{sec:uq}. 
This is then followed by the details of the LES solver in \sect~\ref{sec:LESCFD}.

\subsection{UQ approach}\label{sec:uq}
Consider $p>1$ mutually independent parameters $q$. 
In a UQ forward problem, uncertain $q$ are allowed to vary over a presumed admissible space $\mathbb{Q}\subset \BR^p$, that is comprised of the admissible ranges of the parameters, $\BQ=\prod_{i=1}^p \BQ_i$.
Further assume, each admissible space $\BQ_i$ can be mapped one-to-one to $\Gamma_i=[-1,1]$, so that, there is a one-to-one correspondence between sample $q\in \BQ$ and $Q\in \Gamma =\prod_{i=1}^p \Gamma_i$.

In particular, the generalized polynomial chaos expansion (gPCE), \cite{xiu_gPCE,ghanem:91}, is chosen to construct a surrogate for the model response $f$ as,
\begin{equation}\label{eq:pce}
\tilde{f}(\chi,Q)=\sum_{k=0}^N \hat{f}_{k}(\chi)\Psi_{k}(Q) \, ,
\end{equation}
in which, the whole uncertainty in the function (due to the uncertain parameters) is to be totally expressed by orthogonal bases, $\Psi_k(Q)$, while the fixed parameters (deterministic effects) are enclosed in the expansion coefficients $\hat{f}_k(\chi)$.
These deterministic coefficients are determined by,
\begin{equation}\label{eq:pceCoeff}
\hat{f}_k(\chi)=\frac{1}{\gamma_k}\BE[\tilde{f}(\chi,Q)\Psi_k(Q)] \,,
\end{equation}
where, $\gamma_k = \BE[\Psi_k(Q) \Psi_k(Q)]$.
For any arbitrary $g(Q)$, the expectation is obtained from,
\begin{equation}\label{eq:pceInteg}
\BE[g(Q) \Psi_k(Q)]=\int_\Gamma g(Q) \Psi_k(Q) \rho (Q) \dd Q \,,
\end{equation}
with $\rho (Q)=\prod_{i=1}^p \rho_{i} (Q_i)$ specifying the joint density of the mapped parameters.

In the non-intrusive use of the gPCE, see~\eg~\cite{najm,mariotti17} for review and similar type of application as the current work, $n$ samples must be taken from $\BQ$.
This can specifically be carried out by generating $n_i$ deterministic samples, hereafter called collocation points, from $\BQ_i$, for $i=1,\cdots,p$, which results in $n=\prod_{i=1}^p n_i$ collocation points, in total, throughout $\BQ$.
Upon using tensor-product method to handle the multi-dimensionality of the parameter space, the upper bound in summation (\ref{eq:pce}) becomes $N=n-1$. 
For this particular choice, $\Psi_k(Q)=\prod_{i=1}^p \psi_{k_i}(Q_i)$ in which, $k$ is a unique re-index associated with any combination of $\{ k_i\}_{i=1}^p$. 

For each joint sample $q^{(j)}$, $j=1,2,\cdots,n$, a realization of the system response, $\cR^{(j)}$ is evaluated. 
By assuming $\cR^{(j)}$ to be approximately equal to $\tilde{f}(\chi,q^{(j)})$, the coefficients $\hat{f}_k(\chi)$ can be determined either by solving an $n$-by-$n$ linear system or equivalently, by evaluating integrals (\ref{eq:pceInteg}) by using a quadrature technique in which the mapped collocation points $\{ Q^{(j)} \}_{j=1}^n$ are taken to be the quadrature points.

The main advantage of expansion (\ref{eq:pce}) is that the basis functions are orthogonal with respect to $\rho(Q)$, i.e. $\BE[\Psi_i(Q) \Psi_j(Q)]=\delta_{ij}$. 
This necessitates adoption of appropriate bases for a specific type of distribution of the random parameters, as explained in \cite{xiu_gPCE}. 
In this regard, if all parameters $Q$ have uniform distributions over $\Gamma$, then $\rho(Q)=1/2$, and the Legendre polynomials are the proper choice to preserve the orthogonality requirement. 
If in addition to this, Gauss quadrature points are used as the deterministic samples $Q_i$ in $\Gamma_i$, then the application of expansion (\ref{eq:pce}) becomes equivalent to the classical Gauss-Legendre technique.

Based on the discussion, it is clear that, by construction, expansion (\ref{eq:pce}) at collocation points $Q^{(j)}$ returns the value of observed $\cR^{(j)}$ associated with $q^{(j)}$. 
Once the expansion coefficients, $\hat{f}_k(\chi)$ are determined, expansion (\ref{eq:pce}) acts as a surrogate or meta model of the real model function $f(\chi,q)$ in (\ref{eq:RModelFunc}) that is not identifiable at least due to the excessive computational cost. 
From this aspect, meta model (\ref{eq:pce}) can be used to predict response $\cR$ at any $q\in\BQ$ else than the original collocation points.

The natural question arising here is about the accuracy of such predictions. 
Since, the functional form of the real model response $f(\chi,q)$ is unknown, a-posteriori evaluation of the accuracy is only possible by comparing the predictions made by the meta model $\tilde{f}(\chi,q)$ and the evaluations of the real model $f(\chi,q)$, at a limited number of samples $q^*\in \BQ$. 
One of the factors which influences the quality of the meta model (\ref{eq:pce}) is the maximum polynomial order in the expansion which directly depends on the number of samples, $n$, used to construct the meta model. 
Generally speaking, the higher the $n$, the better the accuracy of (\ref{eq:pce}) is. 
However, in practice, the maximum polynomial order is chosen as a compromise between the computational cost, required to evaluate exact realizations of the responses when constructing the meta model, and the accuracy of expansion (\ref{eq:pce}) in predicting the approximate response surfaces.
In connection to the present study, \sect~\ref{sec:errRates} and \app~\ref{sec:pceConv} are devoted to assessing the accuracy of the meta model.

\subsection{Computational fluid dynamics methods} \label{sec:LESCFD}
The simulations of the present study have been carried out using
version 3.0 of the open source CFD software package \of\footnote{\texttt{www.openfoam.com}},
see~\cite{MaHM:1} and the references therein.
The spatial discretization of the governing equations is carried out with the finite volume (FV) method.
The value of an unknown $\varphi$ (the pressure or a component of the velocity field), associated
with the finite volume cell $V_i$, is thus defined by the volume-average of the quantity over the
cell,
\begin{equation} \label{eq:fvm}
\varphi_i(t) = \frac{1}{V_i}\int_{V_i} \varphi(\mathbf{x},t)\,\mbox{d}V_{\mathbf{x}}.
\end{equation}
The notation $V_i$ is used here both for the cell and its volume.
Note that, to second-order accuracy, this definition is equivalent to using
the value of the field at the cell center position, $\mathbf{x}_i$, of the $i$-th cell, i.e.,
$$
\varphi_i(t) =  \varphi(\mathbf{x}_i,t) + \mathcal{O} (h^2)\,,
$$
where, $h$ is a suitable length scale of the finite volume cell.
Comparing equation~(\ref{eq:fvm}) to the definition of LES filter in equation~(\ref{eq:filter}),
it is seen that the finite volume formulation is equivalent to a filter kernel $G$ which
is equal to one in the cell $V_i$ and zero elsewhere, i.e. a Heaviside function.
Thus, the FV-framework provides an implicit LES filtering.

The computational domain is divided into non-overlapping finite volume cells which, generally,
can be of arbitrary polyhedral shape. 
As described in \sect~\ref{sec:simCase}, only
block-structured grids with (orthogonal) hexahedral cells are used in the present study.
All approximations involved are second-order accurate.
Gauss' theorem is applied to transform a number of terms, in the conservation
form of the momentum equation, from volume integrals to surface integrals.
Linear interpolation is applied to obtain face-center values of the fields,
for the evaluation of the surface integrals, from the cell-center values.
\revCom{The resulting numerical scheme is referred to as linear.}

The spatial discretization of the momentum equation
leads to a large system of ordinary differential equations which are solved
using a second-order backward-difference method (implicit).
The momentum equation is treated in a segregated manner, solving
sequentially the components of the momentum equation.
The coupling between the pressure and momentum equations is handled
using the PIMPLE method which is based on the SIMPLE (Semi-Implicit Method
for Pressure-Linked Equations), see~\cite{Patankar:1},
and the PISO (Pressure-Implicit with Split Operator), \cite{issa86}, methods.

For more detailed descriptions of the techniques described above,
see~\cite{FP:1} and~\cite{MaHM:1}, and the references therein.
A domain decomposition technique, applied to the grid,
in combination with an efficient MPI implementation is used
for running on parallel computers.
In the bulk of the simulation campaign no explicit SGS model
for LES is used.
For one set of simulations, see \tab~\ref{tab:caseSummary} below,
the WALE (Wall-Adapting Local Eddy viscosity) model~\citep{nicoud:99}
was employed in order to investigate the sensitivity of the results to
SGS modeling.

\subsection{Definition of the Responses}\label{sec:Rdefs}
The main aim of the present work is to assess the quality of the LES.
To this end, the responses $\cR$ in (\ref{eq:RModelFunc}), are chosen to be the errors between the QoIs computed by LES and the corresponding DNS values. 
\rev{As it will be discussed in \sect~\ref{sec:simCase}, the QoIs are taken to be the mean (averaged in both time and homogeneous directions) wall friction velocity, and the cross-channel profiles of mean velocity, rms velocity fluctuations, and Reynolds stress of channel flow.}
To normalize the error in quantity of interest $g$, a DNS reference value specified by superscript circle, is used. 
The normalized errors, $\epsilon$, can adopt different forms as discussed below.

If $g$ is scalar, the error is defined as,
\begin{eqnarray*}
\epsilon[g]=(g-g^\circ) / g^\circ \,.
\end{eqnarray*}
If, instead, $g$ is a vector (e.g. discrete \rev{cross-channel} profile \rev{of flow statistics}) over $y$, for $y\in \Omega=[a,b]$, with $a$ and $b$ being two constant real numbers, different forms for error can be defined, including
\begin{eqnarray*}
\epsilon_p[g] = \| g-g^\circ \|_p / \|g^\circ\|_p \,,
\end{eqnarray*}
for $p=1,2,\cdots$, where $\|\cdot\|_p$ specifies ${\rm L}_p$ norm, i.e.,
$$
\|g\|_p = \left [\int_a^b \left| g(y)\right|^p \dd y \right]^{1/p} \,.
$$
By H{\"o}lder's inequality, see~\eg~\cite{kreys}, it can be shown that $\eabs[g]\leq c_1 \el[g]$ for $c_1=\sqrt{b-a}\|g^\circ\|_2/\|g^\circ\|_1 \geq 1$. 
The maximum difference between the LES and DNS profiles can be measured by,
\begin{eqnarray*}
\einf[g] =\|g-g^\circ\|_\infty / \|g^\circ\|_\infty \,,
\end{eqnarray*}
where, $\|g\|_\infty = \max_{\substack{y\in \Omega}}|g(y)|$.
It is straightforward to derive $\eabs[g]\leq c_2 \einf[g]$ for $c_2=(b-a) \|g^\circ \|_\infty/\|g^\circ\|_1 \geq 1$.

In addition to these error measures, two other errors will be employed in the next sections. 
One expresses the difference between the average values of the LES and DNS profiles over $\Omega$ (or equivalently, the difference between the areas under the LES and DNS profiles when plotted against $\Omega$), 
$$
\earea[g] = \left[ \int_a^b \left|g(y)\right|-\left|g^\circ(y)\right| \dd y \right] 
\left[ \int_a^b \left|g^\circ(y)\right| \dd y\right]^{-1} \,.
$$
The other measure returns the normalized error between the peaks of LES and DNS profiles,
$$
\epik[g] =\left[ \max_{\substack{y\in \Omega}}(g(y))-\max_{\substack{y\in \Omega}}(g^\circ(y)) \right]
 \left[ \max_{\substack{y\in \Omega}}(g^\circ(y)) \right]^{-1} \,.
$$


\section{Simulation Case and Pre-processing}\label{sec:simCase}
The methodology described in the previous section is employed to study effects of variation of grid resolution in different directions, on the computed quantities of turbulent channel flow. 
For this purpose, other potentially influential factors are kept invariant, \rev{as described below}.

As summarized in \tab~\ref{tab:caseSummary}, channel flow simulations at two target Reynolds numbers $\reyt=300$ and $550$ are considered, with corresponding DNS data of Iwamoto \et~\cite{iwamoto02}, and Lee and Moser \cite{lee15}, respectively.
These DNS data are used as reference values in the errors defined in \sect~\ref{sec:Rdefs}. 
Moreover, $\reyt^\circ$, the value of $\reyt$ achieved by DNS, is used to express the grid spacings in wall-units, i.e. $\Delta^+ = \Delta/\delta \reyt^\circ$, where $\delta$ denotes the channel half-height. 

When setting up the LES simulations, channel flow bulk velocity $U_b$, kinematic viscosity $\nu$, and $\delta$ are chosen such that the resulting bulk $\rey$-number, $\rey_b=U_b\delta/\nu$, is equal to the value of the reference DNS.

The base lengths of the domain in the streamwise and spanwise directions are considered to be respectively equal to $l_x=2.5\pi \delta$ and $l_z=\pi \delta$ for $\reyt=300$, and $l_x=9 \delta$ and $l_z=4 \delta$ for $\reyt=550$.
Given a grid cell size, these lengths may be required to be slightly modified in order to be discretized by integer number of cells. 

For each channel flow simulation, grid cells are equi-spaced in the streamwise and spanwise directions.
In the wall-normal direction, a value for the inner-scaled distance of the first cell center from the wall, $\dyp_w$, is chosen first \rev{according to the requirement of wall-resolving LES.}
The other grid spacings are then generated from the following function,  
\begin{equation}\label{eq:delY}
\Delta \eta = \min\left[ \max\left( \frac{\dyp_w}{\reyt^\circ},\eta_{\rm in} \eta \right),\frac{1} {\rm M} \right] \, ,
\end{equation}
where, $\eta=y/\delta$ is the normalized distance from the wall, $\eta_{\rm in}$ denotes the beginning of the outer part of the turbulent boundary layer that is taken to be fixed and equal to $0.1125$, and finally $\rM$ specifies the number of equi-spaced cells considered in the outer layer of the TBL (for these points $\dyp=\reyt^\circ/\rM$).
More details regarding function (\ref{eq:delY}) can be found in \cite{saleh:2}.

\begin{table}[!htbp]
\centering
\caption{Summary of the channel flow simulation campaign. For all simulations, $10\leq \dxp\leq 150$ and $7\leq\dzp\leq 70$, where $\Delta^+ = \Delta/\delta \reyt^\circ$.}\label{tab:caseSummary}
\begin{small}
\begin{tabular}{*{5}c}
\toprule\toprule
Simulation Set & Target $\reyt$ & $\reyt^\circ$ & $\dyp_w$  & SGS Model \\
\hline
Set-A & 300 & 297.899, \cite{iwamoto02} & 0.445 & No explicit model\\
Set-B & 300 & 297.899, \cite{iwamoto02} & 1.105 & No explicit model\\
Set-C & 300 & 297.899, \cite{iwamoto02} & 1.765 & No explicit model\\
Set-Bw & 300 & 297.899, \cite{iwamoto02} & 1.105 & WALE, \cite{nicoud:99}\\
Set-D & 550 & 543.496, \cite{lee15} & 0.445 & No explicit model\\
\bottomrule
\end{tabular}
\end{small}
\end{table}

In each of the simulation sets in \tab~\ref{tab:caseSummary}, for the fixed conditions ($\chi$ in \sect~\ref{sec:approach}) that include domain size, numerical schemes, target $\reyt$, SGS model, and $\dyp_w$, the grid spacings in the streamwise and spanwise directions, $\dxp$ and $\dzp$, are allowed to take any values within the prescribed admissible ranges $\Qx=[10,150]$ and $\Qz=[7,70]$, respectively.
These ranges are approximately chosen equal to those assumed by Meyers and Sagaut \cite{meyers07}, to ease comparison.

In order to study their impact, $\dxp$ and $\dzp$ are assumed to be uncertain and random within the UQ framework of \sect~\ref{sec:uq}. 
In particular, they are assumed to be uniformly distributed as $\dxp\sim \mathcal{U}[\Qx]$ and $\dzp\sim \mathcal{U}[\Qz]$.
To span each of $\Qx$ and $\Qz$, 5 Gaussian quadrature points along with Legendre polynomials as basis functions are used in expansion (\ref{eq:pce}).
\rev{Associated with each of the sampled grid resolutions, an LES of channel flow is carried out.}
\rev{Consequently,} 25 channel flow simulations for each set in \tab~\ref{tab:caseSummary} are required, in order to evaluate the unknown coefficients in (\ref{eq:pce}).

The fixed values of $\dyp_w$ for sets A, B, and C in \tab~\ref{tab:caseSummary} are chosen equal to three Gaussian quadrature points in range $\Qy=[0.25,1.96]$.
Consequently, by combining sets A, B, and C (hereafter called Set-ABC), the effect of simultaneous variation of $\dxp\sim \mathcal{U}[\Qx]$, $\dzp\sim \mathcal{U}[\Qz]$, and $\dyp_w\sim \mathcal{U}[\Qy]$ on the QoIs of LES of turbulent channel flow with no explicit SGS modeling at target $\reyt=300$, can be studied with 75 independent channel flow simulations. 
This specific choice of the quadrature points results in the combinations of Legendre polynomials up to the orders $4\times 4 \times 2$ to span the error responses in the parameter space of $\dxp \times \dzp \times \dyp_w$.

Corresponding to fixed conditions $\chi$ and any grid resolution, a turbulent channel flow simulation is conducted using \of.
\rev{A constant time step size, $\Delta t$, is explicitly set in each simulation.}
\revCom{
In particular, for all simulations at target $\reyt=300$ and $550$, the time step was chosen as, $\Delta t=10^{-3} \delta/U_b$, and, $\Delta t=5\times 10^{-4}\delta/U_b$, respectively. 
The resulting average value of the CFL number of the simulation with the highest resolution was observed to be between $0.02$ and $0.045$ at $\reyt=300$ and between $0.02$ and $0.06$ at $\reyt=550$.} 
In all simulations, a synthetically-perturbed laminar flow is used as the initial condition, see \cite{deVilliers}. 
Due to the periodic condition in the streamwise direction, a pressure source term is used in the momentum equation to enforce the mass flux corresponding to the prescribed bulk velocity, $U_b$, at each time step. 
The flow undergoes transition to turbulence and consequently when statistically-stationary condition is achieved, averaging in time and homogeneous directions, represented with operator $\langle \cdot \rangle$, is conducted over at least $\approx 115$ flow-through times \rev{($\approx 115 l_x/U_b$)}.
The potential uncertainty in the conclusions due to the length of time-averaging is discussed in \app~\ref{sec:timeAvg}.

\rev{By post-processing the channel flow simulation results and comparing them to the DNS data, different error responses introduced in \sect~\ref{sec:Rdefs} are evaluated.}
To compute the integrals appeared in the errors defined in \sect~\ref{sec:Rdefs} \rev{ over the channel half-height, i.e. over $\Omega=[0,\delta]$}, composite Simpson's rule is employed. 
For this purpose, both LES and DNS profiles are first linearly interpolated to a set of $5000$ equi-distance nodes, distributed between \rev{the wall} and the channel centerline.

To construct the meta model associated to each error response, the coefficients in the expansion (\ref{eq:pce}) are calculated using the open source library, \dak~\cite{dakotaMan}.
The linking between \of~and \dak~is made non-intrusively through a driver, following the schematic depicted in \fig~\ref{fig:uqFWDSchem}.

By excluding and including explicit SGS modeling, respectively corresponding to Set-B and Set-Bw, the numerical and the combined numerical-modeling contributions in the errors can be identified.
Neglecting explicit SGS modeling is based on the assumption that the numerical errors can mimic the effects which would otherwise be induced to the flow by a SGS model. 
From this point of view, the simulations with no explicit SGS modeling can be seen as coarse DNS (see~\eg~\cite{meyers07}) conducted by a nominally 2nd-order accurate FV-based solver.

\section{Results and Discussion}\label{sec:results}
In the following sections, the errors in different quantities of the channel flow, \rev{associated with the} simulation sets of \tab~\ref{tab:caseSummary}, are thoroughly discussed.
The main focus of each section is tried to be kept on a particular aspect of the results.

For sake of brevity and clarity, by error in a channel flow quantity, a relevant form of the normalized error $\epsilon$ defined in \sect~\ref{sec:Rdefs} is meant.
When computing the errors, the average value of the QoIs of both halves of the channel is considered. 
  
The error isolines \rev{(or equivalently, error response surfaces, or, ``error portraits")} are constructed in the admissible space of the grid cell spacings using the meta model (\ref{eq:pce}).
\rev{Hereafter, the admissible space of $\dxp$ and $\dzp$ is referred to as the $\dxp\dash\dzp$ plane.}
\rev{In this plane, the quadrature points, each associated with a channel flow simulation, are specified by solid markers.}
Unless otherwise mentioned, the values of the labels on the error isolines are expressed in percentage.
When the error between the LES results for a specific quantity and the corresponding DNS value tends to zero, we use the word  ``convergence".

In the following discussions, the QoIs of the channel flow are non-dimensionalized. 
In particular, wall friction velocity, $\ut$, and the mean streamwise velocity profile $\U$ are non-dimensionalized by $U_b$.
Similarly, for non-dimensionalizing the second-order statistical moments of velocity, $U_b^2$ is considered. 
For convenience, the word ``non-dimensional" is dropped in the following sections.

\subsection{Relative importance of the grid spacings in different directions}\label{sec:ResComp3Resolut}
\begin{figure}[!htbp]
\centering
\begin{tabular}{ccc}
   \includegraphics[scale=0.26]{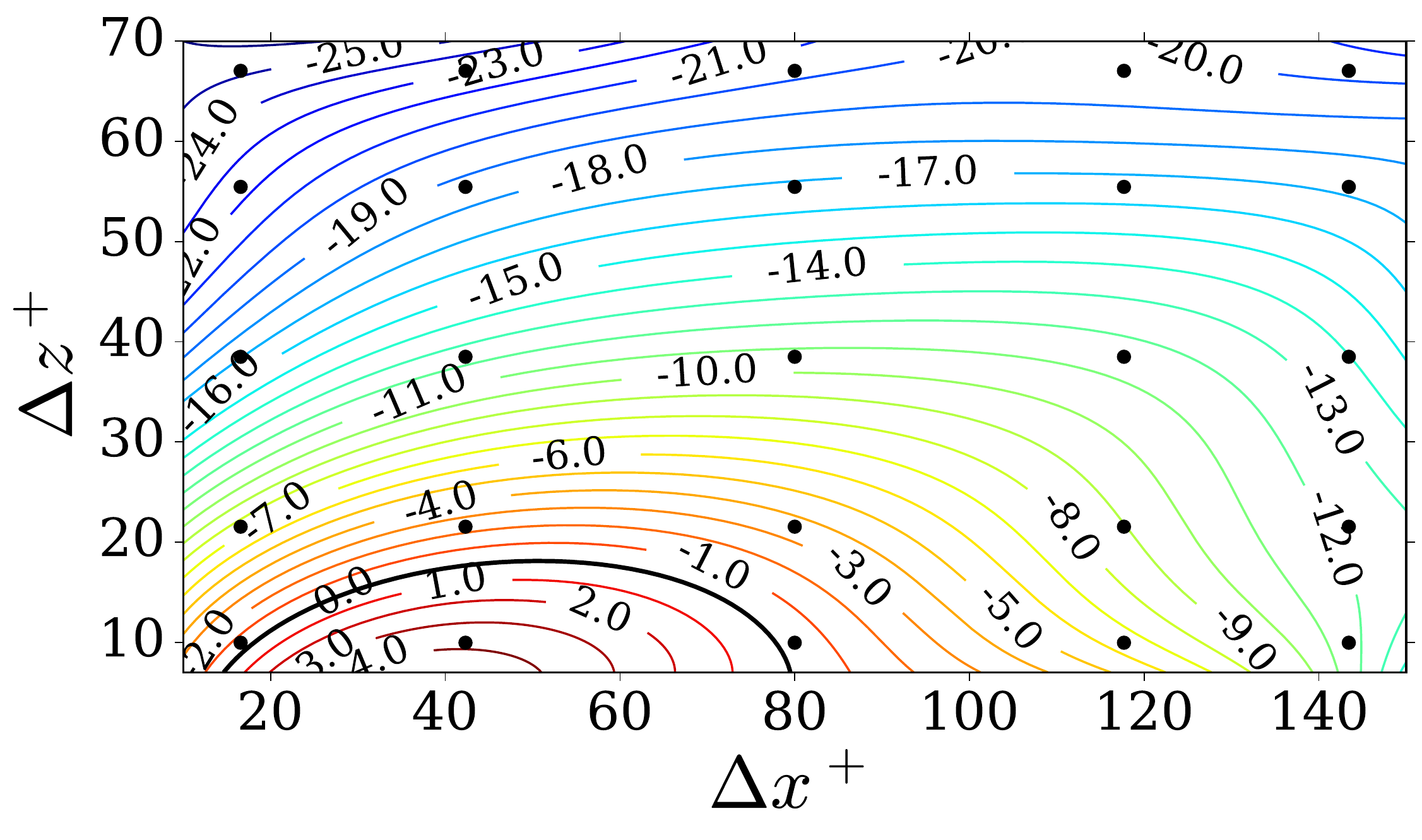} &
   \includegraphics[scale=0.26]{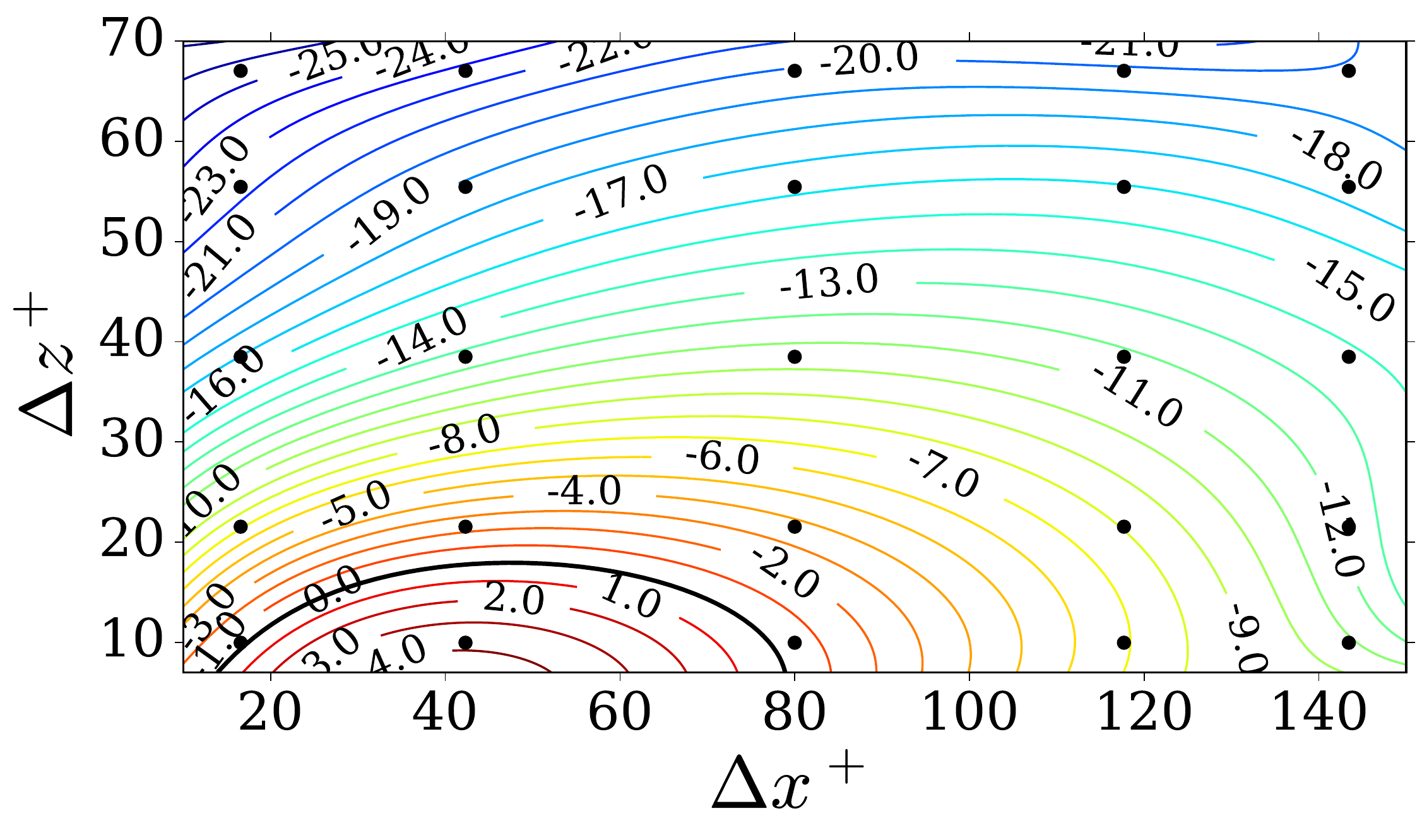} &
   \includegraphics[scale=0.26]{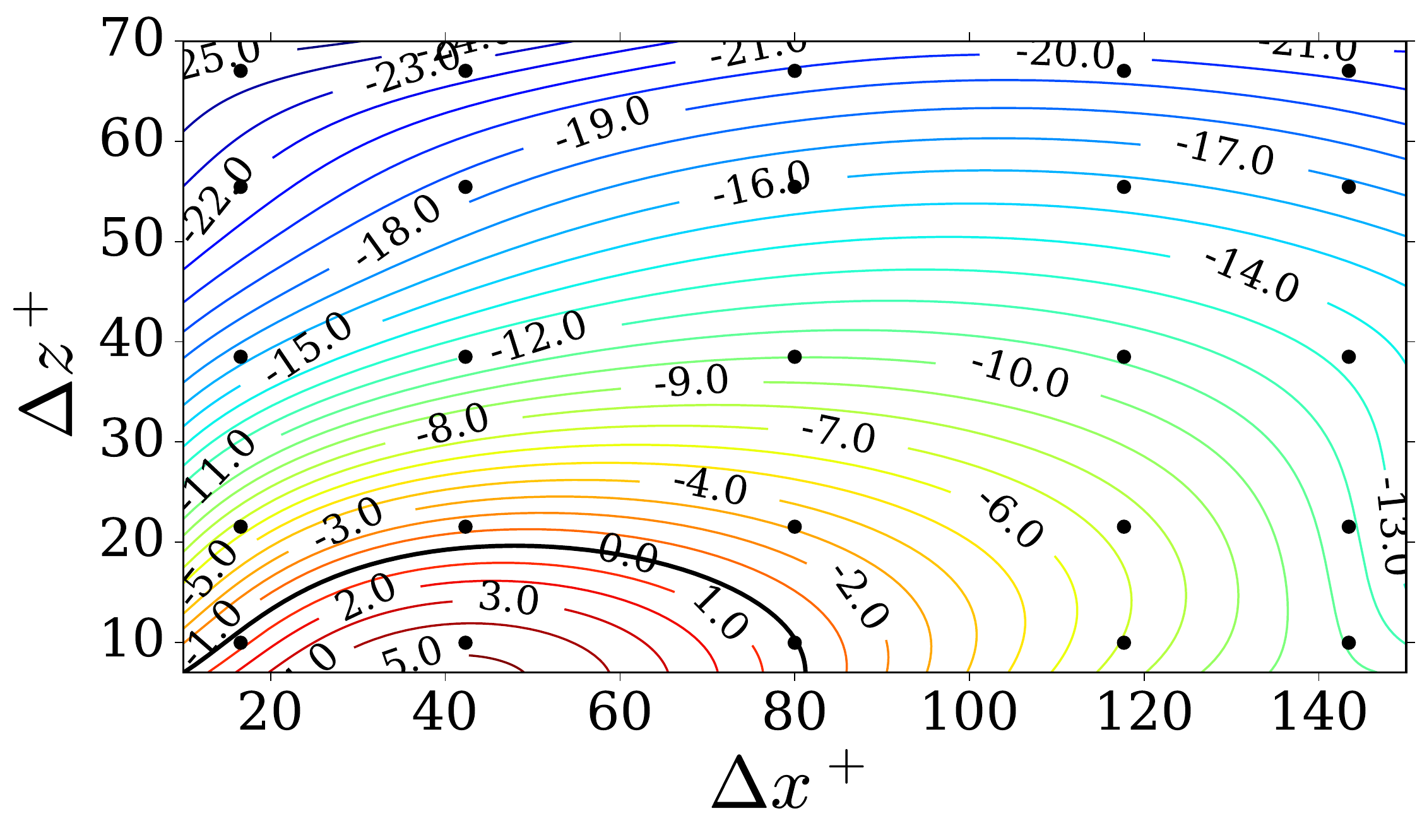} \\
   {\small (a)} & {\small (b)} & {\small (c)} \\
\end{tabular}
\caption{Isolines of $\epsilon[{\ut}]$ in the $\dxp\dash\dzp$ plane at target $\reyt=300$, and for $\dyp_w=$ 0.445 (Set-A) (a), 1.105 (Set-B) (b), and 1.765 (Set-C) (c).}\label{fig:ABC_duTau}
\end{figure}

The focus is on Set-ABC in which the target $\reyt$ is 300, and no explicit SGS model is used. 
Thus, the effects of simultaneous variation of $\dxp$, $\dyp_w$, and $\dzp$ over their associated admissible ranges can be studied.
\fig ~\ref{fig:ABC_duTau} illustrates the isolines of the error in the wall friction velocity, $\epsilon[\ut]$, in the $\dxp\dash\dzp$ plane for three different values of $\dyp_w$. 
\rev{The wall friction velocity is computed as, $\ut=\sqrt{\rey_b^{-1} \left(\dudy \right)|_w}$, using the velocity gradient at the wall. 
Note that, $\eta=y/\delta$ is the outer-scaled distance from the wall.}
In all three cases, a locus of zero error in $\ut$ can be identified, represented in black. 
These loci separate two regions over which friction velocities of LES can be over- and under-predicted compared to the DNS.
A simple interpretation is that given $\dyp_w$ and target $\reyt$, there are many possible combinations of $\dxp$ and $\dzp$, for which the LES friction velocity is as accurate as the DNS value.

For $\dyp_w\sim \mathcal{U}[0.25,1.96]$, no significant difference between the isolines of $\epsilon[\ut]$ is observed.
The only difference is that the peak of the zero-$\epsilon[{\ut}]$ curve slightly shifts toward higher values of $\dzp$, as $\dyp_w$ increases from 0.445 to 1.765. 
The less influence of $\dyp_w$ compared to $\dxp$ and $\dzp$, is also observed for other quantities of the channel flow (not shown here), see \sect~\ref{sec:GSA}.

The clear difference between the isolines of $\epsilon[\ut]$ at $\dyp_w=0.445$ with those in the study of Meyers and Sagaut \cite{meyers07}, reveals the relatively expected fact that
employing different numerical schemes and wall-normal grid distribution, while keeping other simulation conditions the same, \revCom{may result} in different error response surfaces \rev{or ``error portraits"}.
\revCom{
Further simulations analyzed in \app~\ref{sec:SetALAm}, demonstrate that the numerical scheme can have a large impact on the error portraits, while the role of grid distribution is found to be much less significant.
}

In particular, Meyers and Sagaut \cite{meyers07} observed that the loci of $\epsilon[\ut] = 0$ could occur for particular combinations $\dzp \geq 37$ and $\dxp \lesssim 82$, while here, for some $\dzp\lesssim 17$ and almost the same range of $\dxp$, accurate prediction of $\ut$ is achieved.
The \rev{non-invariance with respect to method makes} suggesting any \rev{universal} guideline for grid resolutions aimed for minimizing the error between LES results and DNS, difficult.
\rev{However, for the numerical settings used in the present study, see \sect~\ref{sec:LESCFD}, appropriate grid resolution for WRLES of channel flow is suggested in \sect~\ref{sec:gridSuggest}.}

\subsection{Convergence of friction velocity}\label{sec:uTauConv}

\revCom{The characteristics of the simulations with resolutions $\dxp\dash\dzp$ on the $\epsilon[\ut]=0$ curve, can be further investigated.
In~\cite{meyers07}, for such simulations all three components of rms velocity fluctuations are over-predicted {\revProof{near the wall}} compared to DNS. 
This excessive level of fluctuations is argued to be responsible for compensating the loss of fine-scale strain rates due to the coarse resolutions. 
However, such a balancing mechanism for $\ut$ may not be applicable to the present study. 
Because, for the simulations with $\epsilon[\ut]=0$, the
spanwise and wall-normal rms velocity fluctuations are found to be lower than the reference DNS, while the streamwise velocity fluctuations are over-predicted except for sufficiently fine $\dxp\dash\dzp$ resolutions. 
This trend for velocity fluctuations is not uniquely observed for the resolutions associated to $\epsilon[\ut]=0$ and may be the case for other resolutions as well, as discussed in \sect~\ref{sec:yDepQoIs} and also in other studies, see~\cite{celik:05,klein:05,bae:17} and the references therein.
}

\begin{figure}[!htbp]
\centering
\begin{tabular}{cc}
   \includegraphics[scale=0.36]{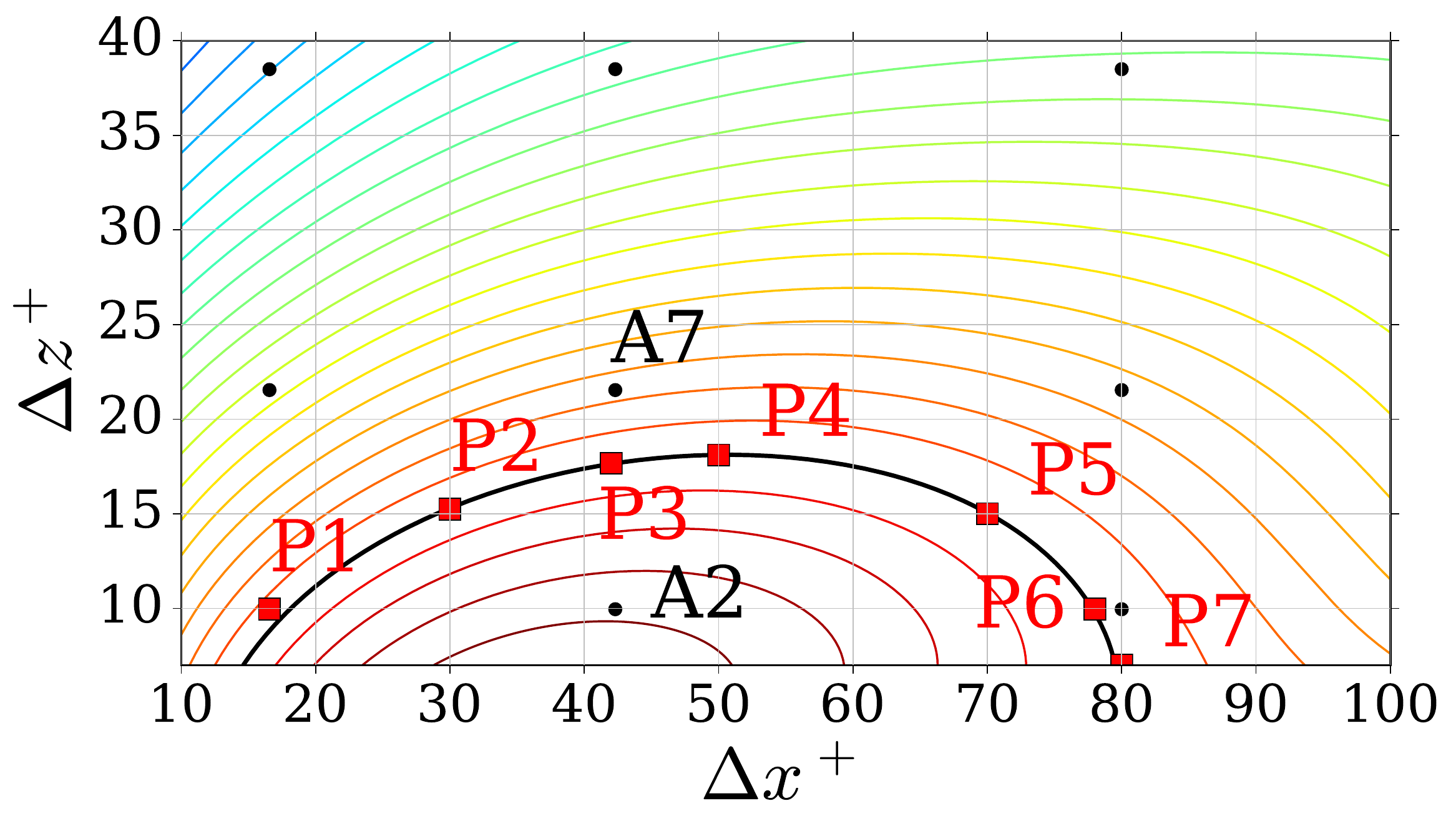} &
   \includegraphics[scale=0.36]{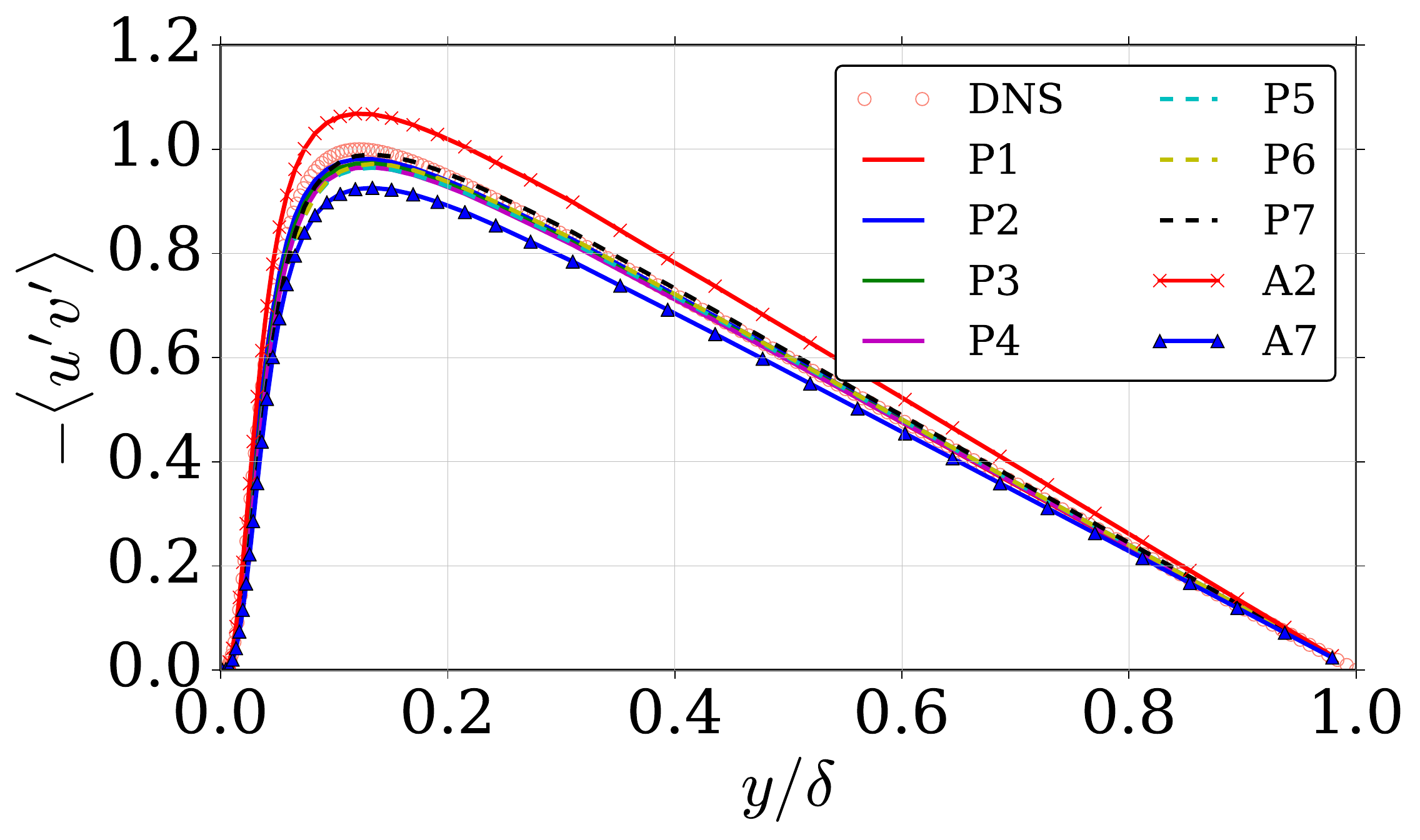} \\
   {\small (a)} & {\small (b)} \\
\end{tabular}
\caption{\revCom{Selected resolutions, see \tab~\ref{tab:integLengths}, on the $\epsilon[\ut]=0$ curve of Set-A (a), with corresponding $-\uv$ profiles accross the channel half-height (b). DNS data of~ \cite{iwamoto02} are used as the reference.}}\label{fig:AuTauLoci}
\end{figure}

\revCom{
It is interesting to note that for all simulations corresponding to $\epsilon[\ut]=0$, the profiles of Reynolds stress $\uv$ are approximately similar to each other and close to (but not the same as) the reference DNS. 
Moreover, the over- and under-prediction of $-\uv$ are accompanied by the over- and under-prediction of $\ut$, respectively, see \fig~\ref{fig:AuTauLoci}.}

\revCom{
In order to assess the resolutions resulting in zero $\epsilon[\ut]$, two-point velocity correlations and integral length scales corresponding to a subset of P1 to P7 are studied in \app~\ref{sec:twoPointAppendix}. 
It is observed that this is only for $\dxp=16.56$ and $\dzp=9.96$, denoted by P1, that the computed two-point velocity correlations agree well with the reference DNS~\cite{iwamoto02}.  
In fact, most of the resolutions P2 to P7 in \fig~\ref{fig:AuTauLoci} are found to be too coarse to resolve the integral length scales and hence the near-wall structures in the streamwise and spanwise directions. 
Besides these, no relation between two-point correlations associated to different resolutions with $\epsilon[\ut]=0$ is found.
}

\revCom{
This analysis along with the observations in \app~\ref{sec:SetALAm}, indicate that the reason behind having curves of $\epsilon[\ut]=0$ in the $\dxp\dash\dzp$ plane is less likely to be physical. 
The existence of such loci can be left totally to the artefacts due to the numerical method and cancellation of the errors.
However, providing a clear-cut explanation on why the particular pattern for the $\epsilon[\ut]=0$ curve is observed in \fig~\ref{fig:ABC_duTau} is not an easy task, considering the complexity of deriving modified equation and truncation errors corresponding to the discretized form of equations (\ref{eq:LEScont}).
}

\begin{figure}[!htbp]
   \centering
    \begin{tabular}{cc}
       \includegraphics[scale=0.30]{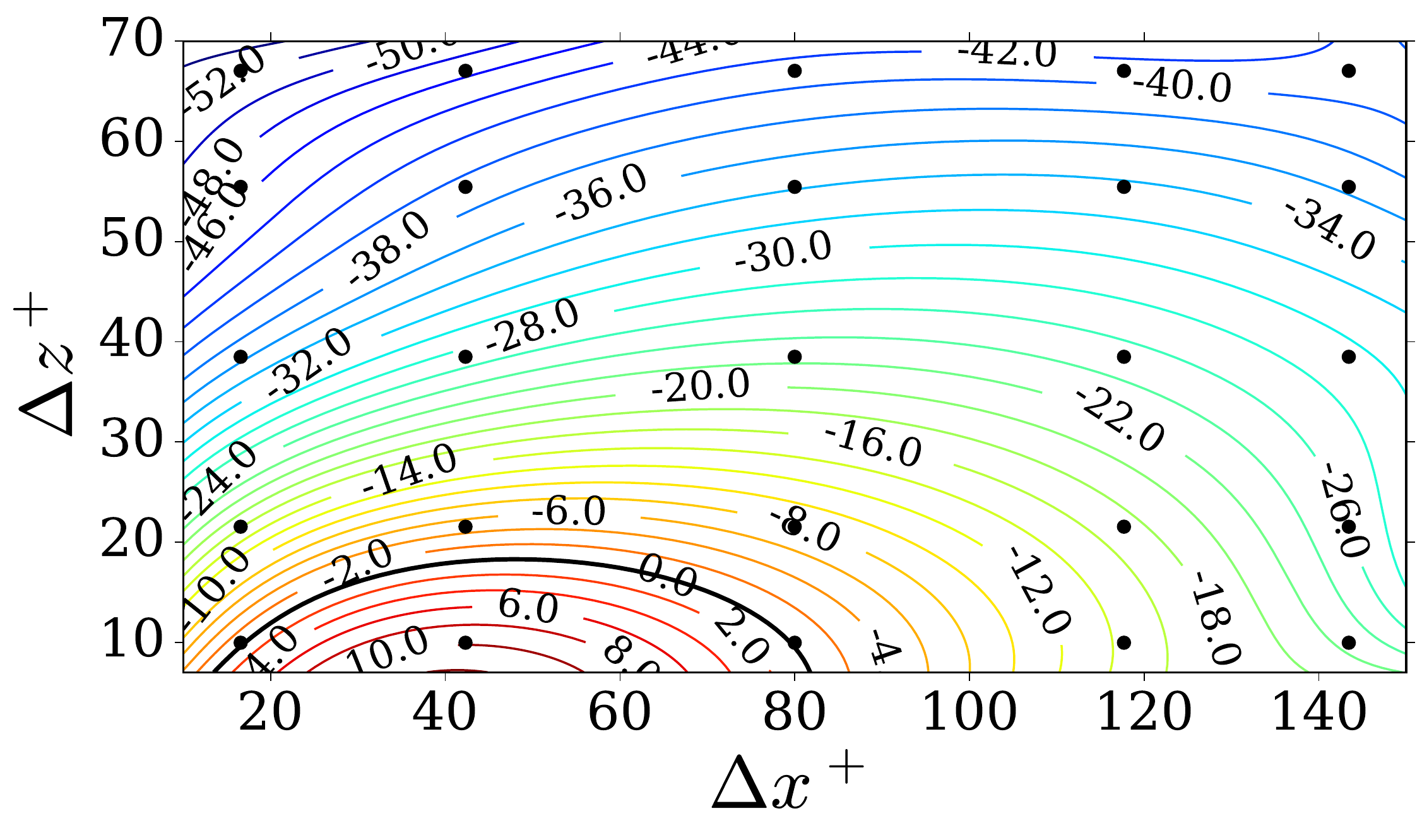} &
       \includegraphics[scale=0.30]{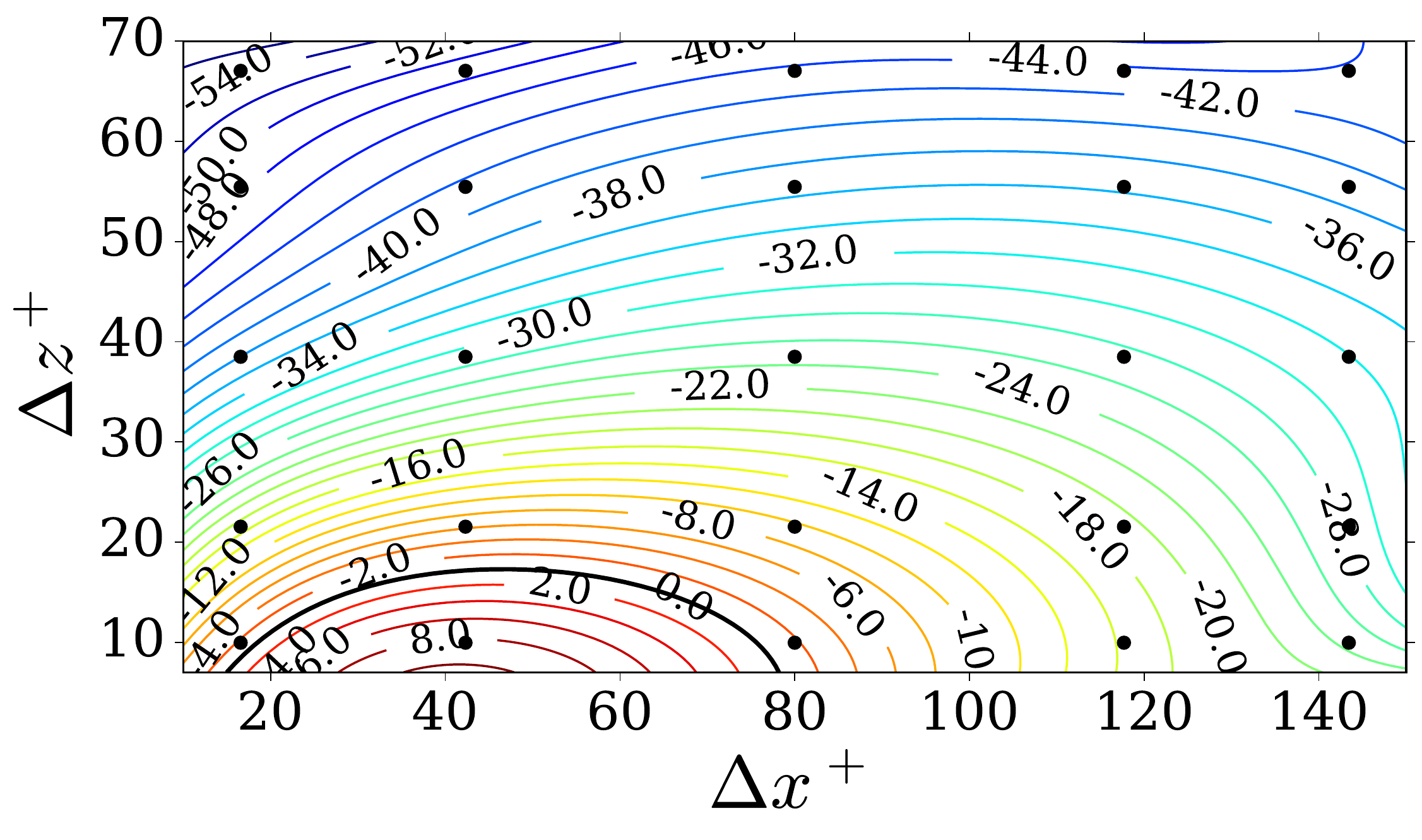} \\    
       {\small (a)} & {\small(b)} \\
    \end{tabular}          
    \begin{tabular}{cc}
       \includegraphics[scale=0.30]{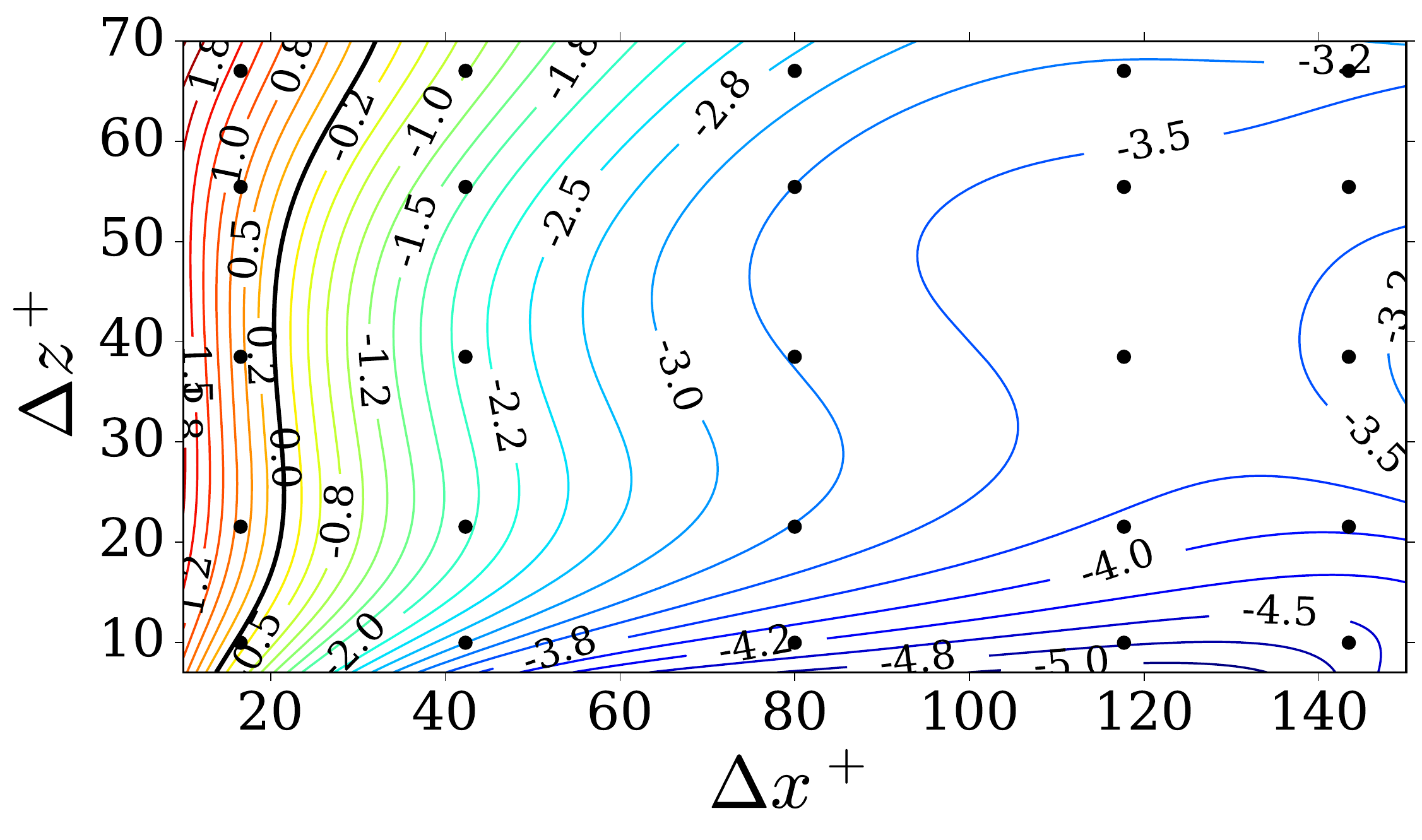} &
       \includegraphics[scale=0.30]{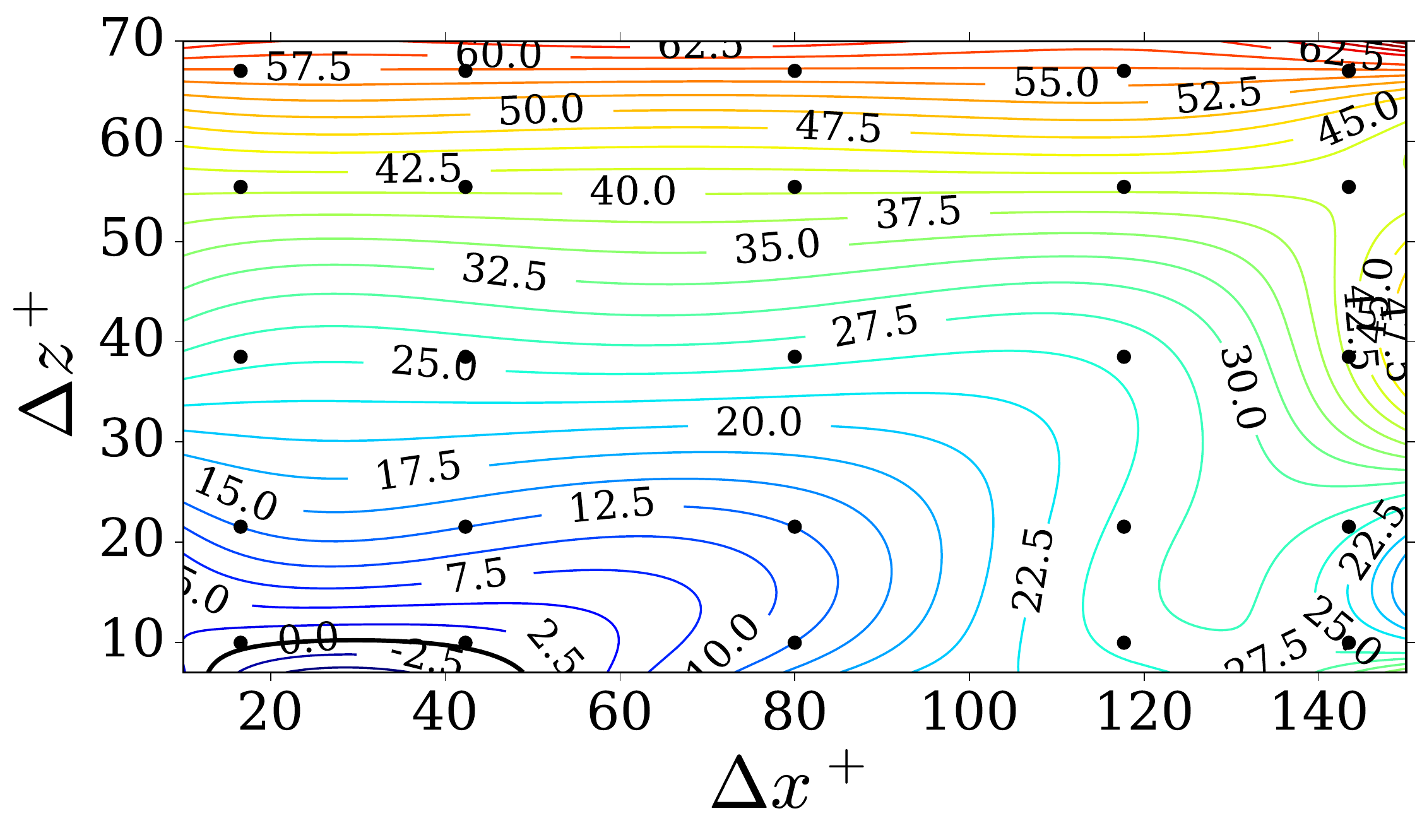} \\    
       {\small (c)} & {\small(d)} \\
    \end{tabular}          
\caption{Isolines of $\earea[|\uv|]$ (a), $\epik[\uv]$ (b), $\epsilon[\U_c]$ (c), and $\epsilon[\eta(\uv_{\rm max})]$ (the error in the $\eta=y/\delta$ at which the peak value of $\uv$ occurs) (d), plotted in the $\dxp\dash\dzp$ plane for Set-B.}\label{fig:B_duv}    
\end{figure}

\revCom{
Hereafter, the convergence such as that of $\ut$ is referred to as} numerical convergence, in contrast to the physical convergence that would occur if for a specific grid resolution, the loci of zero numerical error in different quantities coincide.
As shown in \fig~\ref{fig:B_duv}(a-b), for those grid resolutions that the computed $\ut$ converges to the DNS value, the area under $\uv$ profile plotted against wall-normal coordinate (or equivalently the value of $\uv$ averaged over the channel half-height), as well as the peak of $\uv$, become the same for both LES and DNS. 
But, the LES cross-channel profile of $\uv$ does not necessarily converge, in the pointwise sense, to that of DNS, \revCom{as illustrated in \fig~\ref{fig:AuTauLoci}(b)}.
In particular, it is only for the fine enough resolutions in the $\dxp\dash\dzp$ plane that the $\eta$ of the peak of $\uv$ profile becomes exactly the same as the corresponding DNS value, see \fig~\ref{fig:B_duv}(d). 
It is noteworthy that, although the plots in \fig~\ref{fig:B_duv} belong to Set-B, similar observations are made for the other sets listed in \tab~\ref{tab:caseSummary} (not shown here).


Ideally, to formulate the physical convergence, a physical-based relation connecting two or more QoIs of the channel flow is required. 
To this end, a possible starting point is the streamwise momentum equation which is averaged in both time and over homogeneous directions, see~\eg~\cite{pope},  
\begin{equation}\label{eq:xChanMomentum}
u_\tau^2 (1-\eta) = \frac{1}{\rey_b}  \frac{\dd\langle u\rangle}{\dd\eta} -  \uv \, .
\end{equation}
Integrating this equation over the channel half-height and then subtracting the resulting expression evaluated by the LES and DNS values, the following relation between the errors in $\ut$, the mean centerline velocity $\U_c$, and $\earea[|\uv|]$ is derived, 
{\revProof{
\begin{eqnarray}\label{eq:xi}
\xi:=\lambda_{\ut} \epsilon[\ut]
-\lambda_{\uv}  \earea[|\uv|]
+\lambda_{\U_c} \epsilon[\U_c] 
\,, \quad 0\leq \eta\leq 1 \,,
\end{eqnarray}
}}
in which,
\begin{eqnarray*}
&\lambda_{u_\tau} = \reyt^\circ \left(\ut+\ut^\circ\right)/2 \,,& \\
&\lambda_{\uv} = \left({\reyt^\circ}/{\ut^\circ} \right)  \int_0^1 \left|\uv^\circ \right| \dd \eta \,, &\\
&\lambda_{\U_c}=-\U_c^\circ \,.& 
\end{eqnarray*}
The absolute value of Reynolds stress $\uv$ is used in (\ref{eq:xi}), since $\uv$ is always non-positive in the lower half of the channel \rev{and, as mentioned earlier, the average of the profiles of both halves of channel is considered to evaluate all expressions.}
{\revProof{Observe that $\lambda_{\ut}>\lambda_{\uv}>0>\lambda_{\U_c}$ (in particular, for $\reyt=300$, $\lambda_{\ut}\approx 2\lambda_{\uv}$, $\lambda_{\uv}=7.697$, and $\lambda_{\U_c}=-1.154$ based on the DNS data of \cite{iwamoto02}).}}


To use (\ref{eq:xi}) as a condition to check physical convergence, we seek for a grid resolution which results in simultaneous zero $\epsilon[\ut]$, $\earea[|\uv|]$, and $\epsilon[\U_c]$.
For Set-ABC, the overlap of the zero isolines of these three errors occurs at the finest resolution considered in the $\dxp\dash\dzp$ plane for any $\dyp_w\in[0.25,1.96]$, as it can be inferred from \figs~\ref{fig:ABC_duTau} and \ref{fig:B_duv} (a,c).
In addition, according to the following sections, at the mentioned fine resolution, low error (or zero error) in other quantities of the LES of channel flow can be achieved. 
Therefore, (\ref{eq:xi}) can be used a suitable criterion to make sure by grid refinement, the results of channel flow simulations will be accurate compared to DNS.
\rev{Note that the use of this criterion is universal, i.e. independent of the choice of numerical methods and grid construction strategy.}

\subsection{Errors in \revCom{cross}-channel profiles}\label{sec:yDepQoIs}
In this section, the errors of those quantities of the channel flow whose averaged values vary with the wall-normal coordinate are discussed.
\rev{For sake of brevity, only the errors measured by $\ltw$ and $\lnf$ norms are presented. 
However, as previously mentioned in \sect~\ref{sec:Rdefs}, for the cross-channel profile $\langle \varphi \rangle$, the errors measured by $\lon$ are bounded from above by corresponding $\ltw$- and $\lnf$-norm errors, i.e. $\eabs[\langle \varphi \rangle] \leq c_1 \el[\langle \varphi \rangle]$ and $\eabs[\langle \varphi \rangle] \leq c_2 \einf[\langle \varphi \rangle]$, where $c_1,c_2\geq 1$.}

Similar to the scalar quantities, the sensitivity of the errors in the cross-channel profiles with respect to $\dyp_w$ varying over $[0.25,1.96]$, is insignificant.
Hence, for sake of brevity, only the results of Set-B are discussed.

\begin{figure}[!htbp]
\centering
   \begin{tabular}{cc}
   \includegraphics[scale=0.3]{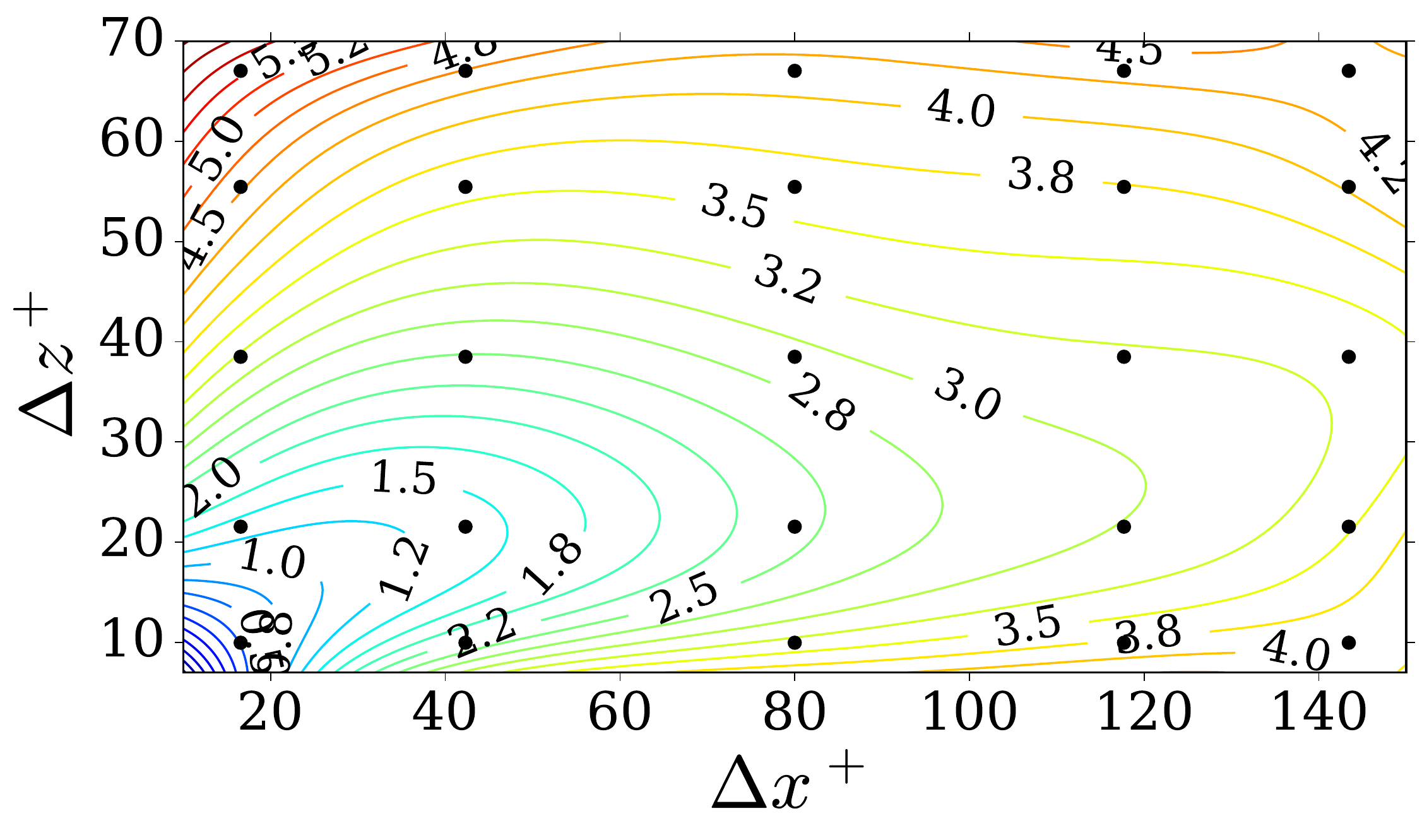} &
   \includegraphics[scale=0.3]{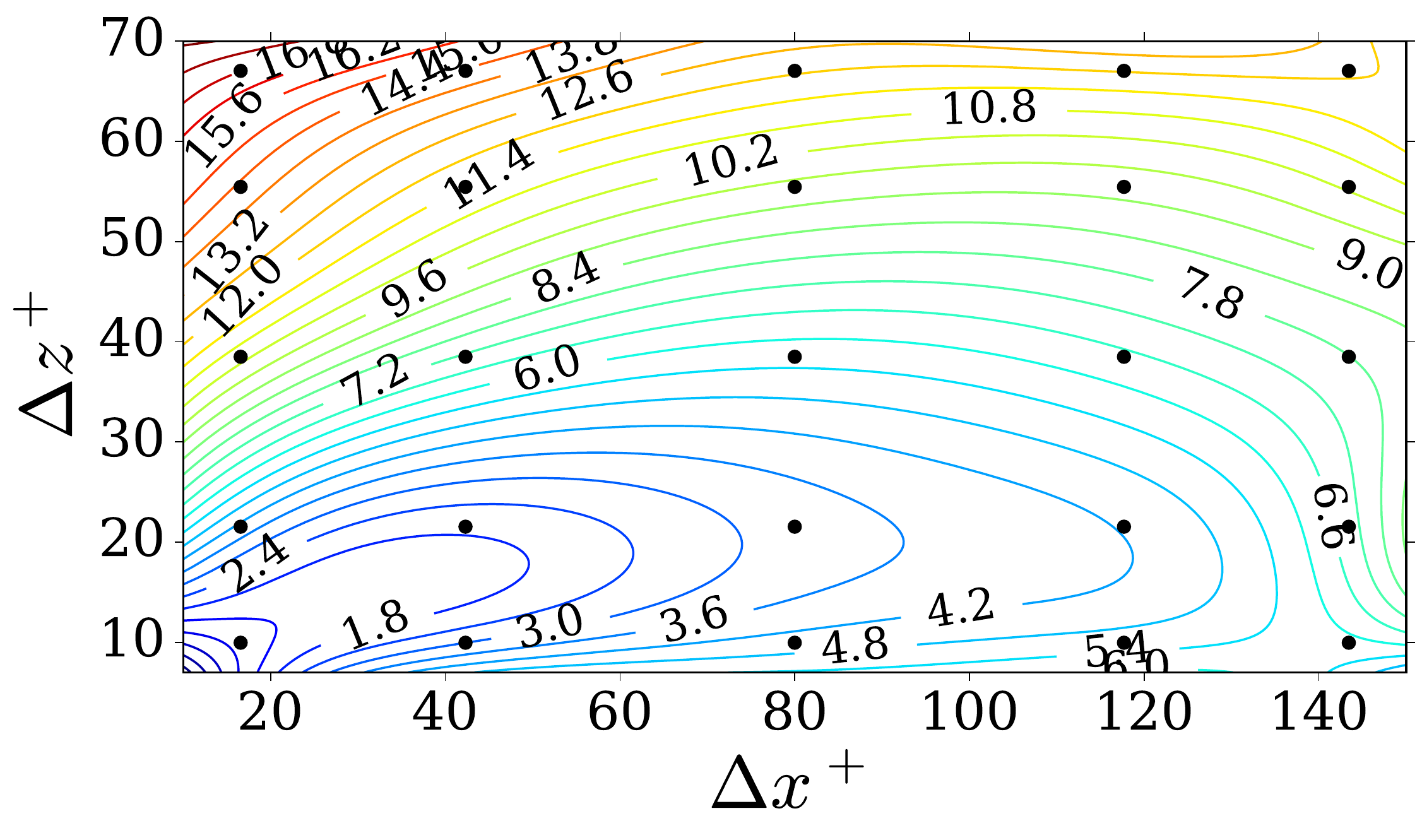} \\
   {\small (a)} &    {\small (b)} \\   
   \end{tabular}
   \begin{tabular}{cc}
   \includegraphics[scale=0.3]{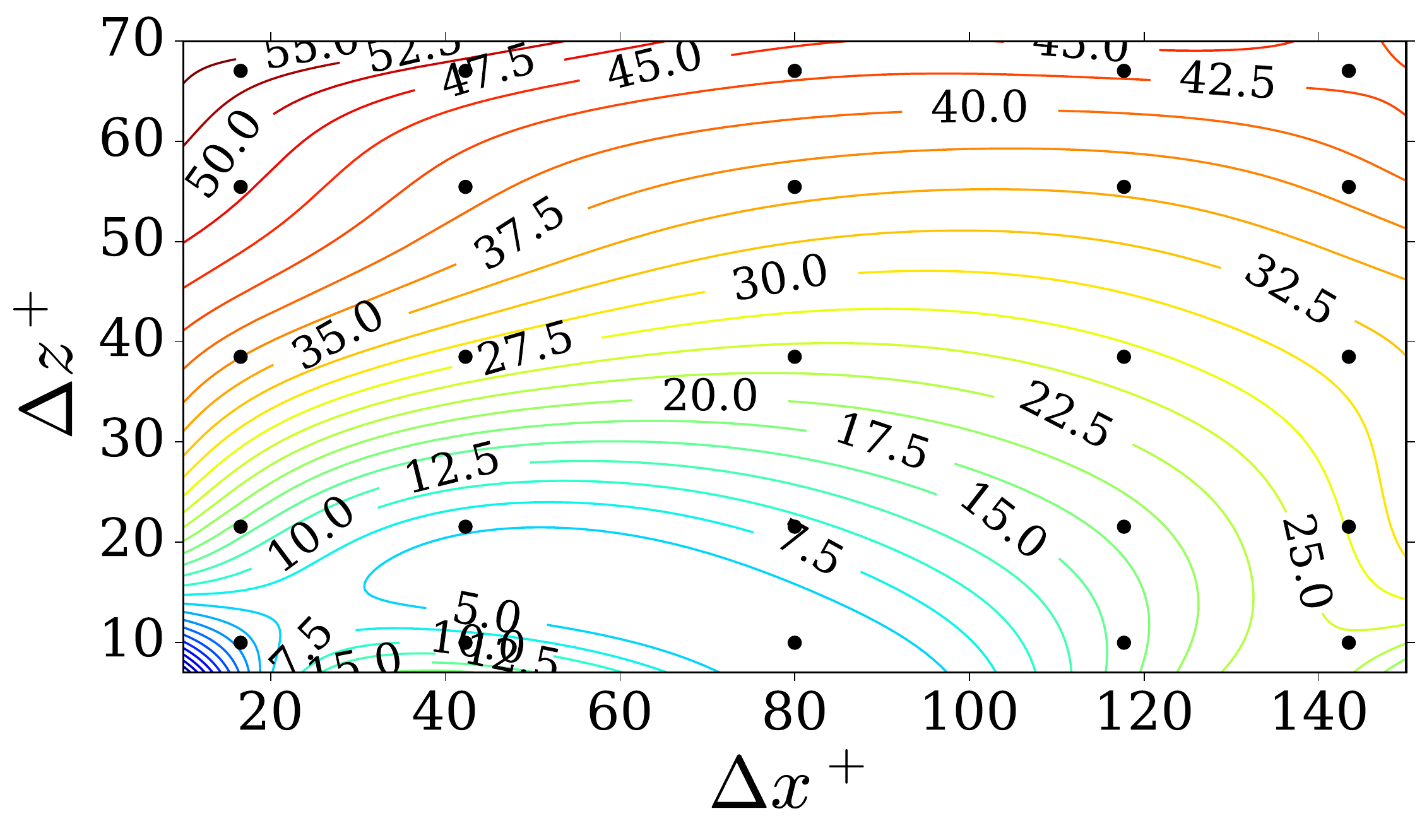} &
   \includegraphics[scale=0.3]{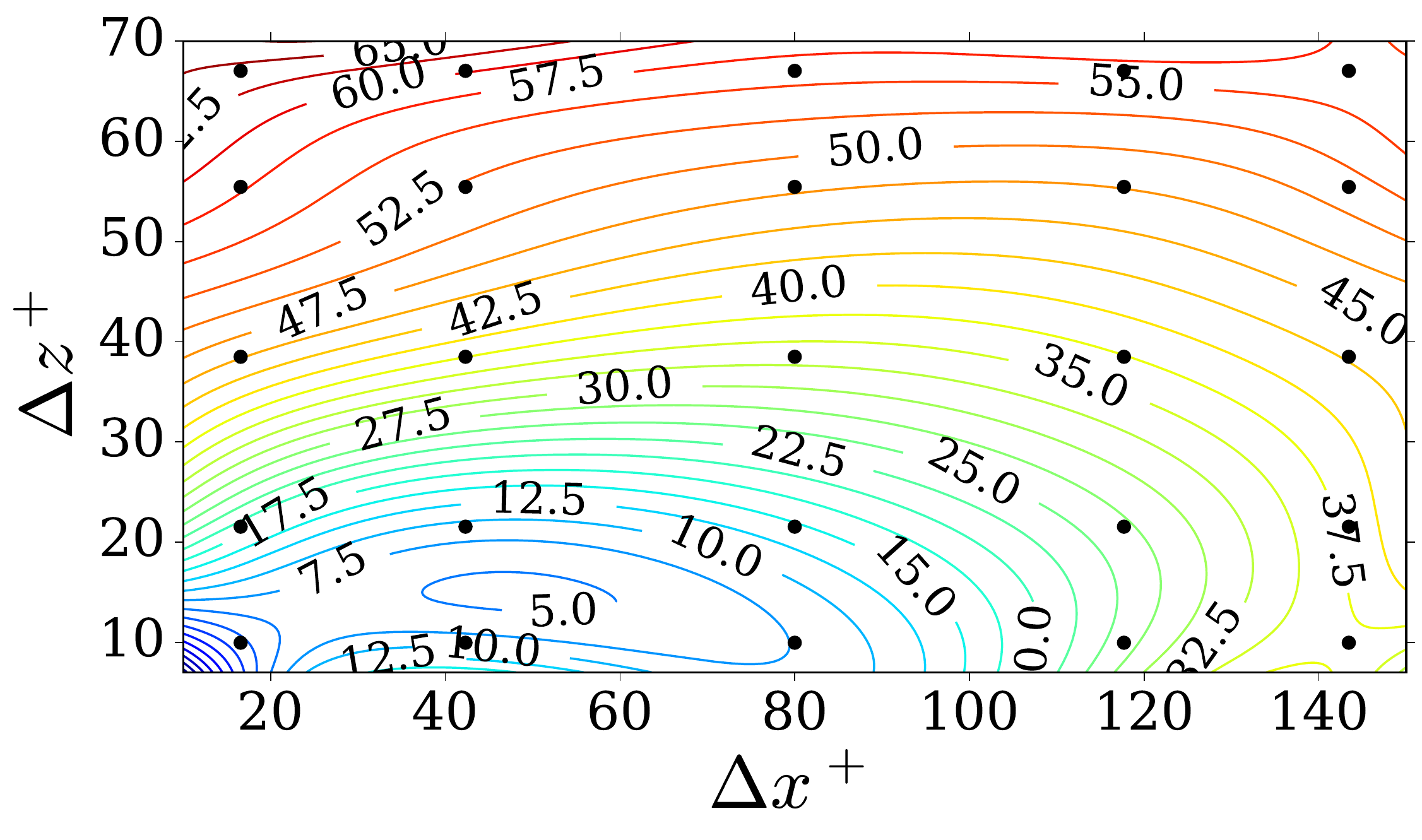} \\
   {\small (c)} &    {\small (d)} \\   
   \end{tabular}
\caption{Isolines of $\el[\U]$ (a), $\einf[\U]$ (b), $\el[\uv]$ (c), and $\einf[\uv]$ (d), plotted in the $\dxp\dash\dzp$ plane for Set-B.}\label{fig:B_Lnorm_1}
\end{figure}

Shown in \fig~\ref{fig:B_Lnorm_1} are the contours of the errors in the mean velocity profile $\U$ and Reynolds stress $\uv$. 
\revCom{It is noted that the error isolines of the mean velocity gradient $\dudy$ and the production rate of turbulent kinetic energy (TKE), $\Pk=-\uv\dudy$ are very similar to those of $\uv$, and not shown here for brevity.}
For fixed target $\reyt$ and $\dyp_w$, the minimum values of $\el$ and $\einf$ of these QoIs are achieved at the finest considered combination of $\dxp$ and $\dzp$. 
This agrees with what was concluded from constraint (\ref{eq:xi}) in the previous section.

The isolines of $\el[\U]$ seem to be approximately symmetric around a straight line passing through the minimum $\dxp$ and $\dzp$ in the $\dxp\dash\dzp$ plane, while the symmetry vanishes for the contours of $\einf[\U]$.
In fact, the isolines of $\einf[\U]$ become less sensitive to the variation in $\dxp$.

Besides the minimum error that can be attained at the finest considered grid resolution, regions of constant low values of $\el$, and more specifically, $\einf$ of $\uv$ \revCom{(and also, $\dudy$, and $\Pk$)} are identifiable in the $\dxp\dash\dzp$ plane.
These plateau-like regions exist around combinations of $\dxp$ and $\dzp$ at which, zero $\epsilon[\ut]$, \rev{$\epik[\uv]$, and $\earea[|\uv|]$} are observed, see \figs~\ref{fig:ABC_duTau} and \ref{fig:B_duv}.
\rev{This observation can be motivated as follows, taking into account the definitions of different error measures.} 
\rev{For specific combinations of $\dxp$ and $\dzp$ (more than one), the error in $\dudy$ at the wall, i.e. the error in $\ut$, vanishes and at the same time, both the peak value of, and, the area under the LES $\uv$ profile tend to the corresponding DNS values.
Nevertheless, the LES profiles of $\U$, $\dudy$, and $\uv$ do not converge in a pointwise sense to the corresponding DNS profiles. 
From this discussion, the crucial role of the error measures such as $\eabs$, $\el$, and $\einf$ for studying the deviation between the LES and DNS profiles is clarified.  
}

\begin{figure}[!htbp]
\centering
   \begin{tabular}{cc}
   \includegraphics[scale=0.3]{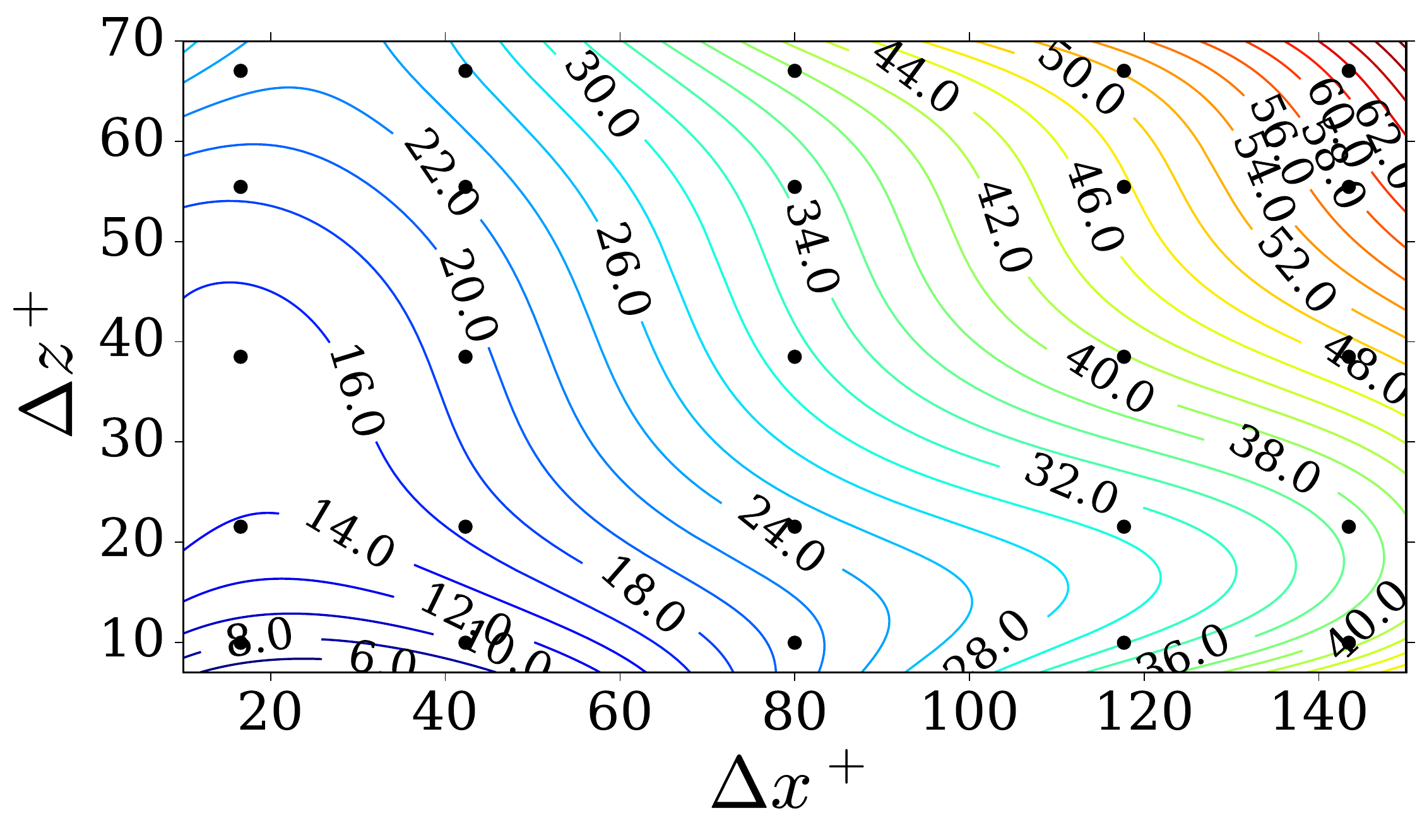} &
   \includegraphics[scale=0.3]{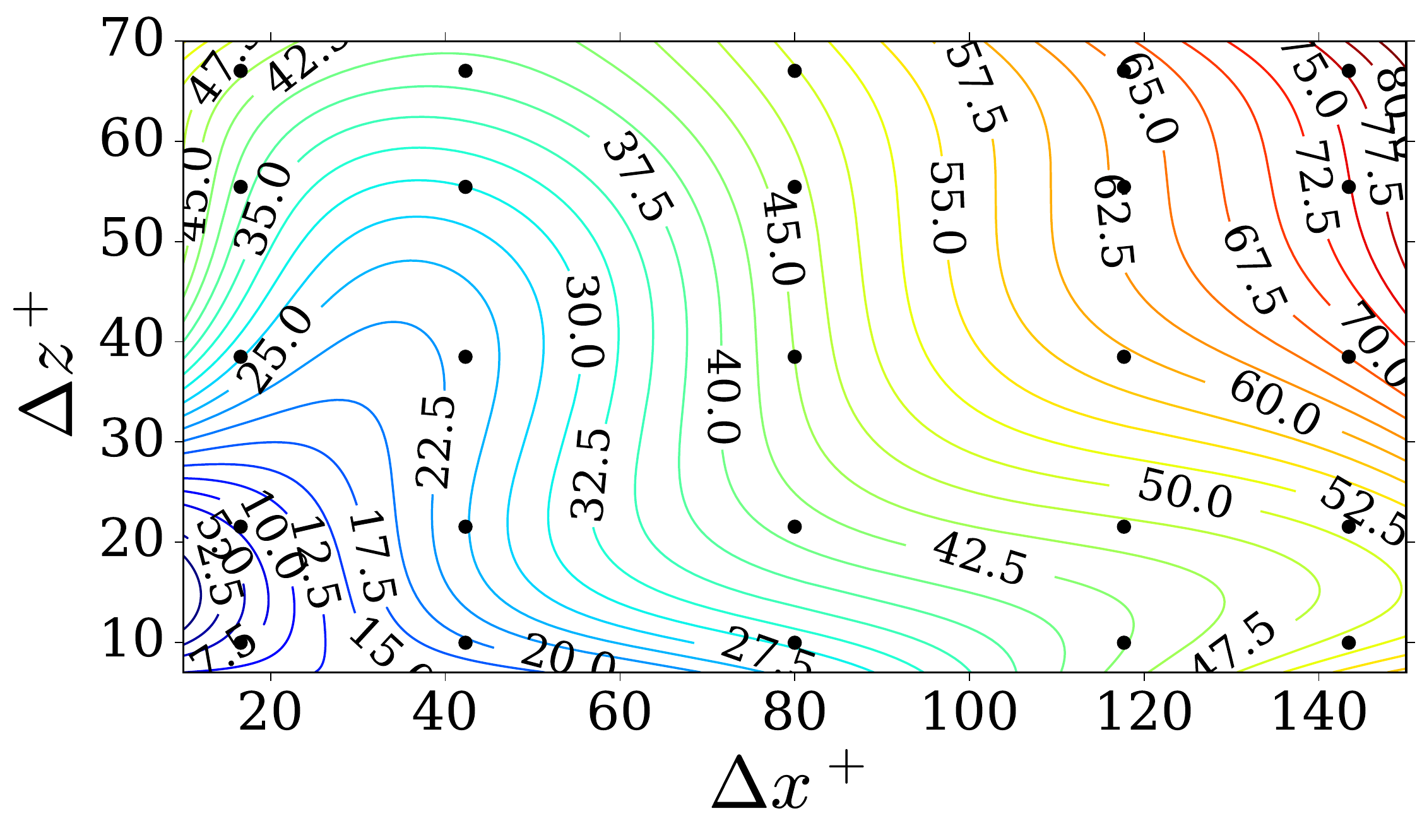} \\
   {\small (a)} &      {\small (b)} \\
   \end{tabular}
   \begin{tabular}{cc}
   \includegraphics[scale=0.3]{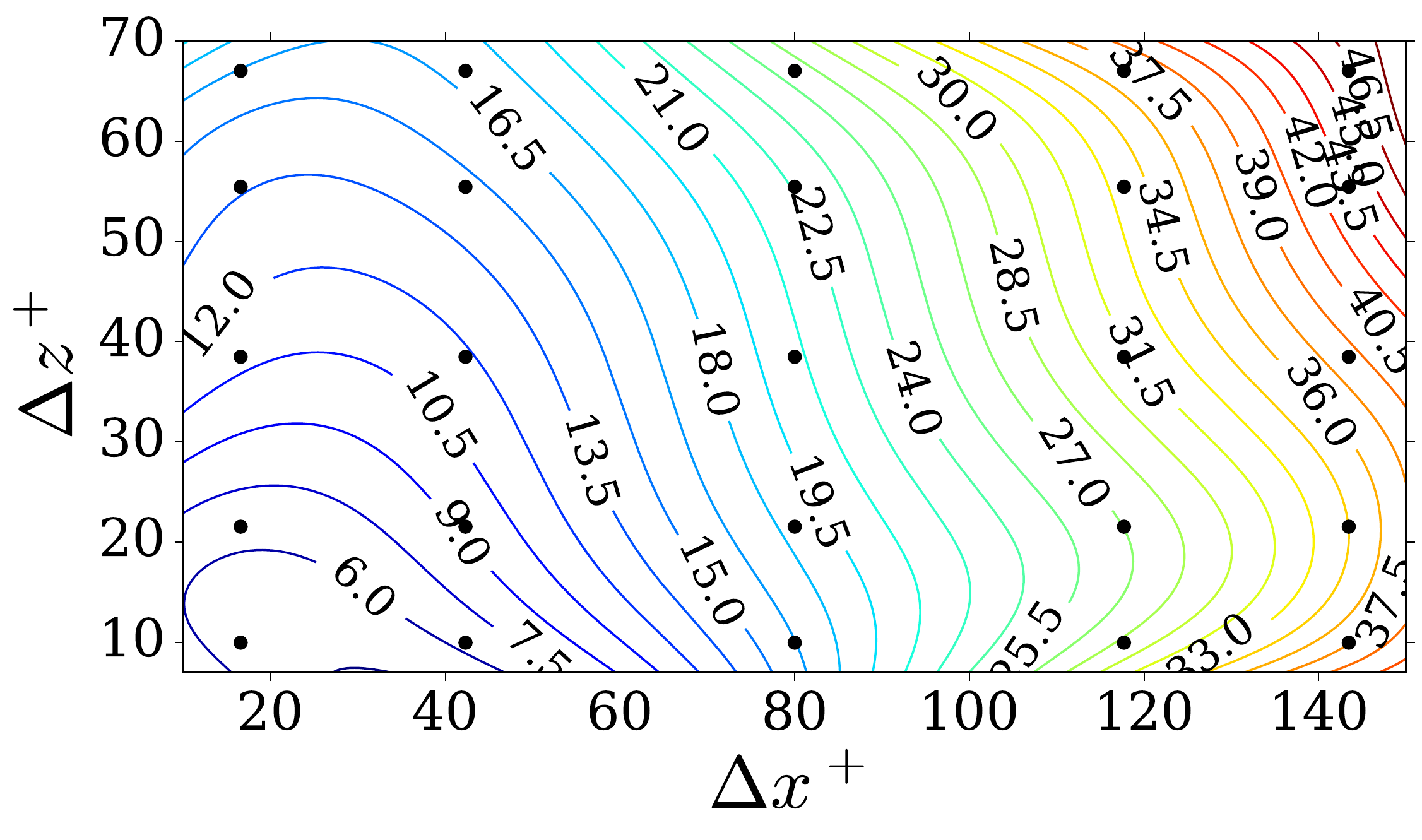} &
   \includegraphics[scale=0.3]{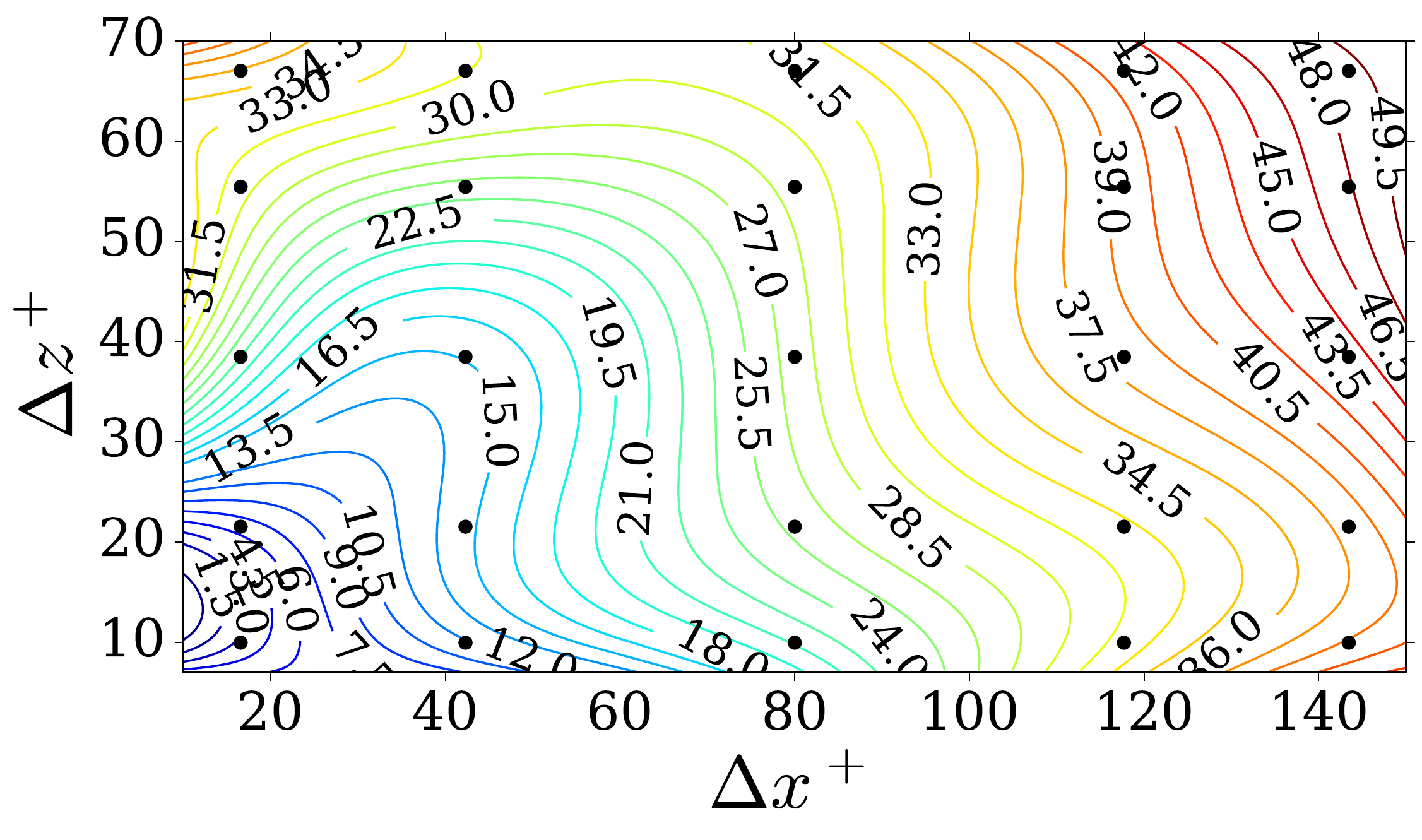} \\
   {\small (c)} &      {\small (d)} \\   
   \end{tabular}
   \begin{tabular}{cc}
   \includegraphics[scale=0.3]{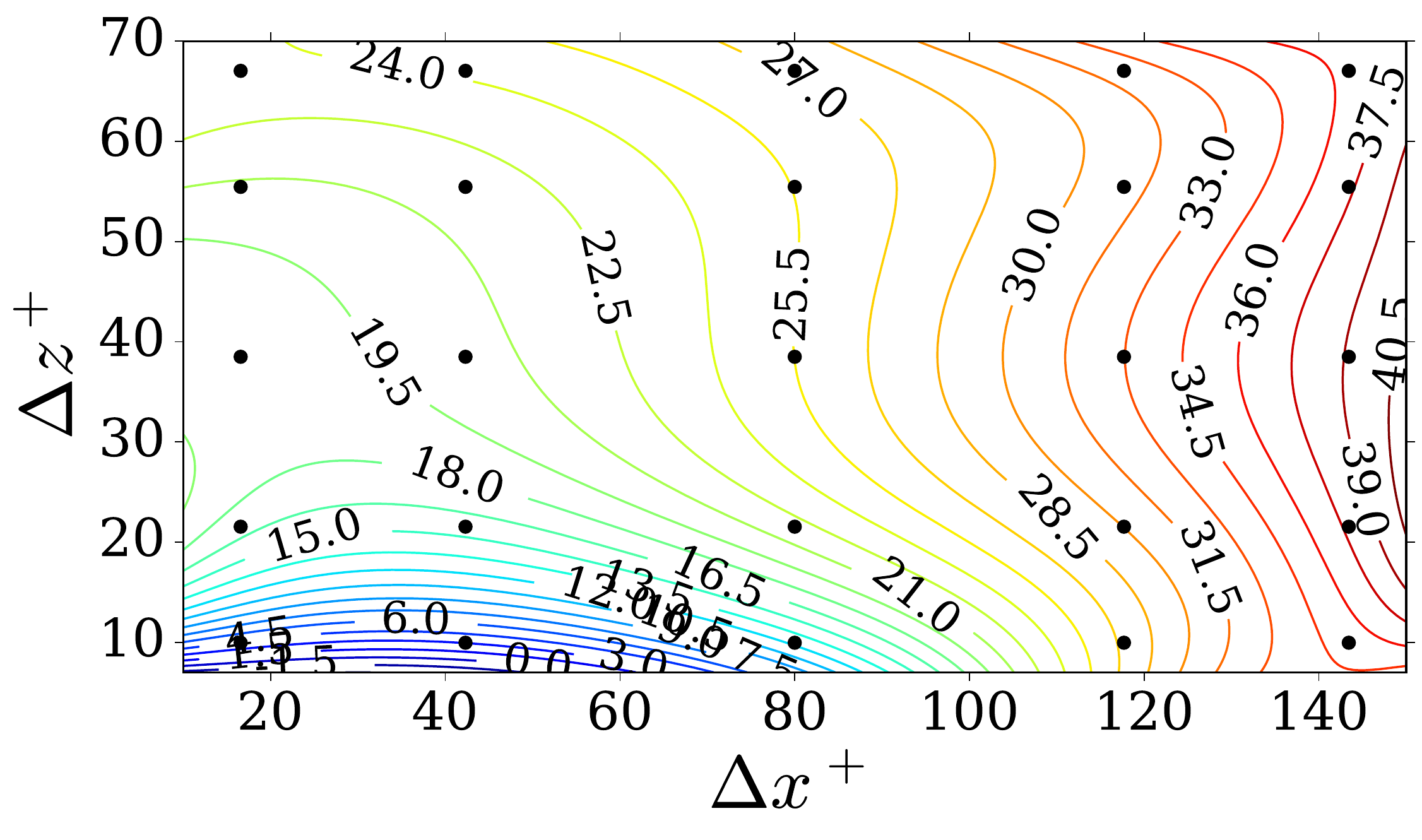} &
   \includegraphics[scale=0.3]{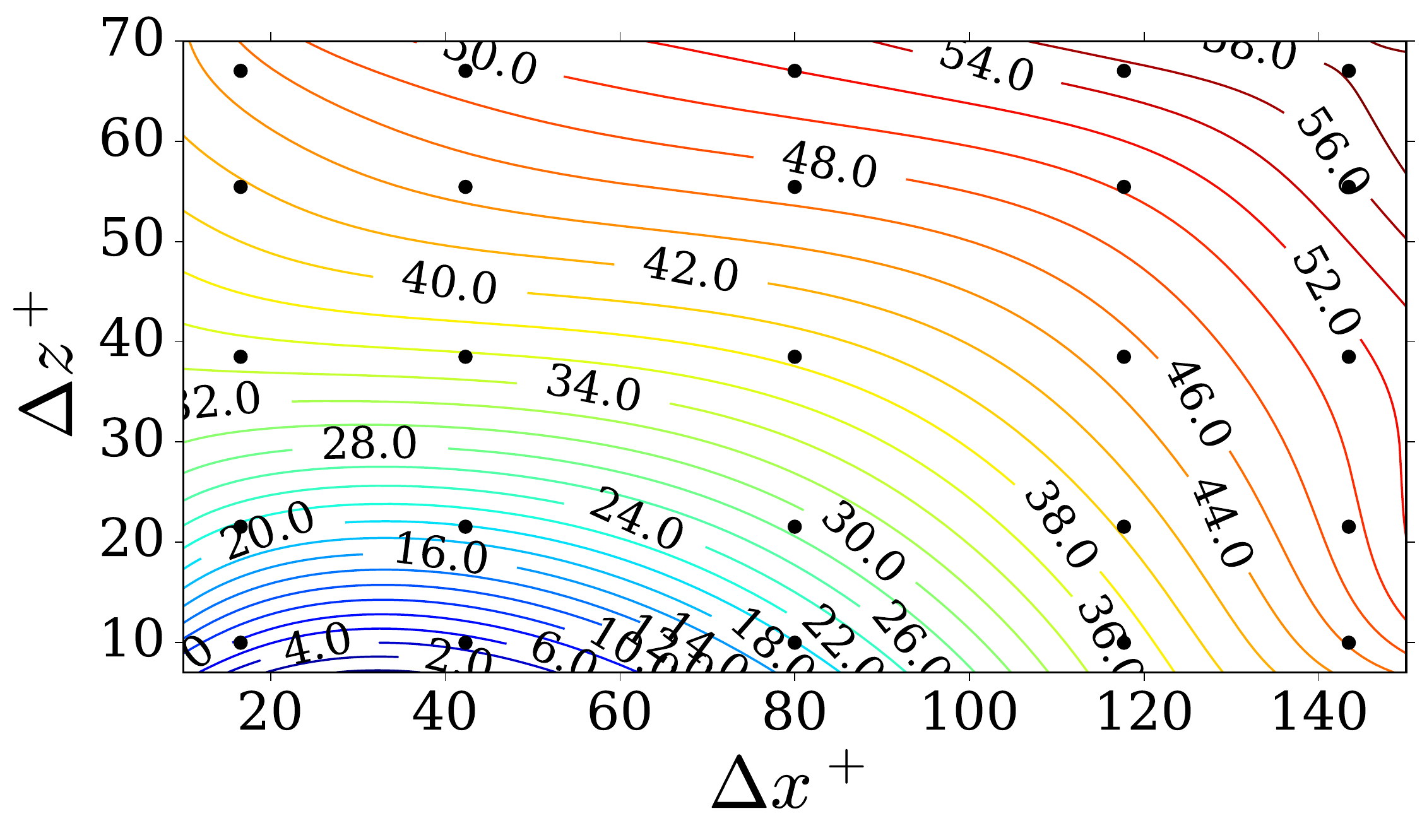} \\
   {\small (e)} &      {\small (f)} \\   
   \end{tabular}
   \begin{tabular}{cc}
   \includegraphics[scale=0.3]{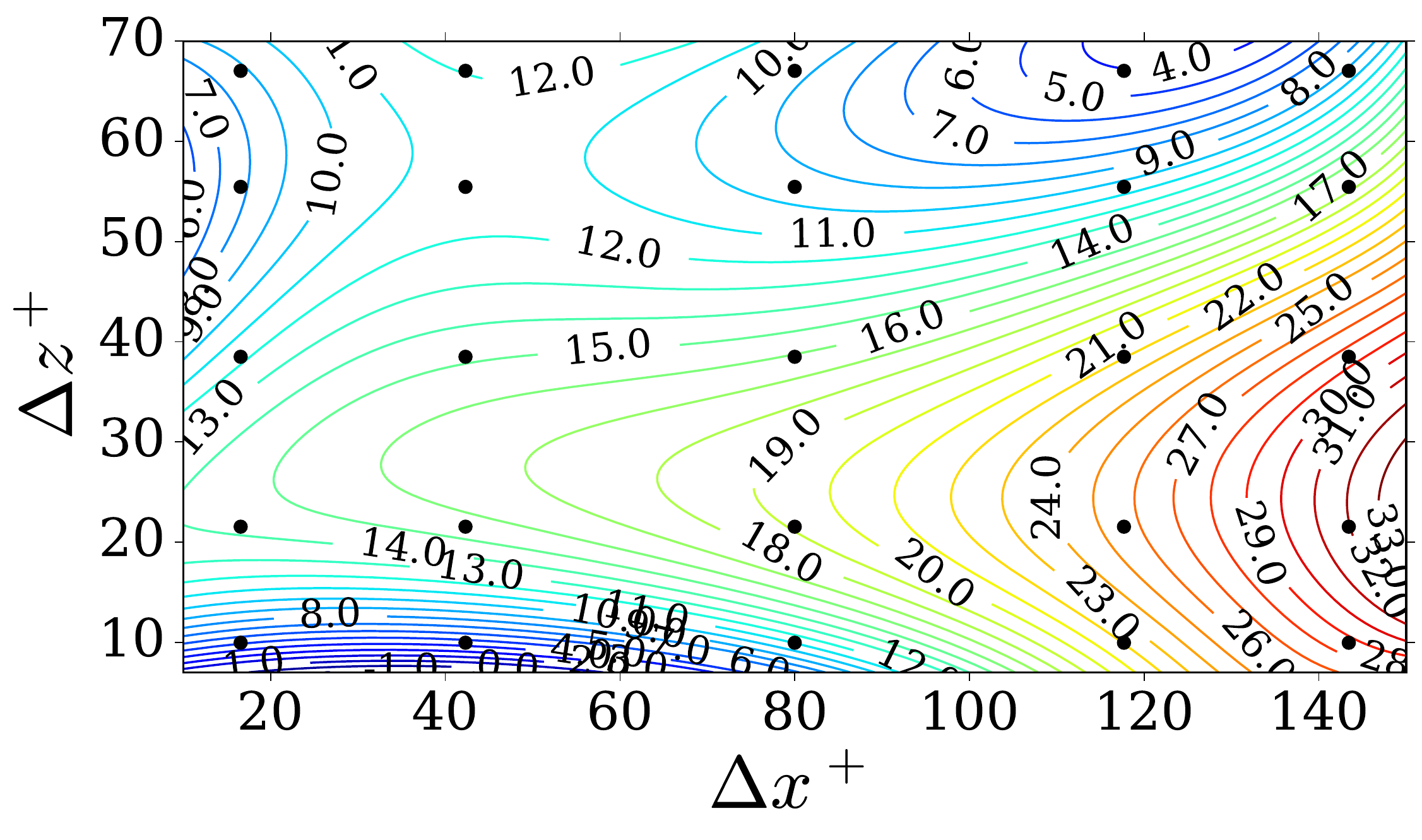} &
   \includegraphics[scale=0.3]{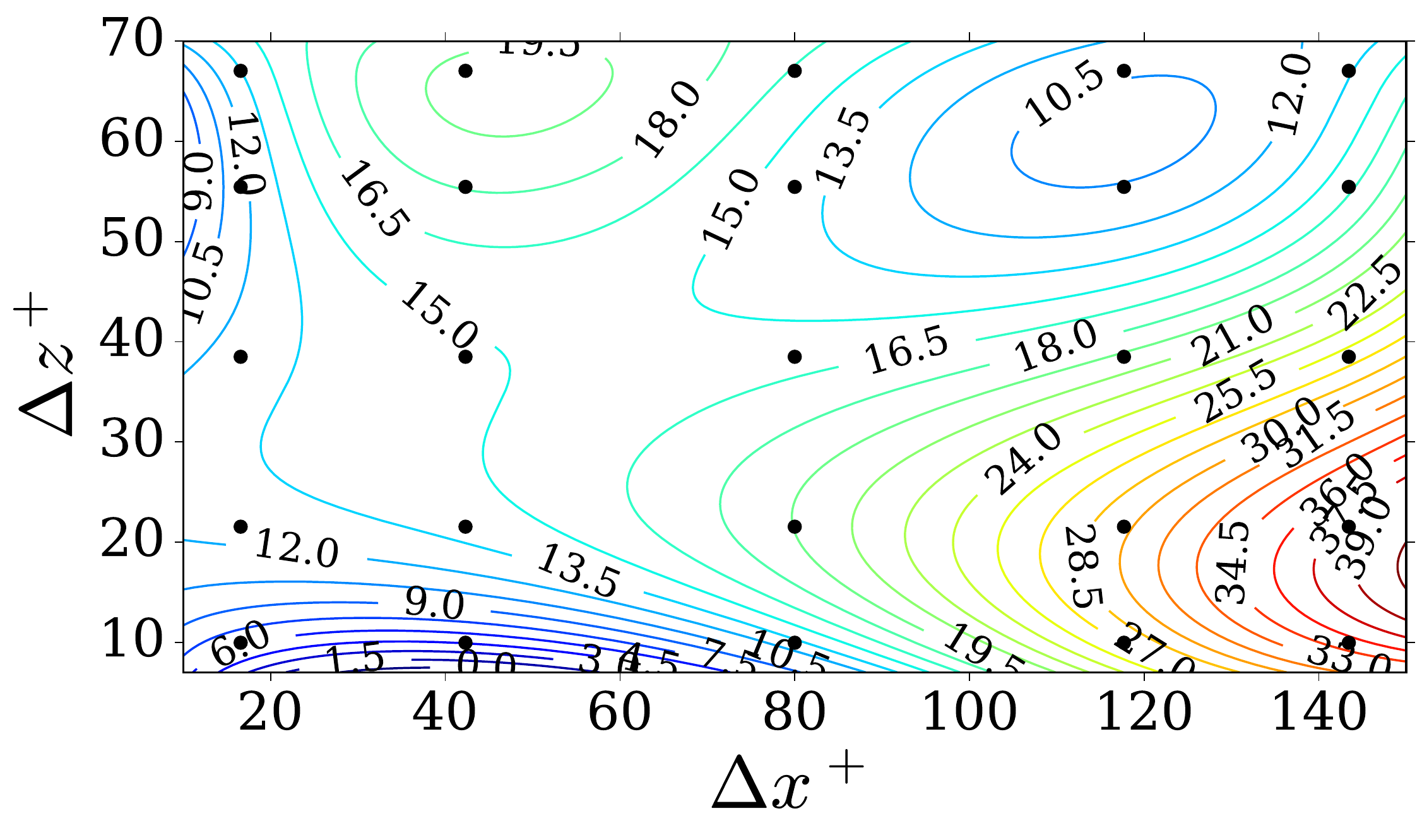} \\
   {\small (g)} &      {\small (h)} \\   
   \end{tabular}   
\caption{Isolines of $\el$ (left column) and $\einf$ (right column) of TKE (a,b), $\urms$ (c,d), $\vrms$ (e,f), and $\wrms$ (g,h), plotted in the $\dxp\dash\dzp$ plane for Set-B.}\label{fig:B_Lnorm_2}
\end{figure}

Next quantities whose error isolines are investigated are the turbulent kinetic energy, $\tke=\frac{1}{2}\langle u'_i u'_i\rangle=\frac{1}{2}(\langle u_i\rangle^2-\langle u_i^2\rangle)$ and the contributing \rev{rms velocity fluctuation components $u_{i, {\rm rms}}=\sqrt{\langle u'_i u'_i\rangle}$}, see \fig~\ref{fig:B_Lnorm_2}. 
\rev{The isolines of $\el[\tke]$ and $\einf[\tke]$ are} the indicators of the difference between the resolved kinetic energy $\tke_{\rm res} = \frac{1}{2}\langle \bu'_i \bu'_i \rangle$ computed by LES, and the total kinetic energy $\tke_{\rm tot}$ which is captured by DNS and is supposed to be the existing value in reality.
The relatively more significant impact of the streamwise rms velocity fluctuations on TKE, compared to the wall-normal and spanwise components, can be clearly understood from the similar patterns of the isolines of $\el[\urms]$ and $\el[\tke]$ and also $\einf[\urms]$ and $\einf[\tke]$, as illustrated in \fig ~\ref{fig:B_Lnorm_2} (a-d).
As shown here for Set-B, and as it persists for Set-A and Set-C, \rev{for none of the considered grid resolutions, the errors between the LES and DNS profiles of rms velocity fluctuations, and consequently, the errors between $\tke_{\rm res}$ and $\tke_{\rm tot}$ completely vanish.}

\begin{figure}[!htbp]
\centering
   \begin{tabular}{c}
   \includegraphics[scale=0.37]{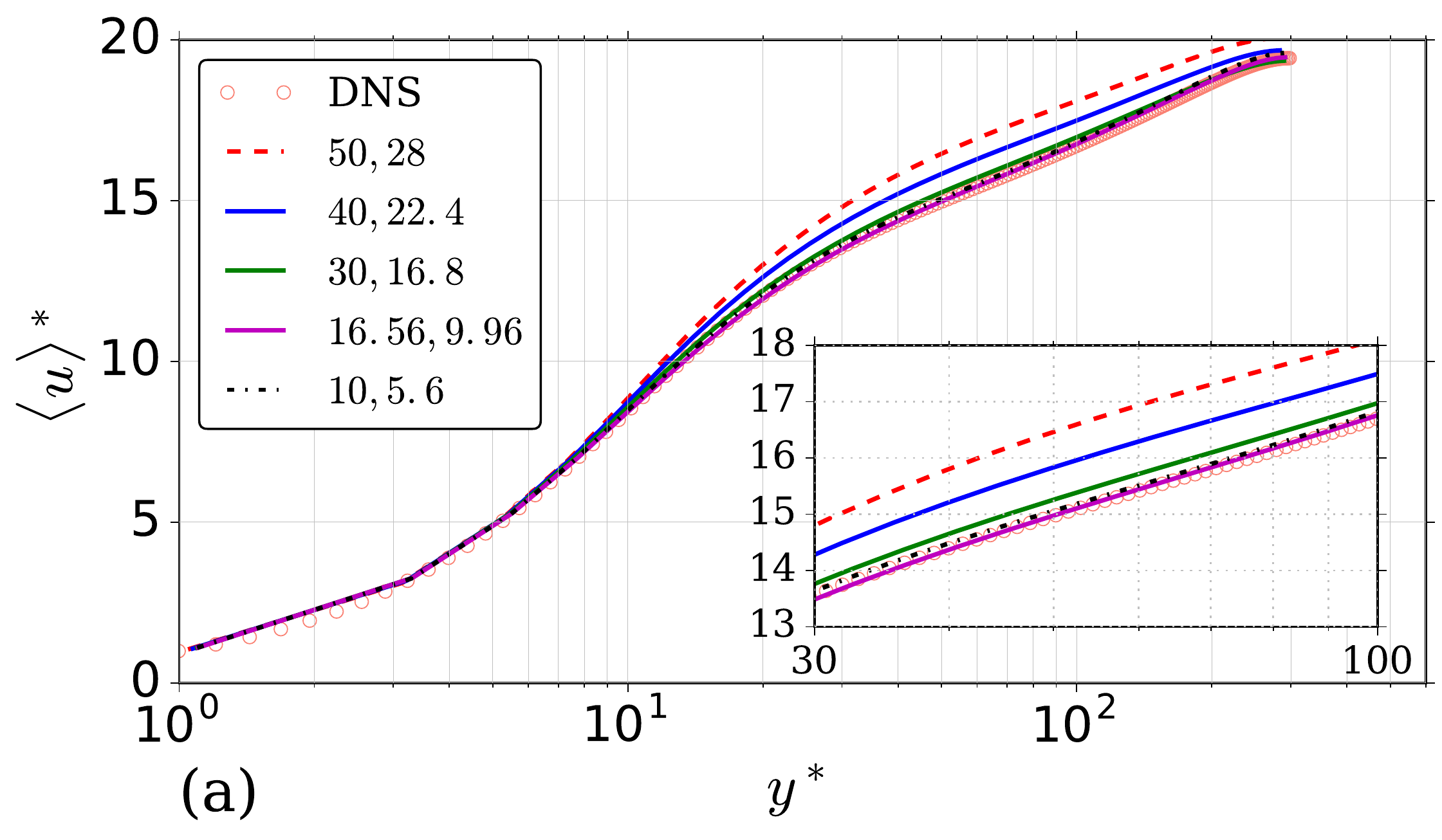}\\
   \end{tabular}
   \begin{tabular}{cc}
   \includegraphics[scale=0.37]{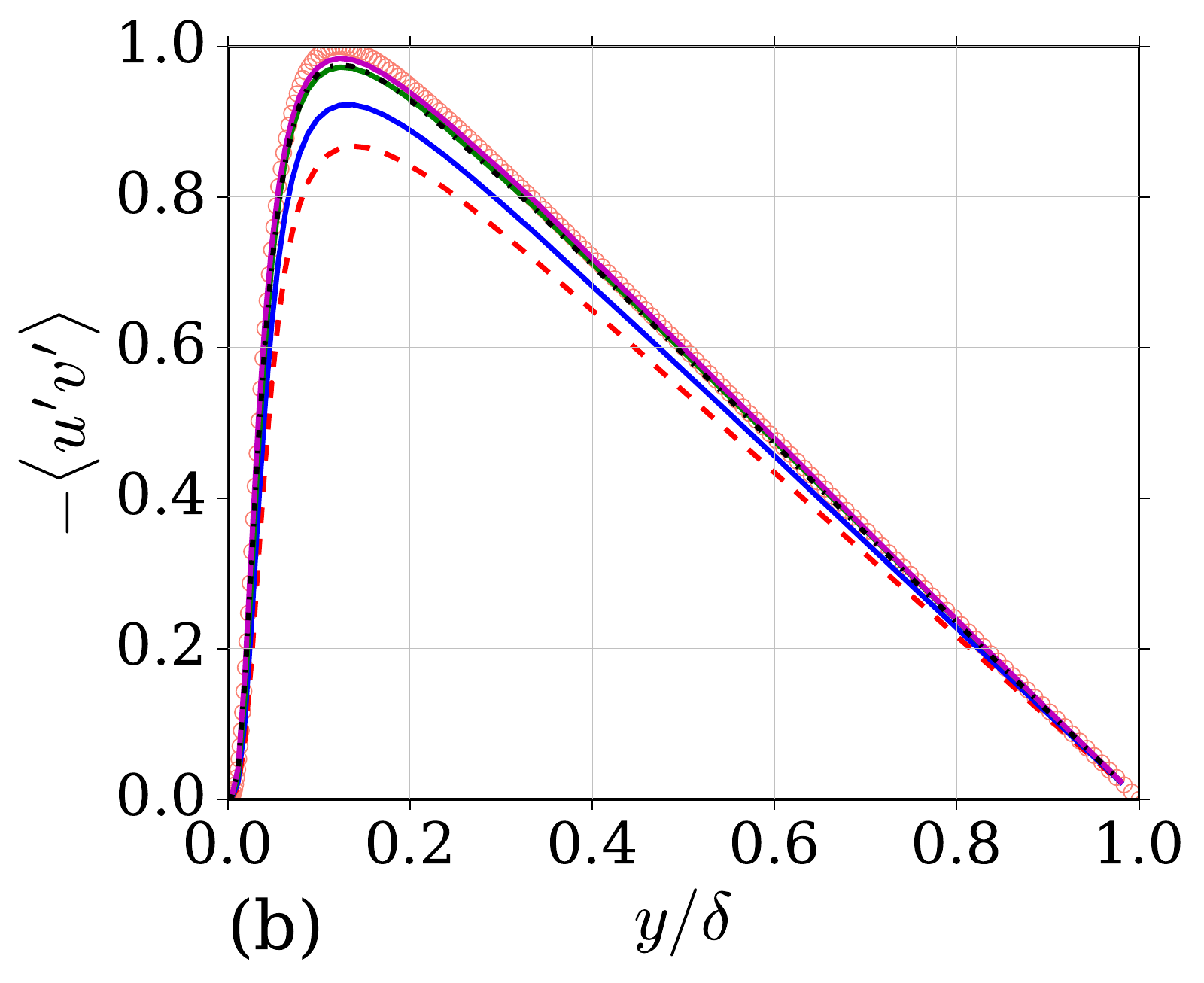} &
   \includegraphics[scale=0.37]{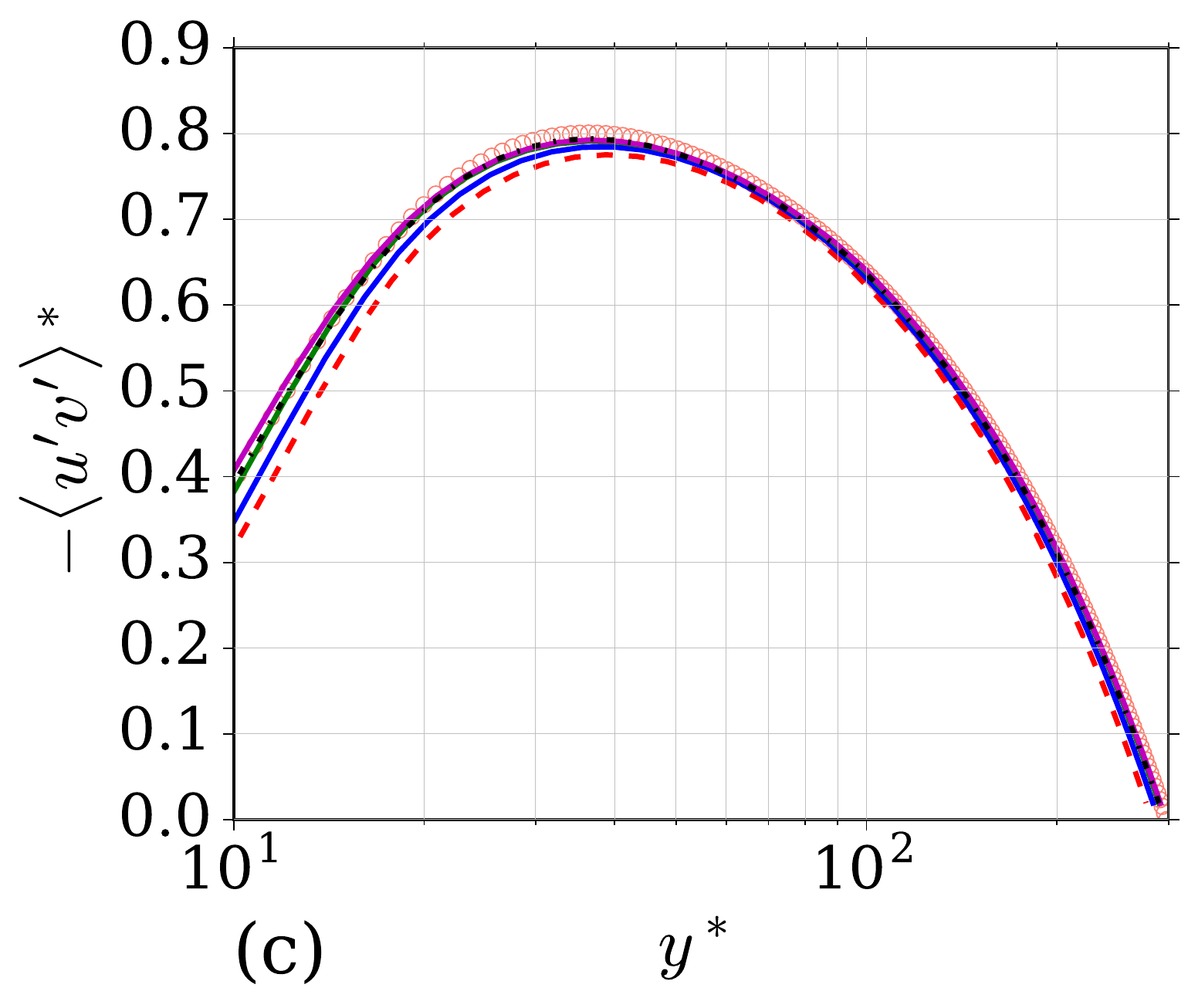} \\  
   \includegraphics[scale=0.37]{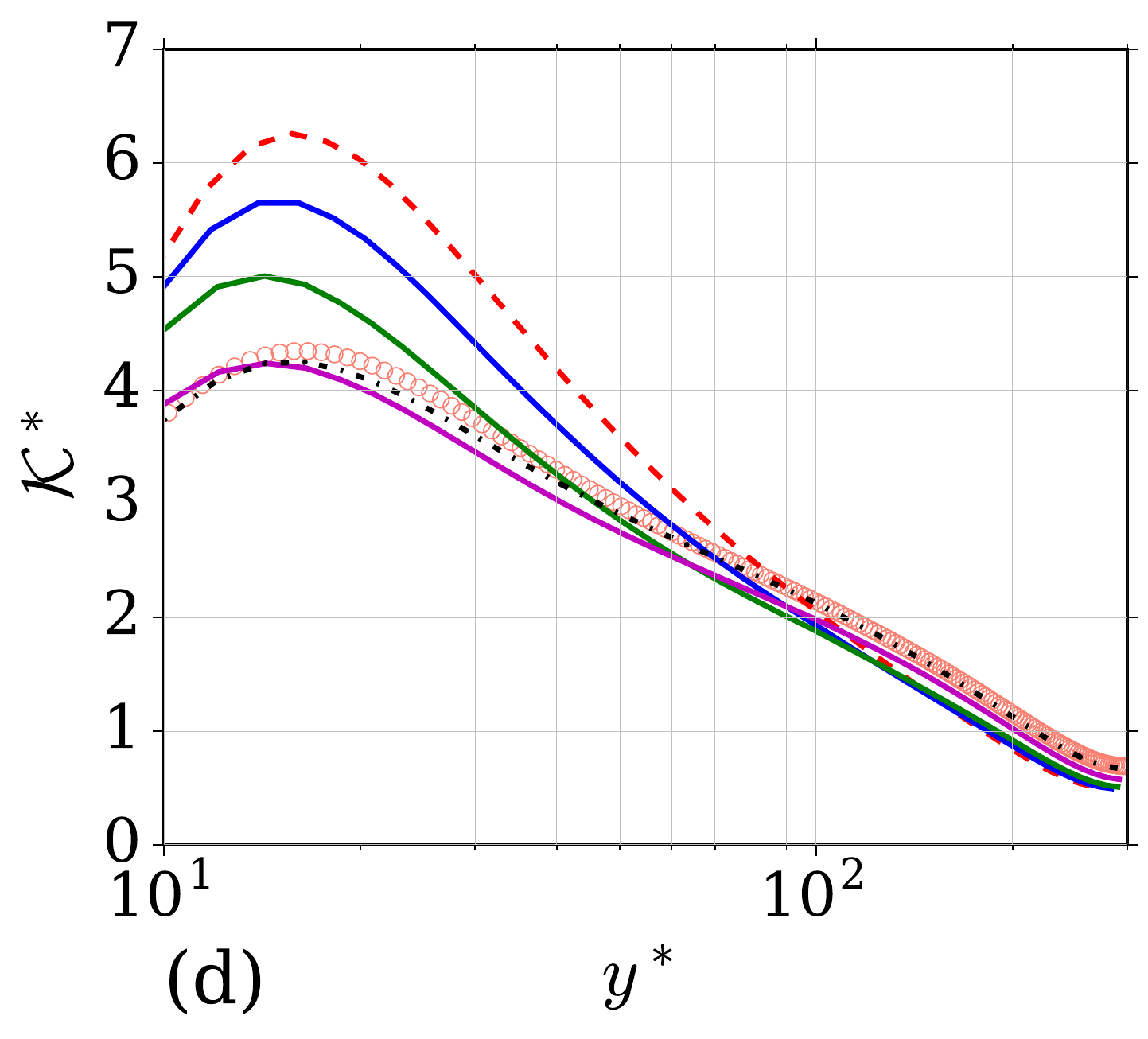} &
   \includegraphics[scale=0.37]{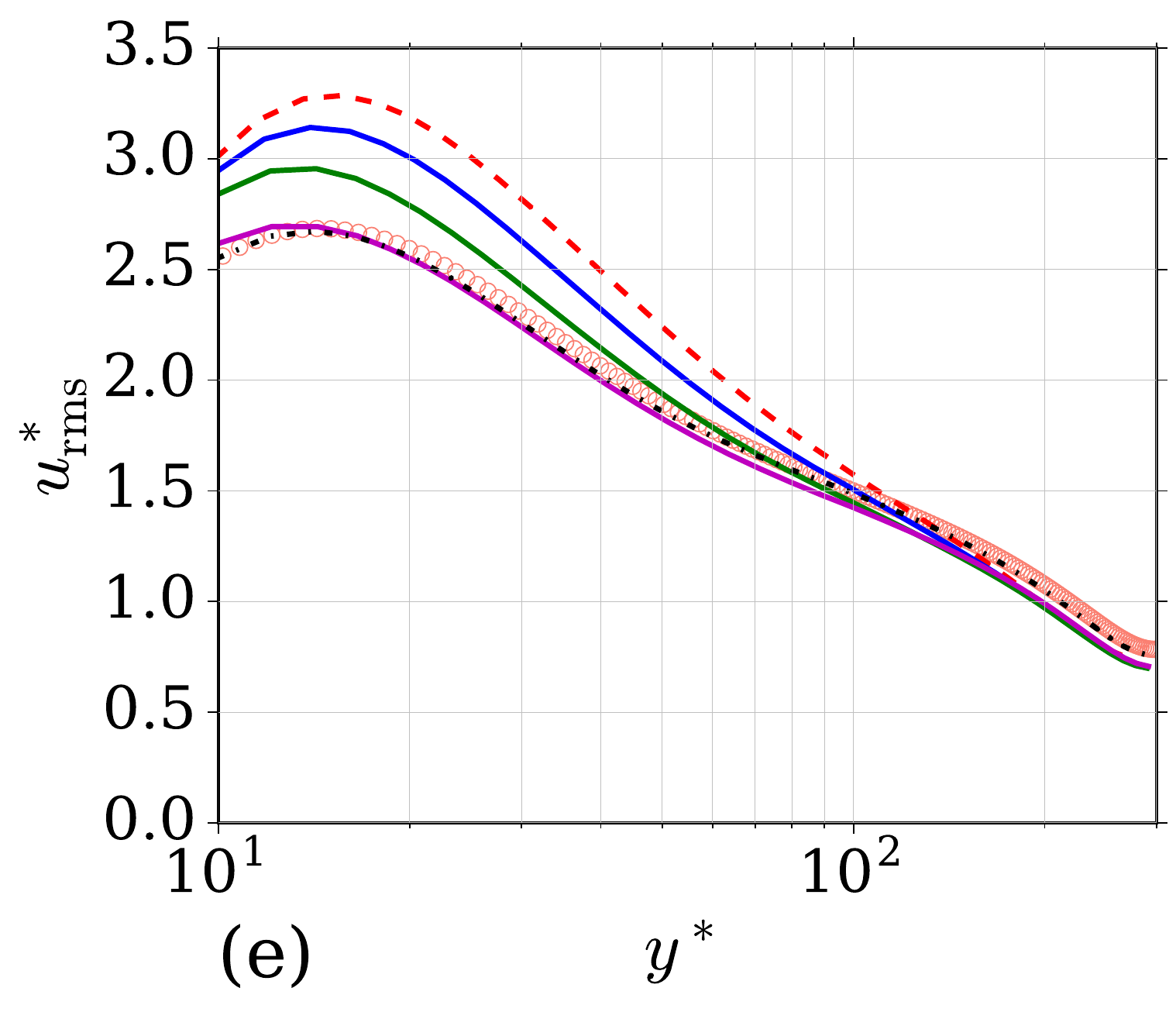} \\
   \includegraphics[scale=0.37]{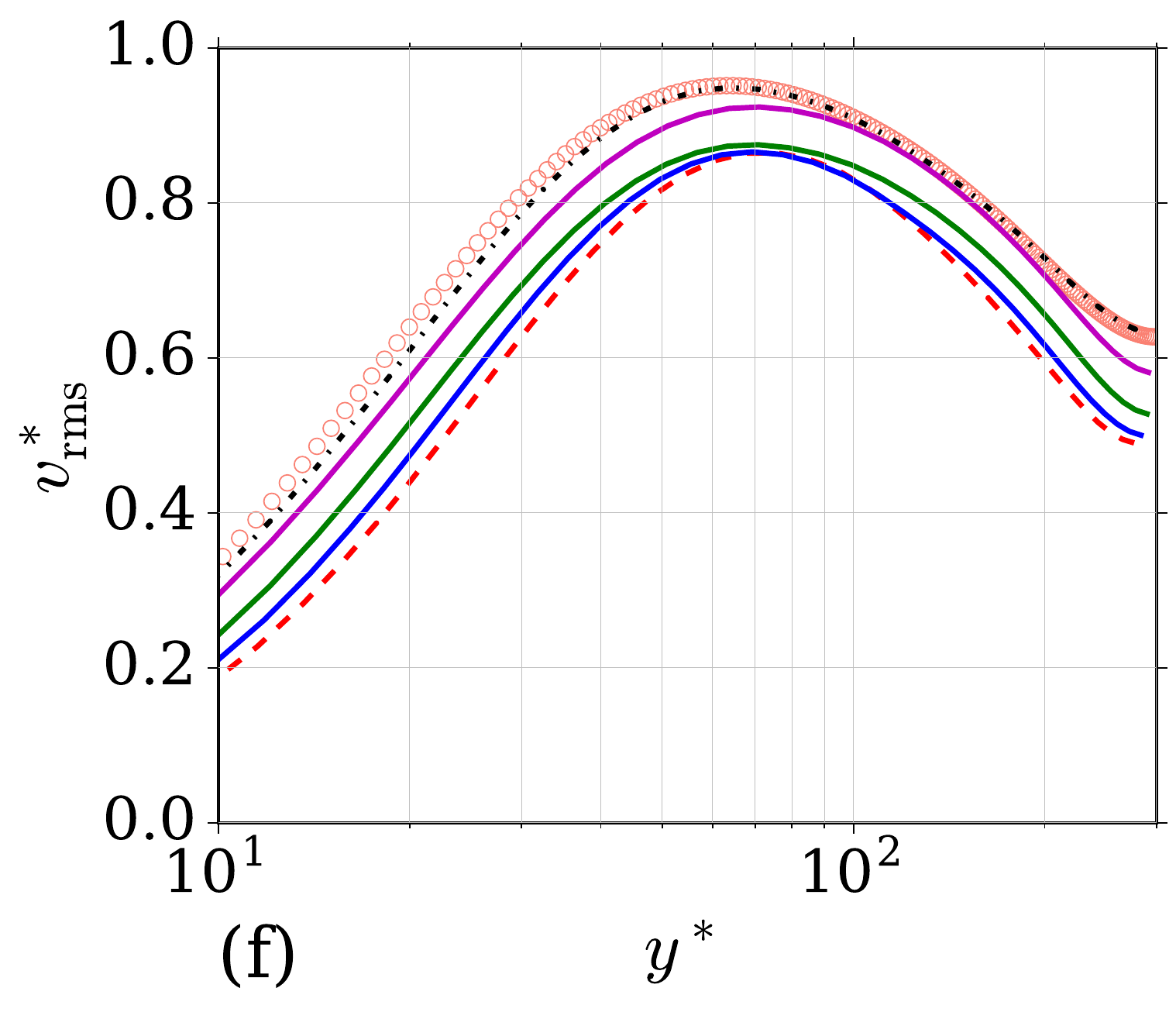} &
   \includegraphics[scale=0.37]{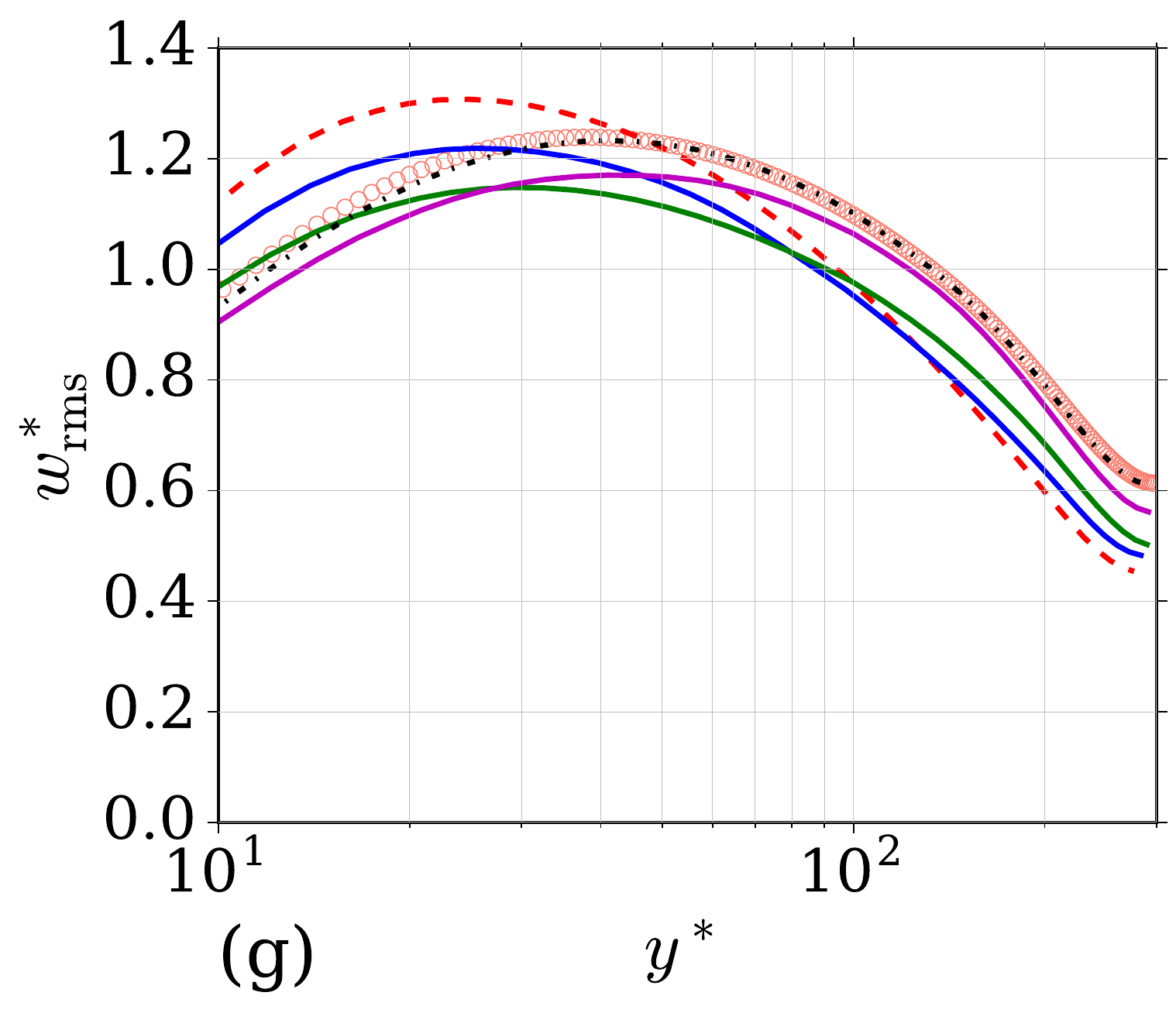} \\
   \end{tabular}
\caption{Profiles of $\U^*=\U/\ut$ (a), \revCom{$-\uv$ normalized by the maximum value of the corresponding DNS profile (b)}, $-\uv^* = -\uv/\ut^2$ (c), $\tke^*=\tke/\ut^2$ (d), $\urms^*=\urms/\ut$ (e), $\vrms^*=\vrms/\ut$ (f), and $\wrms^*=\wrms/\ut$ (g), for channel flow simulations with no explicit SGS model at target $\reyt=300$ and for $\dyp_w=1.105$. \revCom{The inner-scaled distance from the wall is $y^*=y \ut/\nu$. The plots in (b) are added for better visibility. The spacings $\dxp$ and $\dzp$ of the plots are specified in (a).} The DNS data of Iwamoto \et~\cite{iwamoto02} are shown by symbols.}\label{fig:someB_profs}
\end{figure}

In \fig~\ref{fig:someB_profs}, the impact of variation of $\dxp$ and $\dzp$ on the inner-scaled cross-channel profiles is shown.
\rev{The value of $\dyp_w$ and other simulation conditions are chosen to be the same as Set-B in \tab~\ref{tab:caseSummary}.}
The inner-scaled values, denoted by superscript $^*$ are based on the computed $\ut$.
It is clear that the Reynolds stress profile $-\uv^*$ shows the lowest sensitivity to the variation of streamwise and spanwise grid spacings, compared to the other profiles.
\rev{As also observed in \fig~\ref{fig:B_Lnorm_2}, even for $\dxp=16.56$ and $\dzp=9.96$ corresponding to the finest considered resolution in the $\dxp\dash\dzp$ plane, the discrepancy between the LES TKE and rms velocity fluctuation profiles and DNS data does not completely vanish.
However, by further refining $\dxp$ and $\dzp$ to $10$ and $5.6$, respectively, little improvement is achieved despite a significant increase in the computational cost.
}

Another important observation is that for the coarse resolutions, the computed TKE profile, $\tke_{\rm res}$,  is over-predicted compared to the DNS data, $\tke_{\rm tot}$. 
This over-prediction is mainly due to the over-prediction in $\uu$, taking into account higher relative importance of streamwise rms velocity fluctuation component compared to the others. 
Similarly, for other simulated cases in this study (not shown here), continuous reduction of the over-prediction in $\uu$ and $\tke$ with increasing resolution of $\dxp$ and $\dzp$ is observed. 
From the theoretical point of view\footnote{Generally speaking, in a wall-resolving LES at least $80\%$ of the total TKE is required to be resolved everywhere including the near wall region, see \cite{pope}.}, the excessive resolved TKE is not acceptable.
\rev{It is expected that as the LES grid is refined, the uncaptured portion of the energy due to the unresolved scales reduces and eventually completely goes away.}
However, over-prediction of TKE \revCom{due to under-resolving the near-wall region} is a known issue in the numerical simulation of wall-bounded turbulent flows, for instance see the discussions in \cite{celik:05,klein:05,meyers07,bae:17}, and the references therein.

\rev{As a conclusion}, when seeking for a high quality LES through refining the grid, 
\rev{one has to make sure that acceptably-low errors in different quantities are simultaneously achieved.}
In particular, based on the simulations in Set-ABC, it is observed that even for those combinations of grid spacings that $u_\tau$, mean velocity profile, and Reynolds stress profile are accurately computed, the rms velocity fluctuations and kinetic energy profiles still deviate from the reference DNS.

\subsection{Effect of Reynolds number}\label{sec:ReEffects}
\begin{figure}[!htbp]
\centering
   \begin{tabular}{ccc}
   \includegraphics[scale=0.26]{A_DuTau.pdf} &
   \includegraphics[scale=0.26]{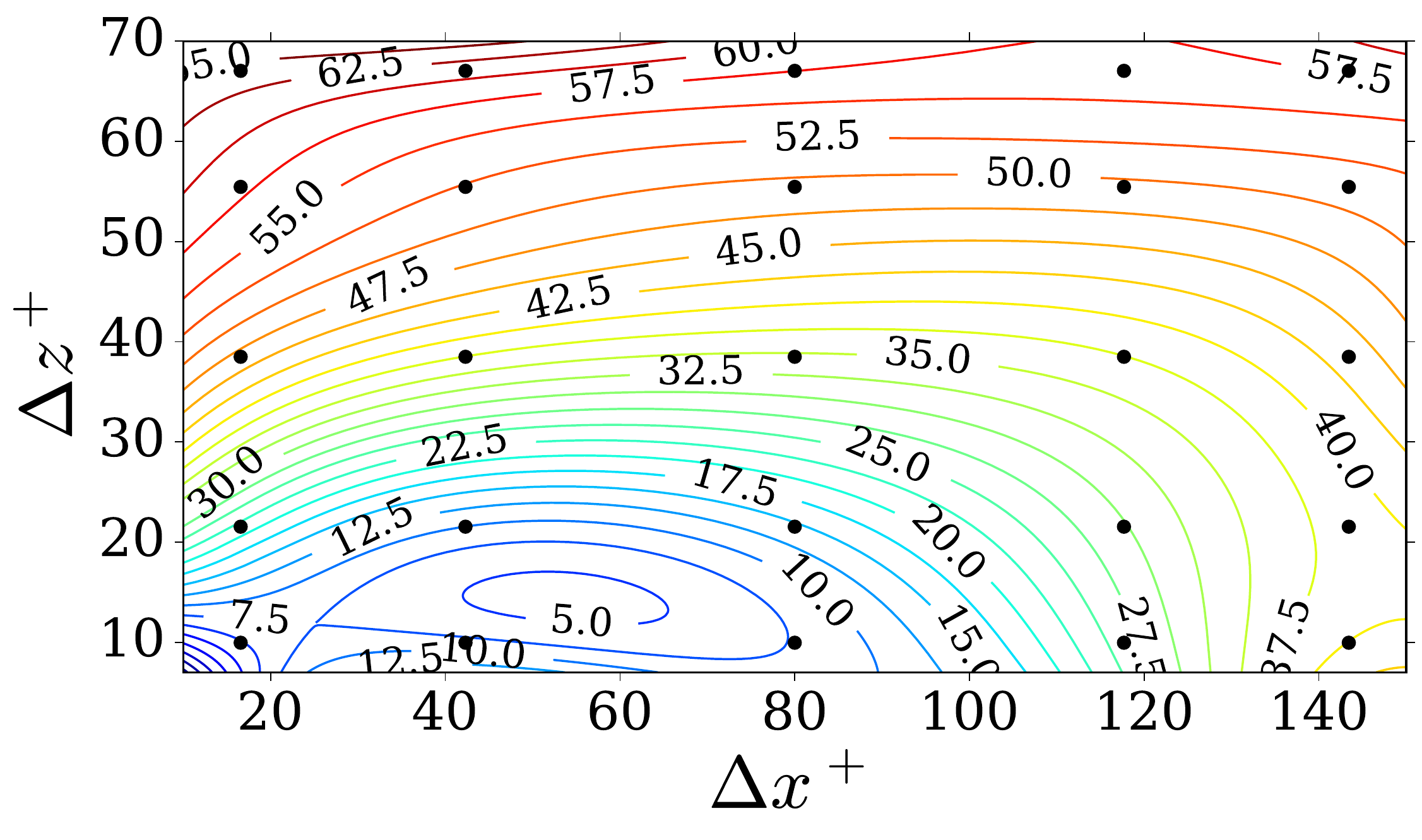} &
   \includegraphics[scale=0.26]{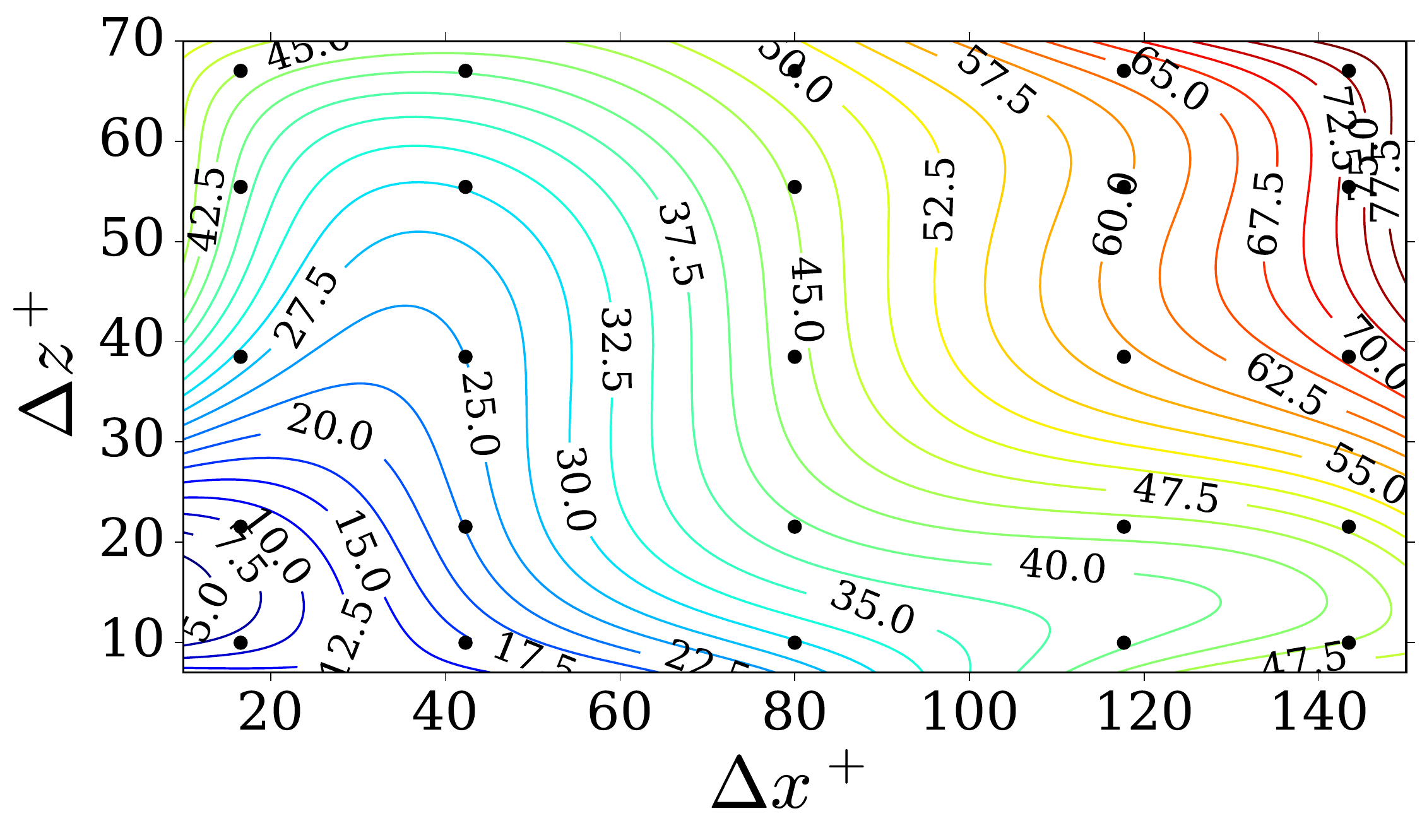} \\   
   {\small (a)} &    {\small (b)} &    {\small (c)} \\   
   \end{tabular}
   \begin{tabular}{ccc}
   \includegraphics[scale=0.26]{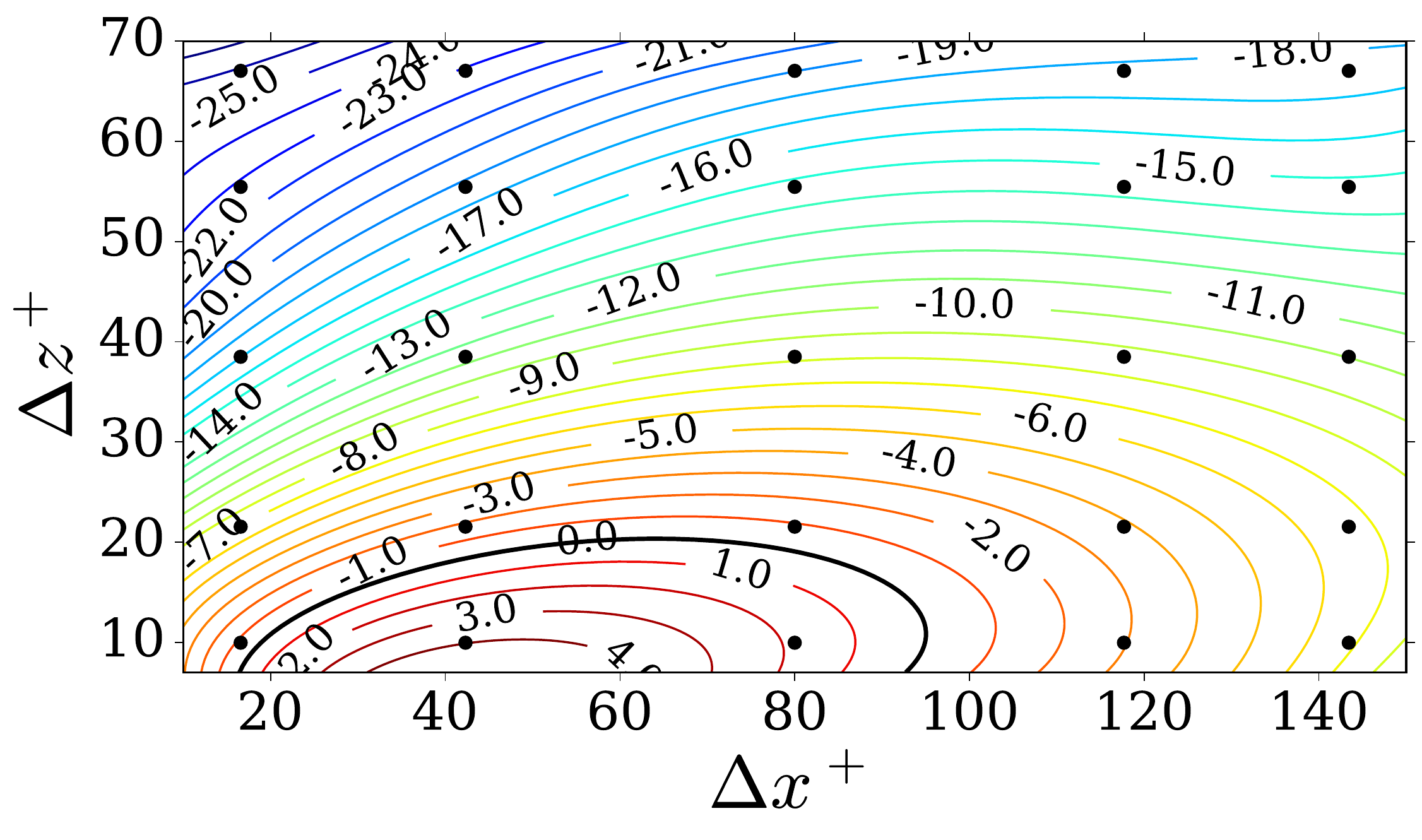} &
   \includegraphics[scale=0.26]{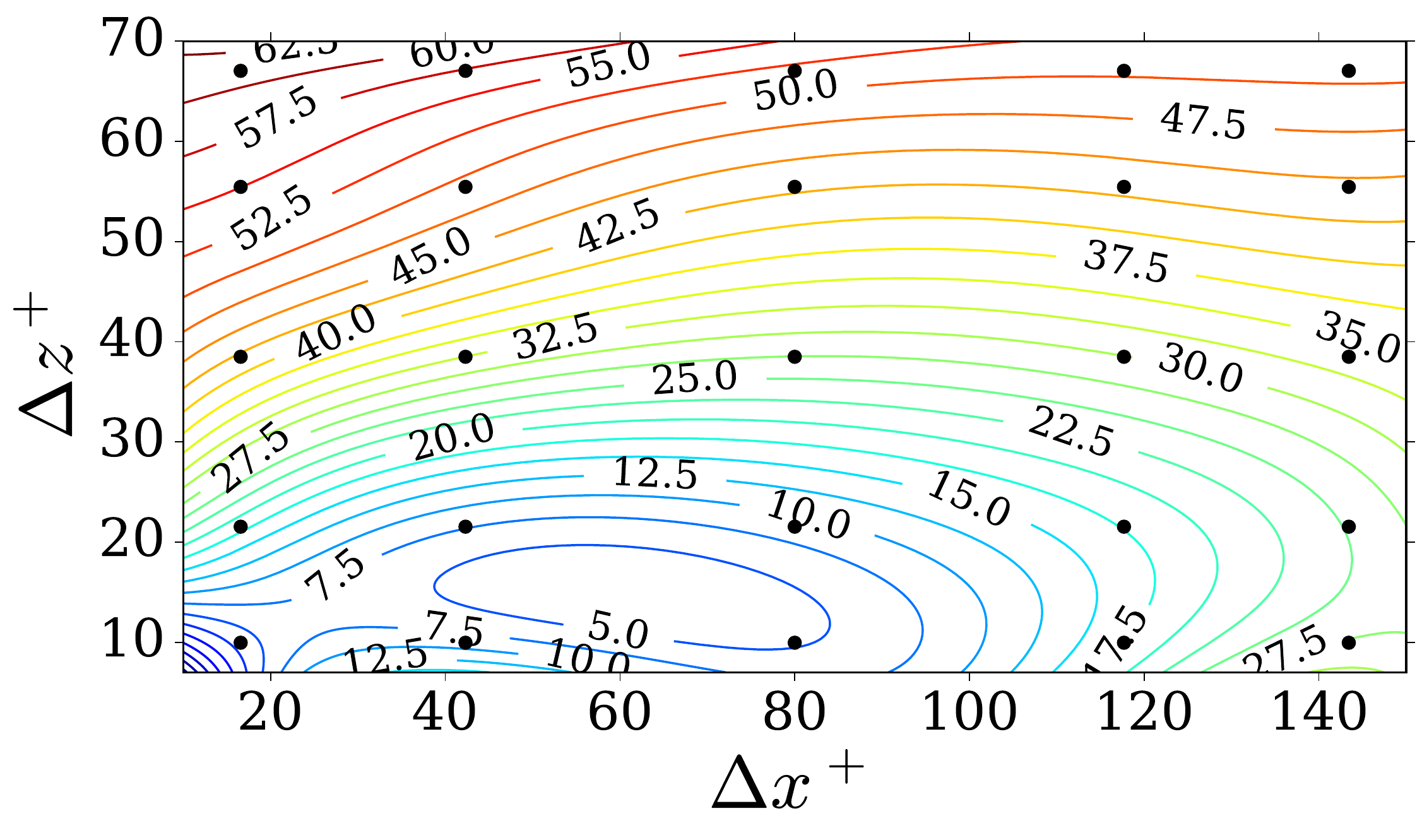} &
   \includegraphics[scale=0.26]{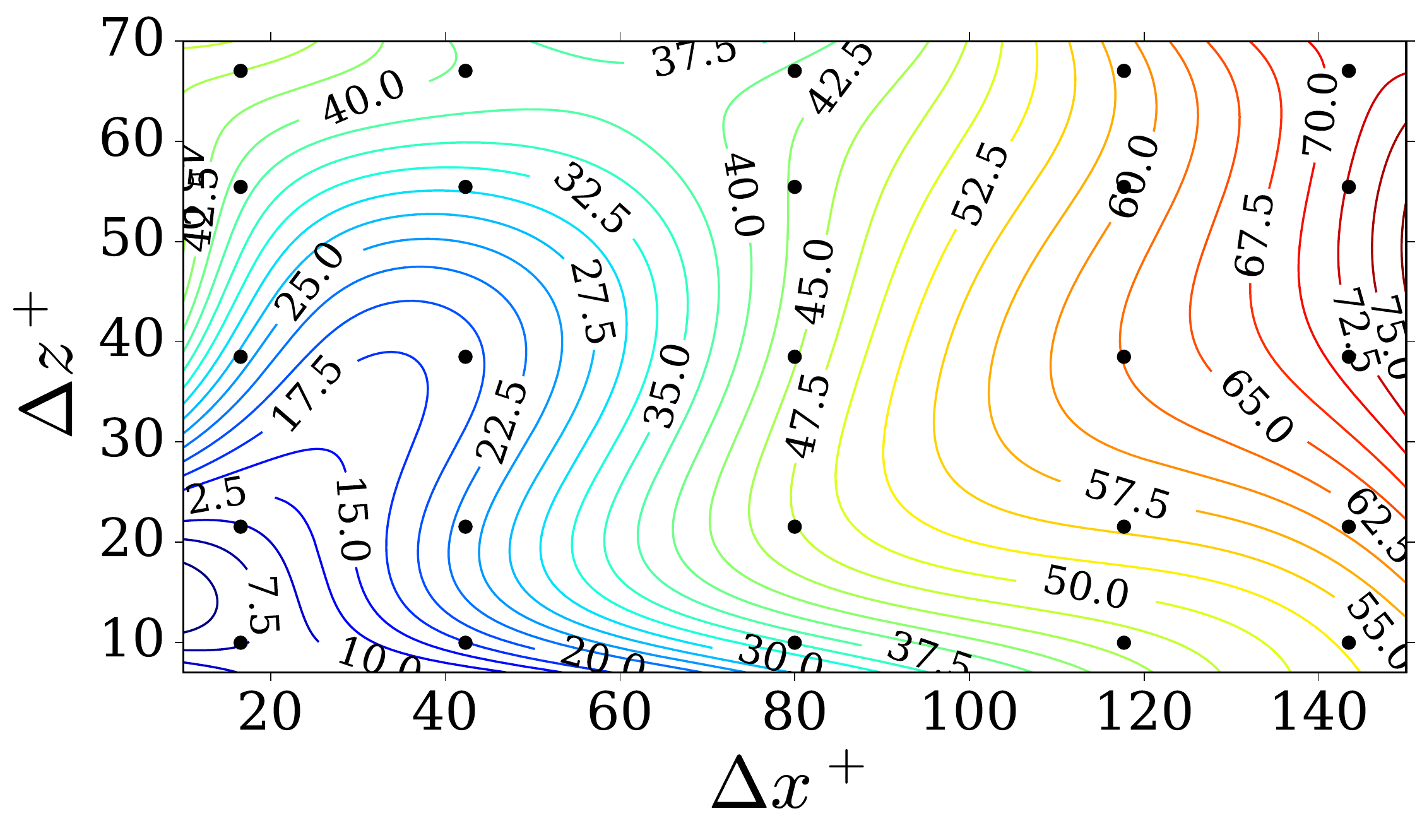} \\   
   {\small (d)} &    {\small (e)} &    {\small (f)} \\
   \end{tabular}   
\caption{Isolines of $\epsilon[\ut]$ (a,d), $\einf[\uv]$ (b,e), and $\einf[\tke]$ (c,f), plotted in the $\dxp\dash\dzp$ plane for Set-A (top row), Set-D (bottom row).}\label{fig:E_contours}
\end{figure}

\rev{To investigate how the error responses are sensitive to variation of $\rey$-number, the focus will be on Set-D with target $\reyt=550$.}
For this set, the loci of zero $\epsilon[u_\tau]$, $\earea[|\uv|]$, and $\epik[\uv]$ in the $\dxp\dash\dzp$ plane are found to be very similar to each other and also to those of Set-A, but encompassing larger values of $\dxp$ and $\dzp$ compared to Set-A, see \fig~\ref{fig:E_contours}(a,d).  
In addition, the isolines of $\einf[\uv]$, and $\einf[\tke]$, as shown in \fig~\ref{fig:E_contours}(b-f), approximately look like the corresponding error lines of Set-A. 
Through observing relatively similar pattern in the error isolines of the other quantities of Set-D and Set-A, the validity of the discussions made in the previous sections is confirmed for $\reyt=550$.

Now, it is interesting to investigate if the insignificant influence of the Reynolds number on the patterns of the error isolines is also expected for other target $\reyt$ between $300$ and $550$.
To this end, it is assumed that the DNS converged value of Reynolds number, i.e. $\reyt^\circ$, is an uncertain parameter, in addition to the grid spacings.
In particular, in the framework of uncertainty propagation discussed in \sect~\ref{sec:uq}, $\reyt^\circ$ is assumed to be a uniformly-distributed random parameter varying over the presumed admissible range $\Qr=[208.63,633.34]$. 
Within this specific range, the DNS Reynolds numbers $297.899$ and $543.496$, respectively associated with target $\reyt=$ 300 and 550, specify the two Gauss quadrature points. 
By this construction, expansion (\ref{eq:pce}) can be used as a meta model to obtain the errors between LES and DNS of channel flow for any combination of $\dxp\sim \mathcal{U}[\Qx]$, $\dyp_w\sim \mathcal{U}[\Qy]$, $\dzp\sim \mathcal{U}[\Qz]$, and $\reyt^\circ \sim \mathcal{U}[\Qr]$.
It is clear that due to the mutual independence of these uncertain parameters, extension of the meta model, \rev{developed for the grid spacings,} to include the effects of the Reynolds number is straightforward.

\begin{figure}[!htbp]
\centering
   \begin{tabular}{cc}
   \includegraphics[scale=0.4]{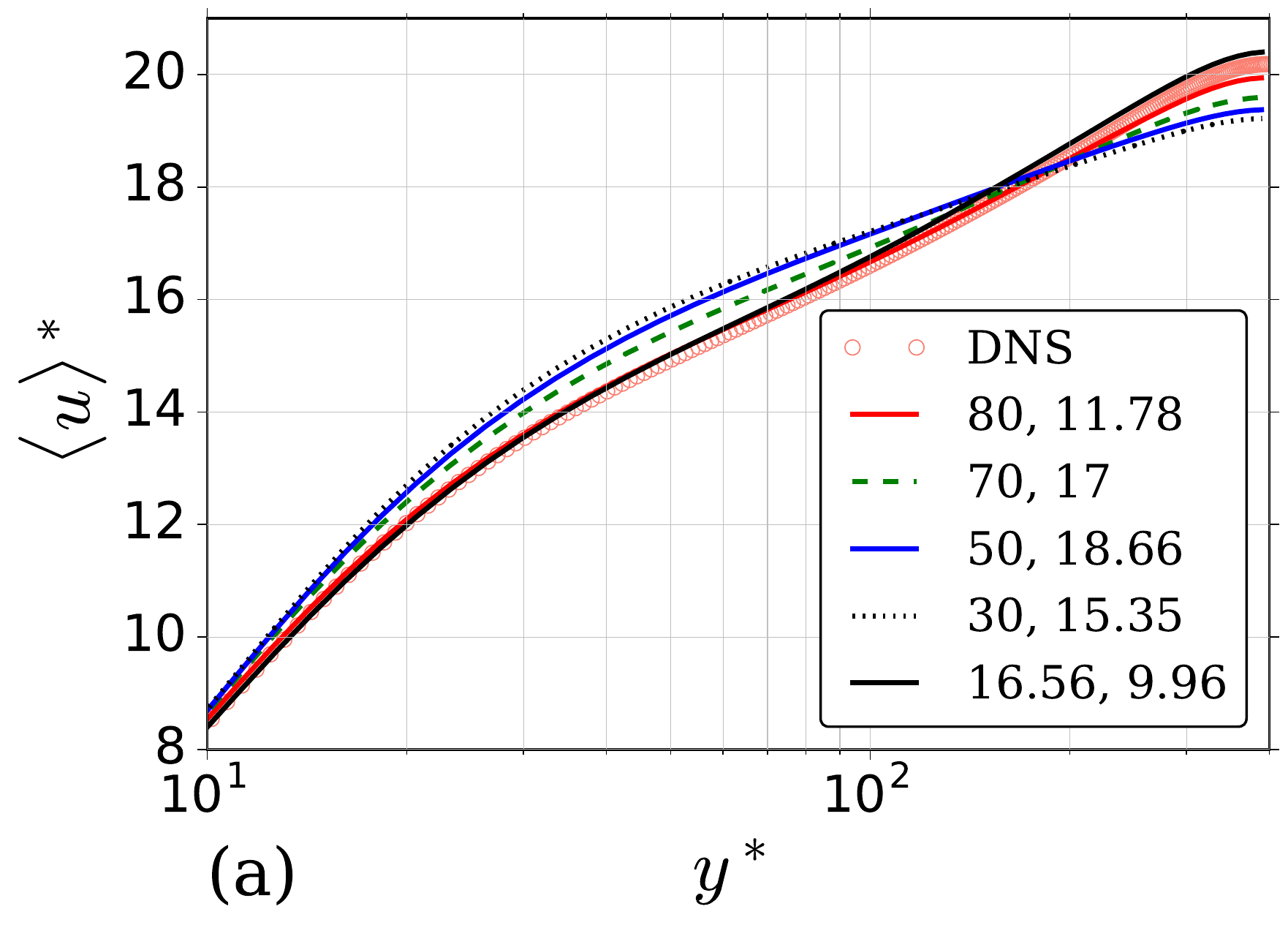} &
   \includegraphics[scale=0.4]{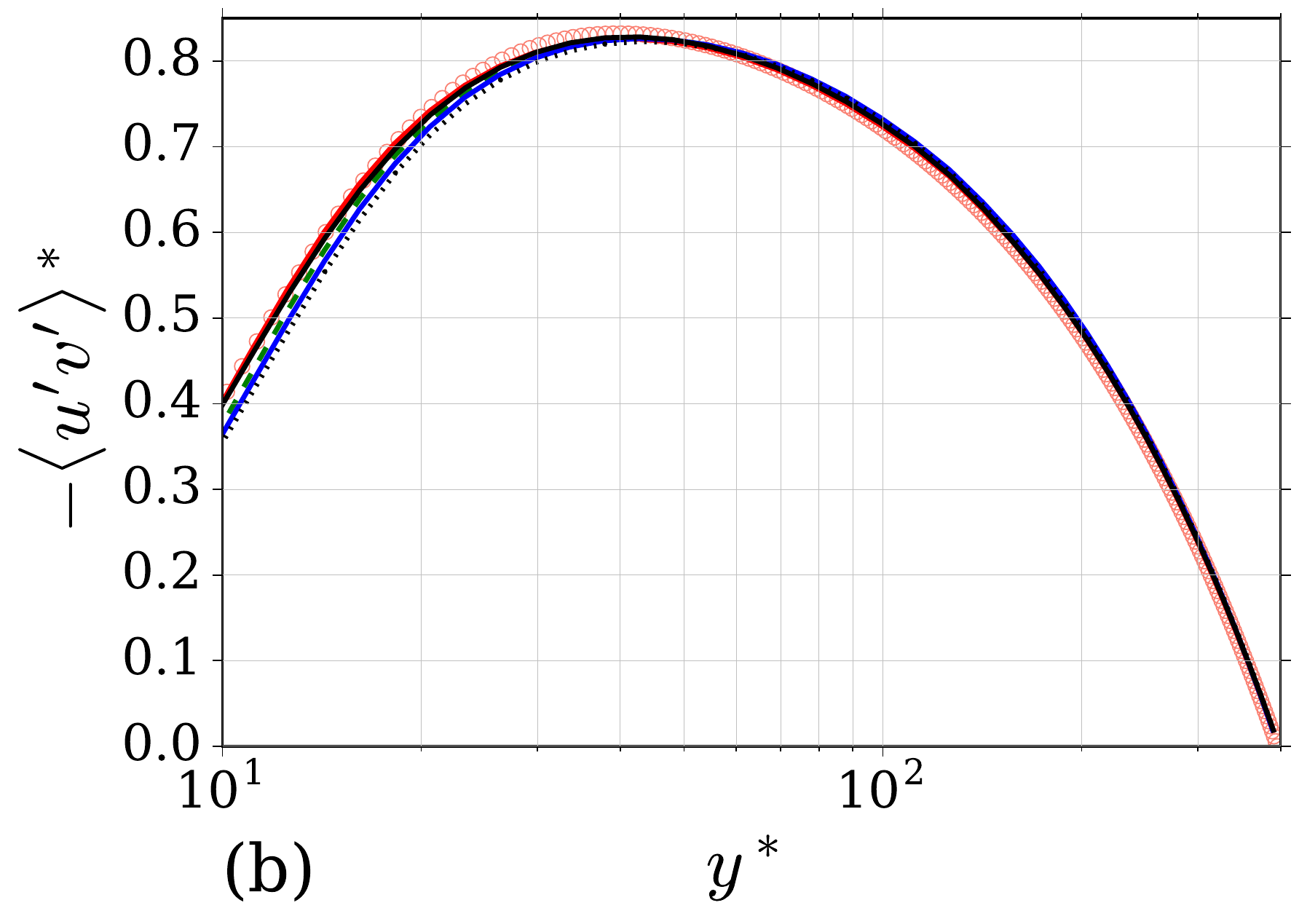} \\
   \end{tabular}
   \begin{tabular}{c}
   \includegraphics[scale=0.4]{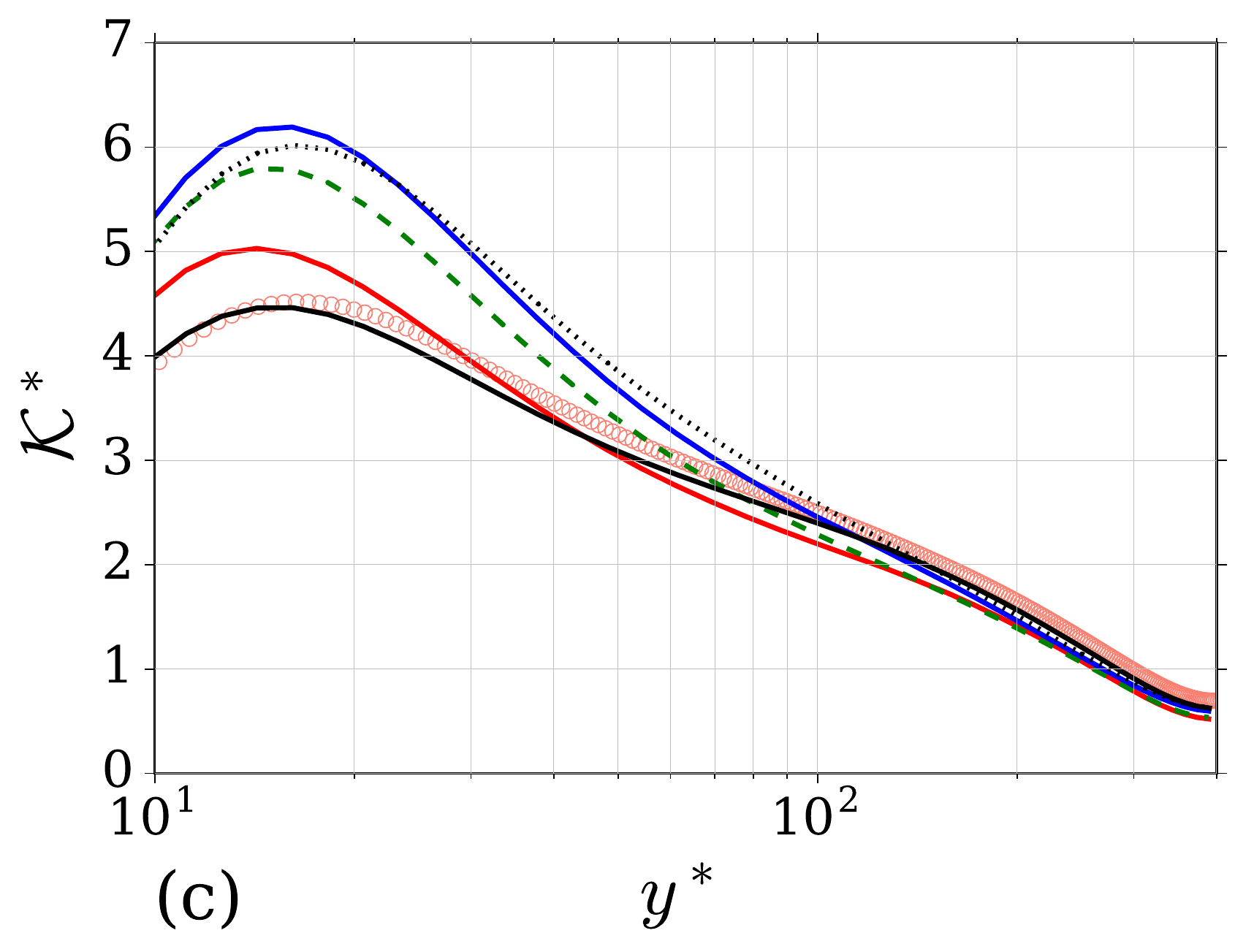} \\
   \end{tabular}   
\caption{Profiles of $\U^*$ (a), $-\uv^*$ (b), and $\tke^*$ (c) plotted against $y^*$ for simulations with no explicit SGS model at target $\reyt=400$, and for $\dyp_w=0.445$. Spacings $\dxp$ and $\dzp$ are specified in (a). The DNS data of Iwamoto \et~\cite{iwamoto02} are shown by symbols.}\label{fig:Q_profs}
\end{figure}

To illustrate an example, the aim is to investigate if a curved loci of zero $\epsilon[\ut]$, such as those in \fig~\ref{fig:E_contours}(a,d), would also exist at target $\reyt=400$. 
The DNS of Iwamoto \et ~\cite{iwamoto02}, with converged $\reyt^\circ=395.760$, is used as the reference. 
\rev{Specifically, for five arbitrarily-chosen $\dxp$, corresponding $\dzp$ values are selected such that $(\dxp,\dzp)$ reside on the curved loci of zero $\epsilon[\ut]$ in the $\dxp\dash\dzp$ plane belonging to Set-A and Set-D (corresponding to $\reyt=300$ and $550$, respectively, with $\dyp_w=0.445$).}
By linear interpolation in $\reyt^\circ$, five sets of $\dxp$ and $\dzp$ are determined for which $\epsilon[\ut]$ at $\reyt^\circ=395.760$ is predicted to be ideally zero. 
In order to cross-validate, the predicted errors by the meta model at the mentioned five resolutions are compared to the actually computed errors between the corresponding LES results and DNS data.
Doing so, $|\epsilon[u_\tau]|$ is found to be less than $0.30\%$ for all five simulations.

Associated with these five resolutions, the computed mean velocity, Reynolds stress, and TKE profiles are plotted in \fig~\ref{fig:Q_profs} in wall-units. 
Similar to what was observed for Set-ABC and Set-D, a low error in $u_\tau$ is not necessarily followed by the error reduction up to the same extent in the other quantities. 
In fact, only for one simulation that has the finest $\dxp$ and $\dzp$ among the other cases, i.e. $\dxp=16.56$ and $\dzp=9.96$, the mean velocity and TKE profiles agree well with the DNS data. 
It is also observed that, for combinations of $\dxp$ and $\dzp$ that $\epsilon[\ut]$ is very close to zero, the inner-scaled $-\uv^*$ profiles nearly (but not completely) match the DNS profile, see \fig~\ref{fig:Q_profs}(b).

\subsection{Effect of SGS modeling}
\begin{figure}[!htbp]
\centering
   \begin{tabular}{cc}
   \includegraphics[scale=0.35]{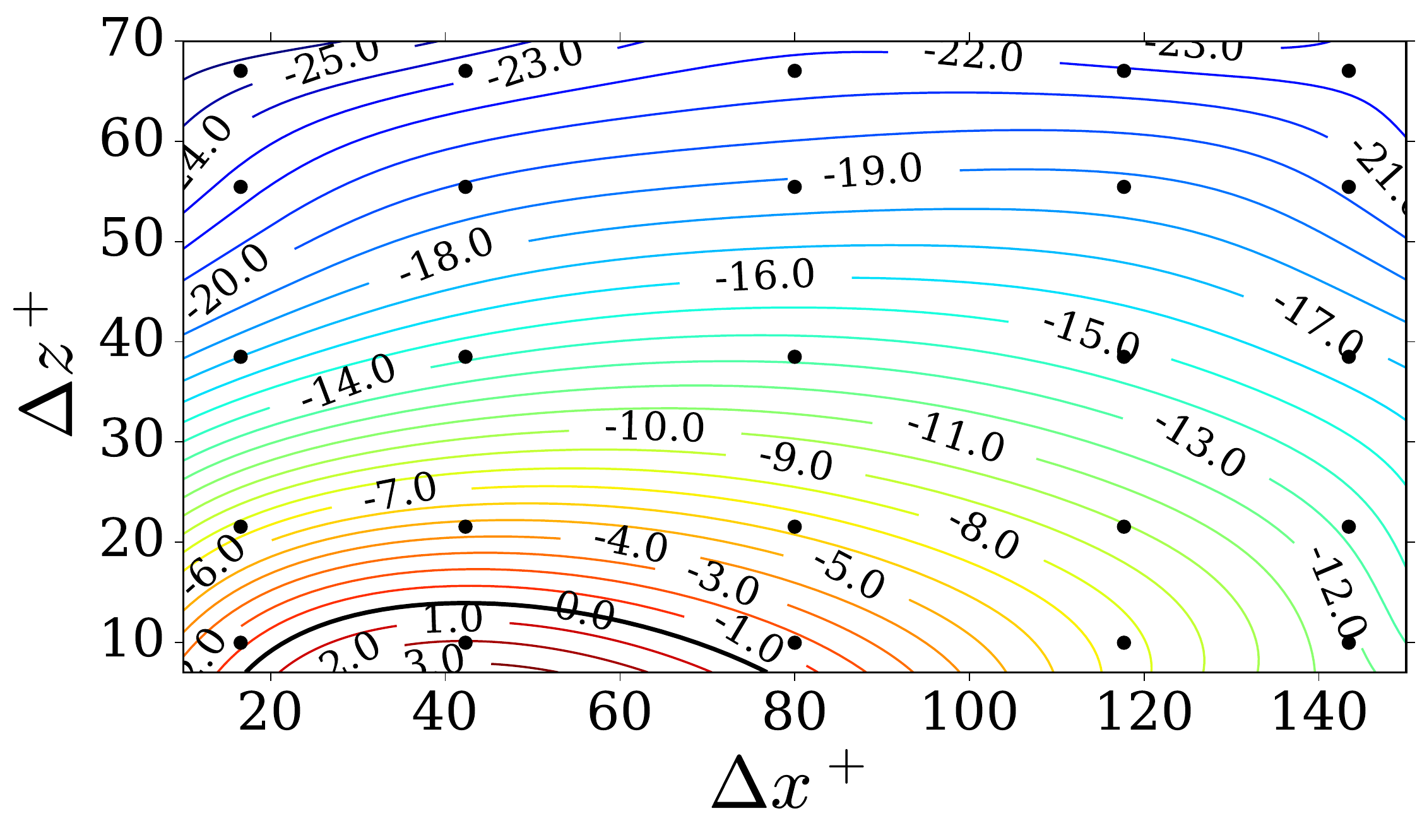} &
   \includegraphics[scale=0.35]{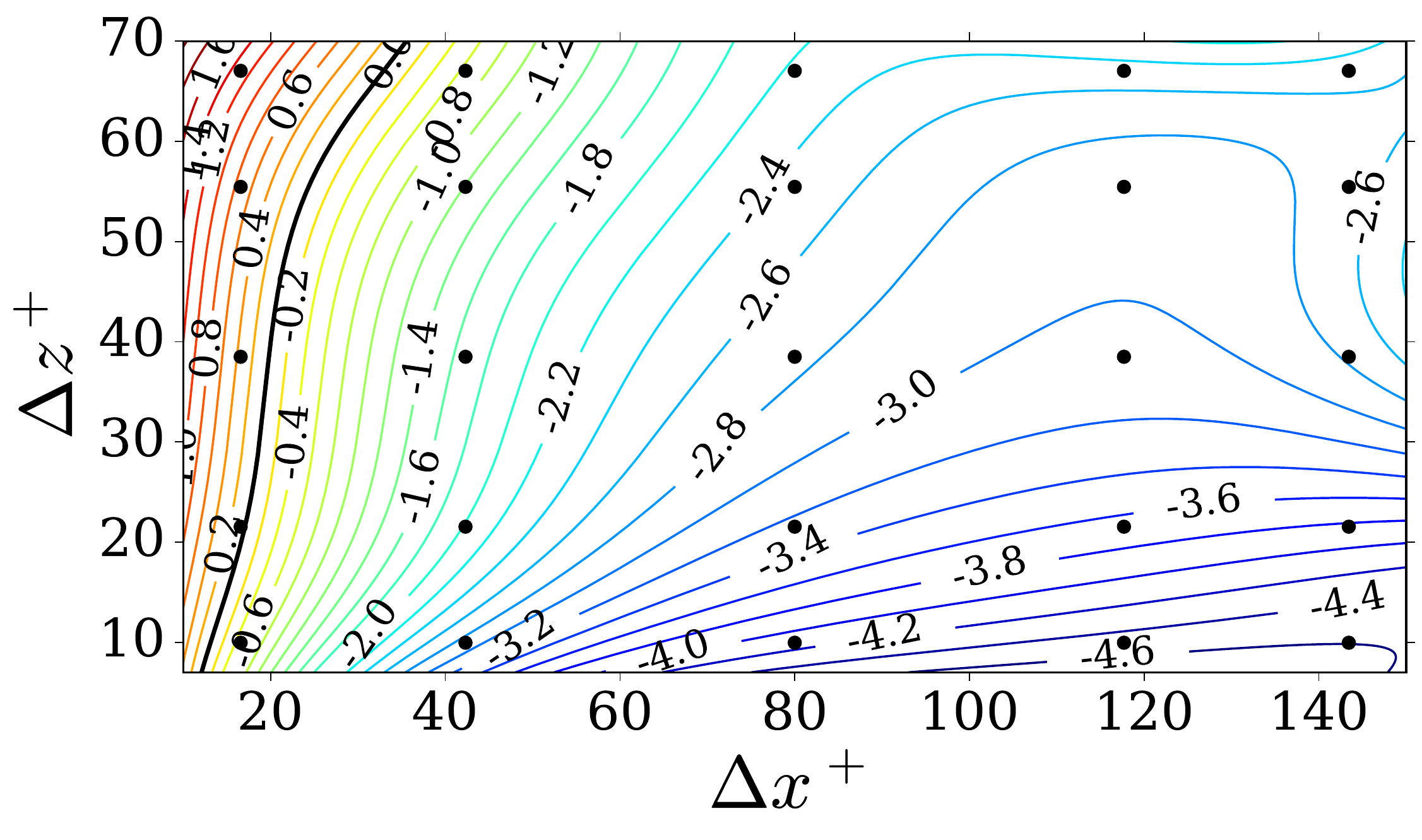} \\
   {\small (a)} & {\small (b)} \\
   \includegraphics[scale=0.35]{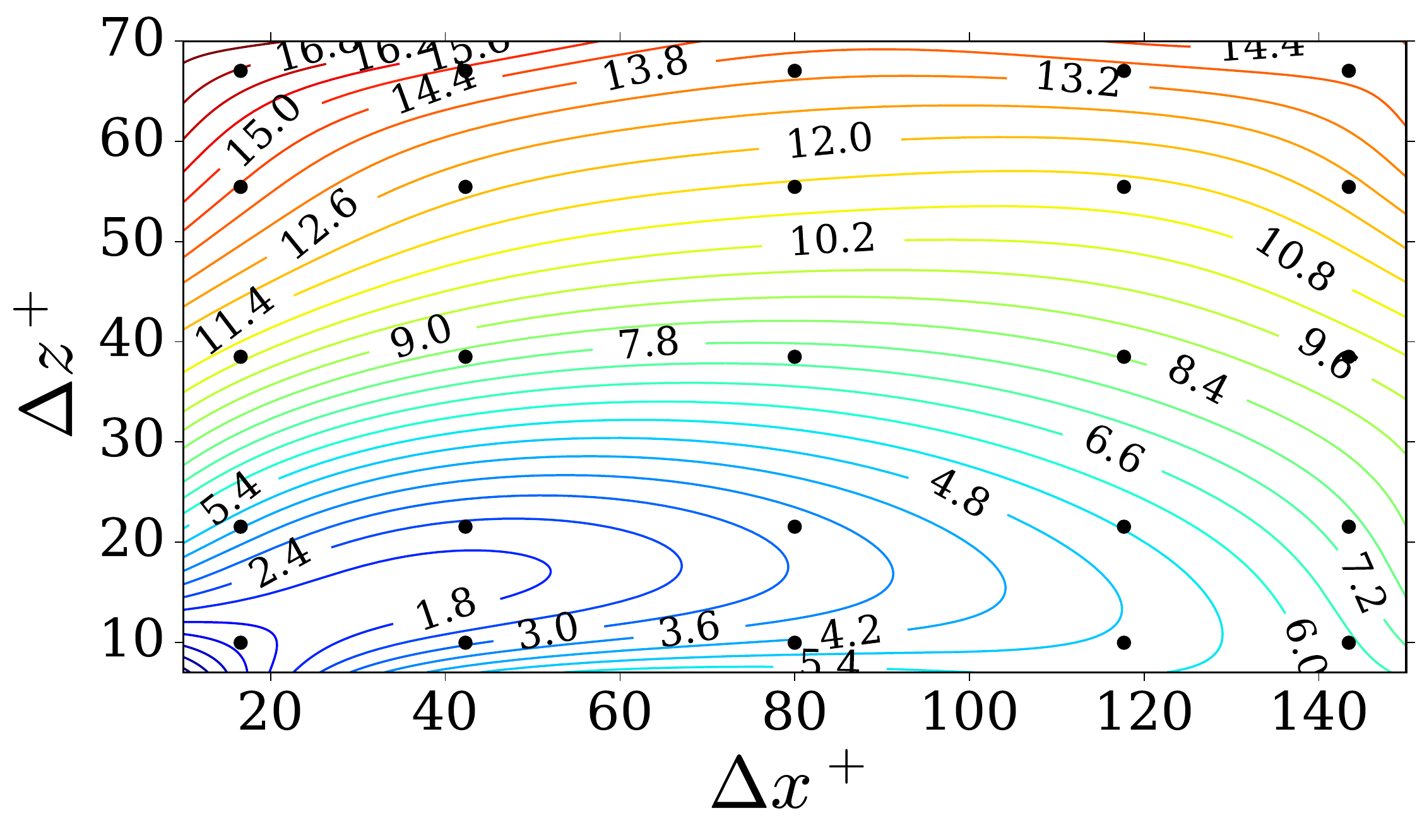} &
   \includegraphics[scale=0.35]{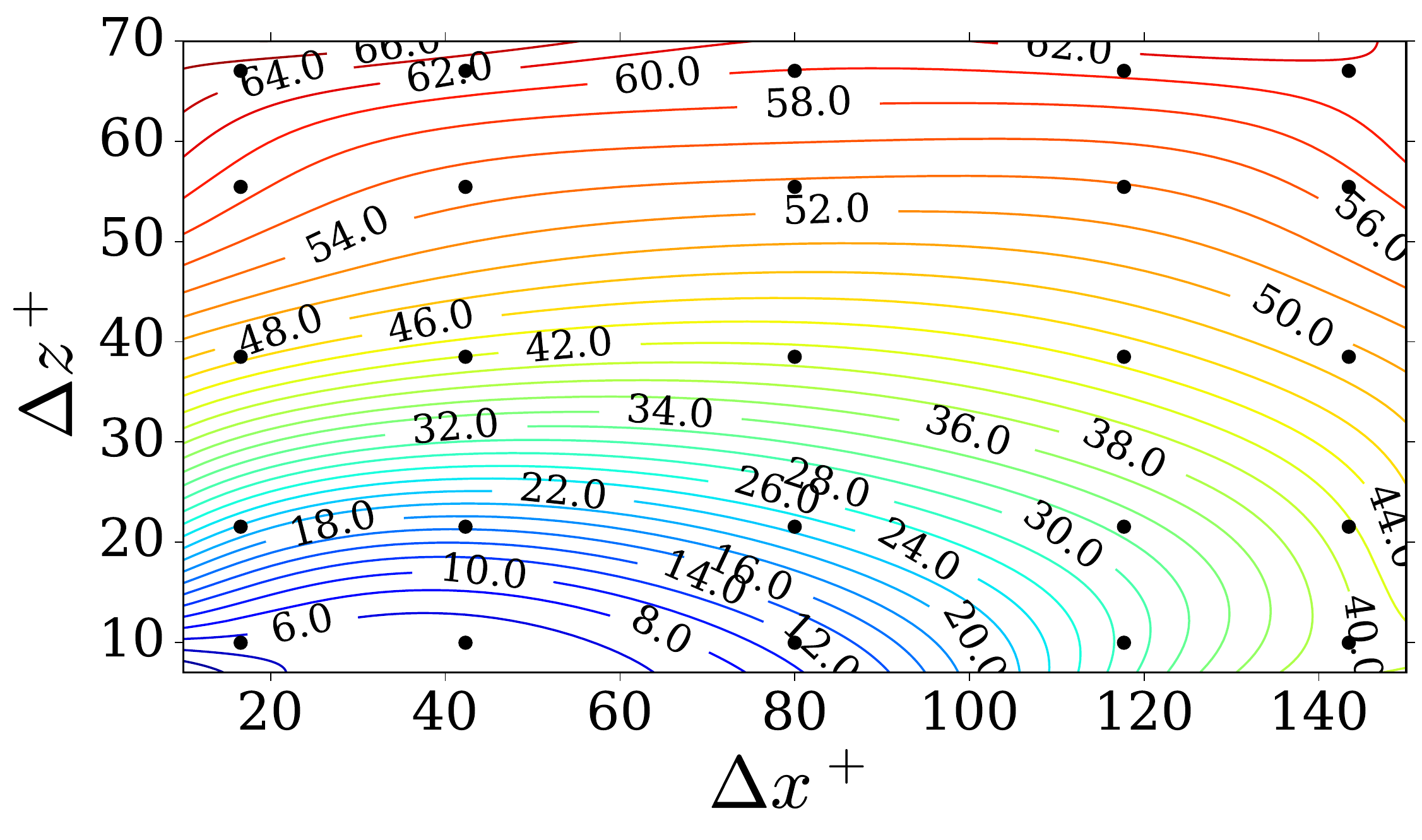} \\
   {\small (c)} & {\small (d)} \\
   \includegraphics[scale=0.35]{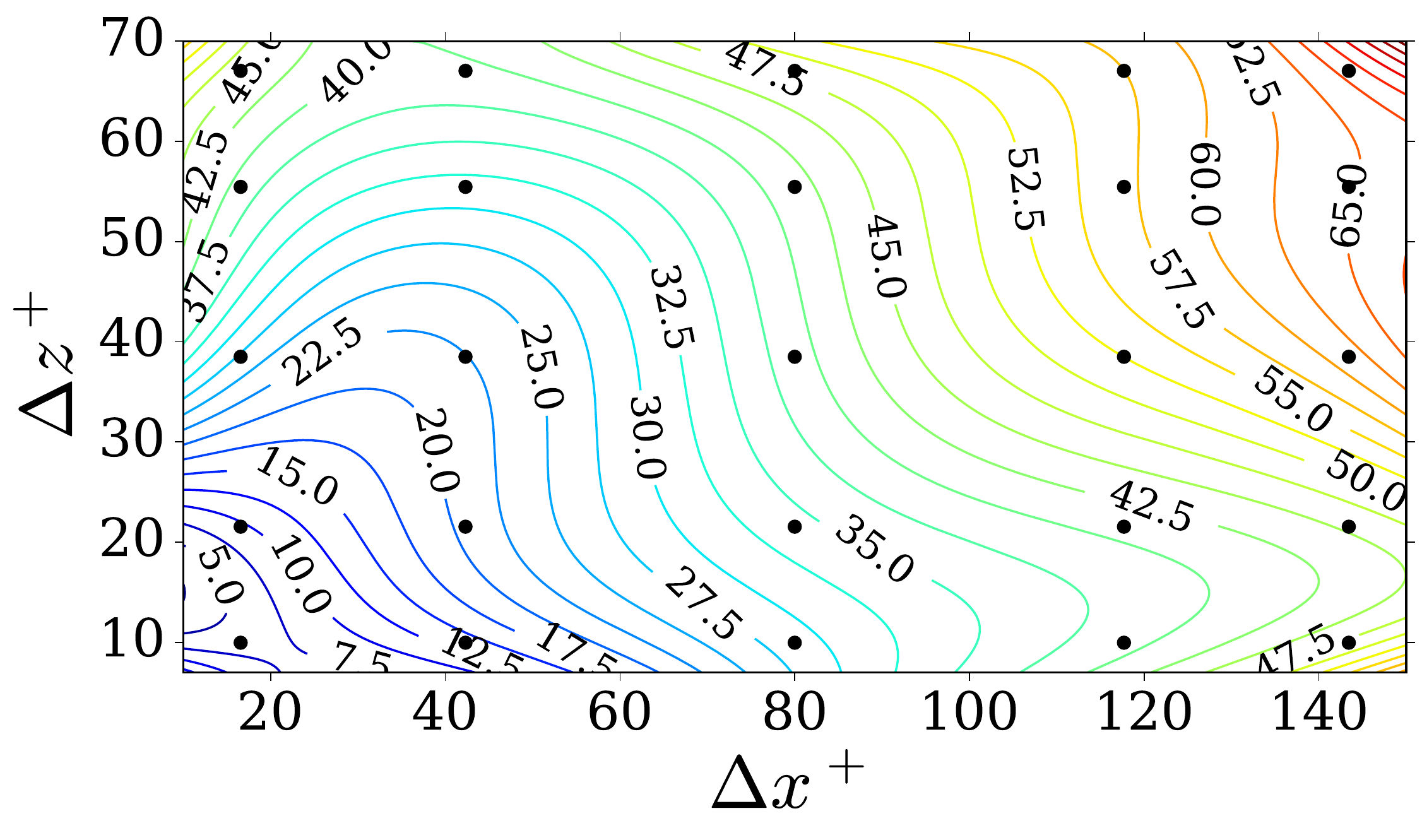} &
   \includegraphics[scale=0.35]{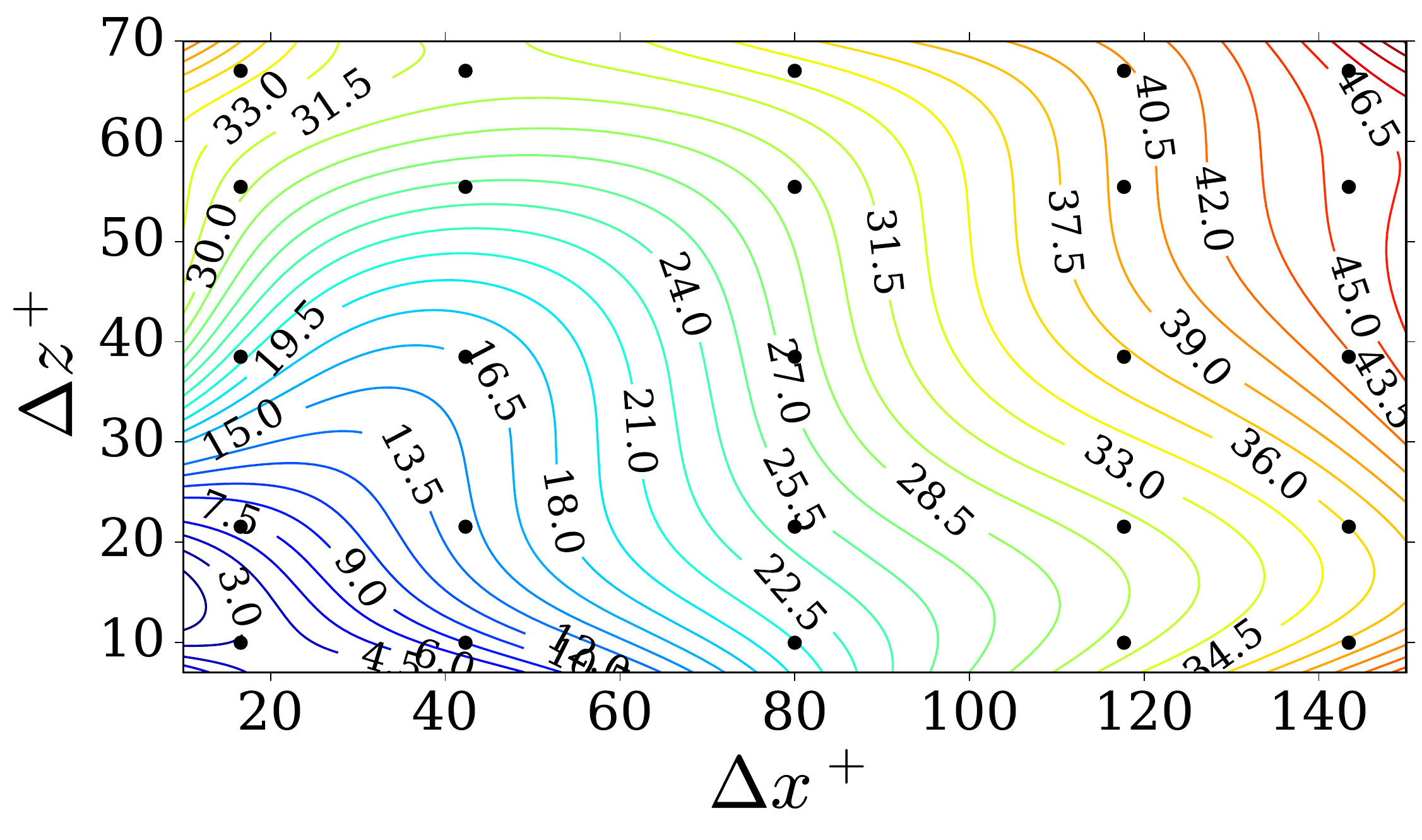} \\   
   {\small (e)} & {\small (f)} \\
   \end{tabular}   
\caption{Isolines of $\epsilon[\ut]$ (a), $\epsilon[\U_c]$ (b), $\einf[\U]$ (c), $\einf[\uv]$ (d), $\einf[\tke]$ (e), and $\einf[\urms]$ (f), plotted in the $\dxp\dash\dzp$ plane for Set-Bw.}\label{fig:Bw_contours}
\end{figure}

In the discussions made up to this point, no explicit SGS modeling was included in the channel flow simulations. 
But, it is important to see how the responses of the LES of channel flow would be influenced by explicit SGS modeling. 
Among many options, WALE model \cite{nicoud:99} with the default coefficient value $0.325$ is chosen to simulate turbulent channel flows at target $\reyt=300$, as denoted by Set-Bw in \tab~\ref{tab:caseSummary}.
The isolines of the errors of this simulation set are plotted in the $\dxp\dash\dzp$ plane in \fig ~\ref{fig:Bw_contours}. 
The small differences between these contours with those in \figs ~\ref{fig:B_Lnorm_1} and \ref{fig:B_Lnorm_2}, belonging to the same simulation settings but without SGS model (Set-B), reveal the fact that for the FV-based implicitly-filtered LES carried out using nominally second-order accurate schemes, see \sect~\ref{sec:LESCFD}, the numerical errors play the dominant role compared to the errors induced by the adopted SGS model.
This is consistent with the results in \cite{vreman96,geurts05,GMR:1,tyacke13}.

Despite this, the errors in different quantities can be slightly affected by SGS modeling. 
In particular, for any $\dxp\in \BQ_{\dxp}$ and $\dzp\in \BQ_{\dzp}$, the errors in $\uv$ are found to be slightly higher in Set-Bw compared to the corresponding values in Set-B.
Accordingly, the area under the zero $\epsilon[\ut]$ curve becomes smaller, meaning that when the WALE SGS model is employed, a finer resolution in the spanwise direction is required in order to achieve the same level of error as it would be reached without any explicit SGS model.
In contrast, the over-prediction in TKE is reduced a little by including the explicit SGS model. 
By more investigation (not shown here), this reduction is found to be mainly due to the reduction of the over-prediction in the streamwise component of the rms velocity fluctuations.



\subsection{Variation of the errors in the $\dxp\dash\dzp$ plane}\label{sec:errRates}
In this section, for a selected simulation set listed in \tab~\ref{tab:caseSummary}, it is shown how variation of the grid spacings in the wall-parallel plane, i.e. $\dxp$ and $\dzp$, may influence the rate of variation of the errors \rev{in different QoIs computed by LES.}

\begin{figure}[!htbp]
\centering
   \begin{tabular}{ccc}
   \includegraphics[scale=0.38]{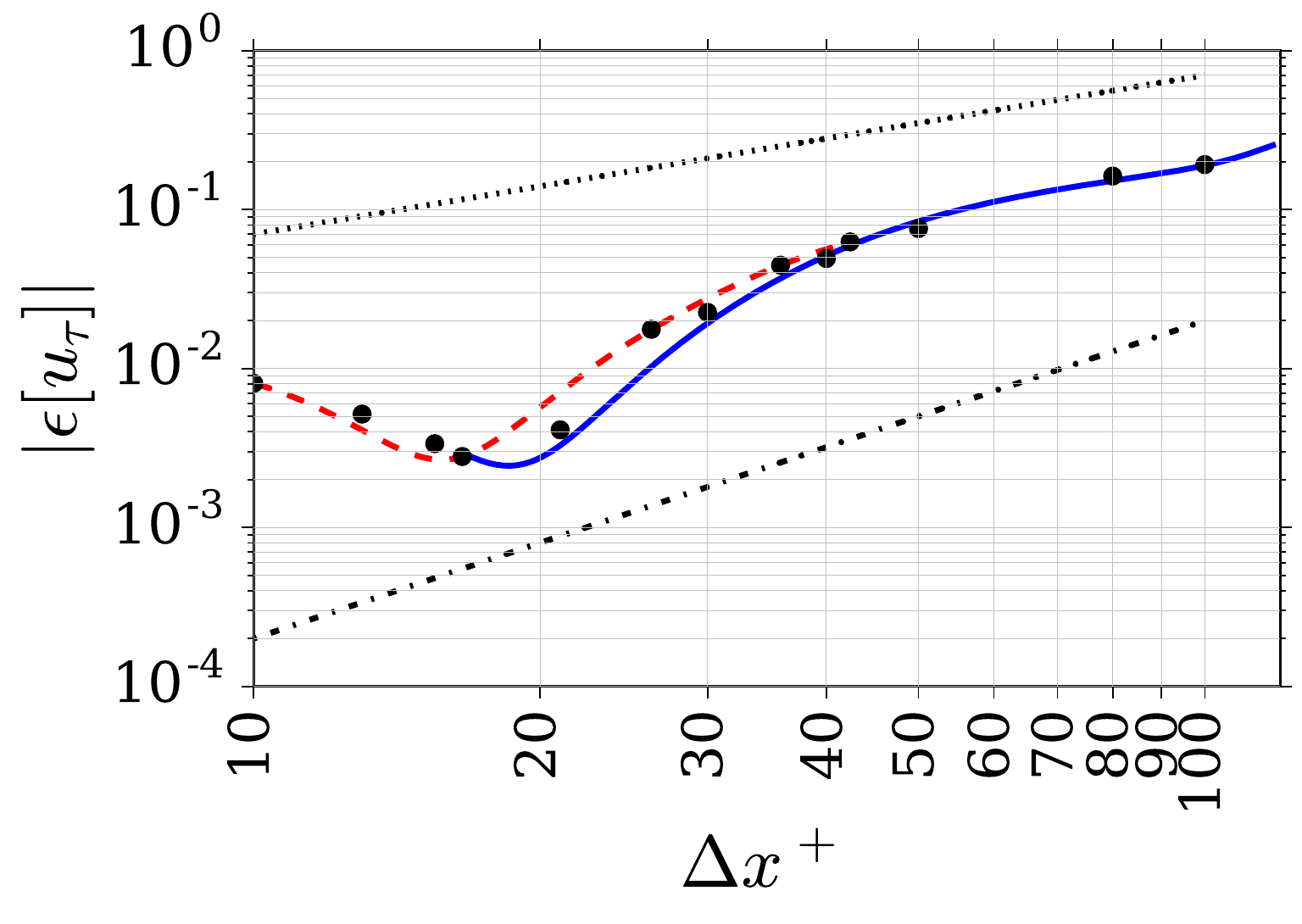} &
   \includegraphics[scale=0.38]{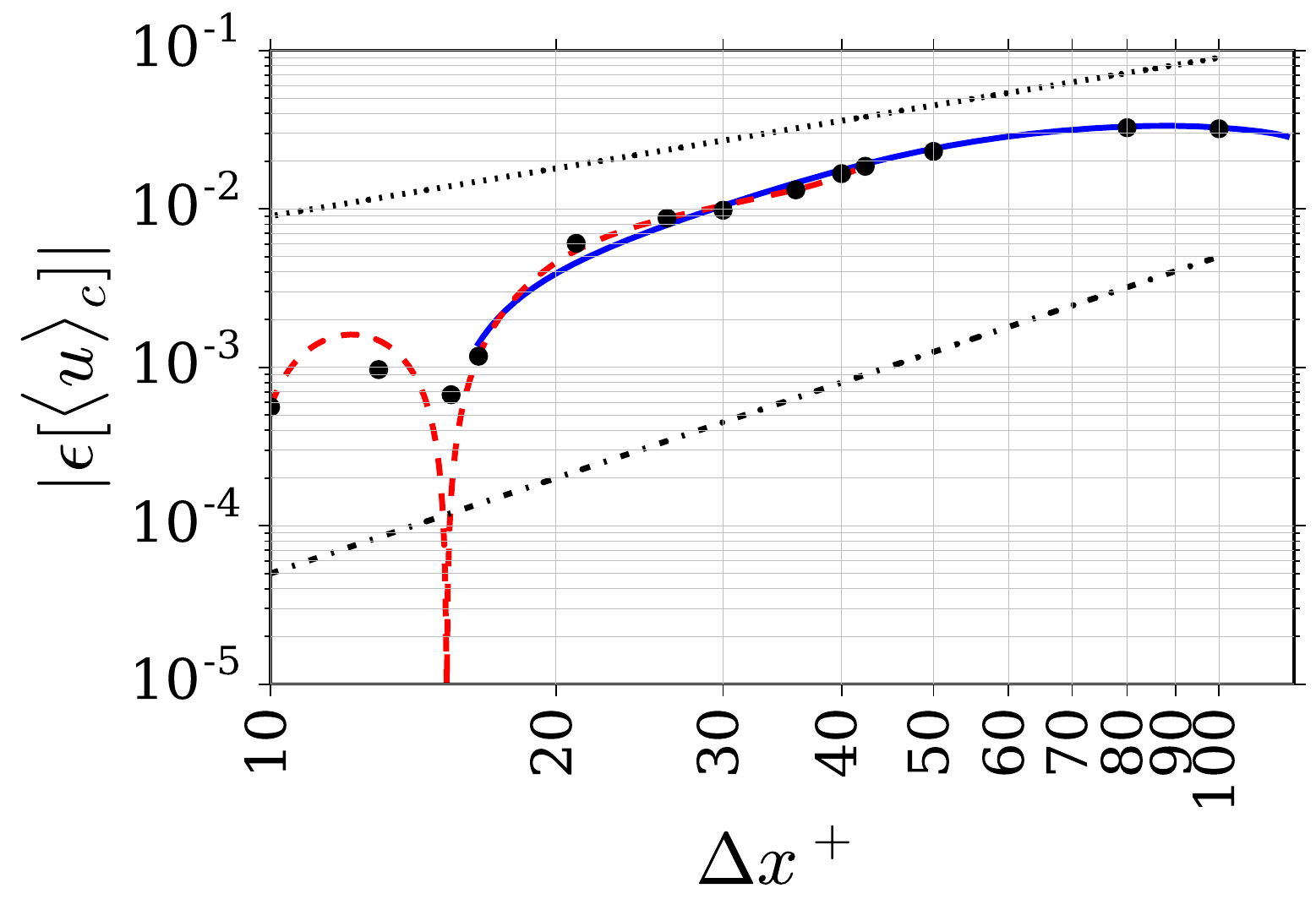} &
   \includegraphics[scale=0.38]{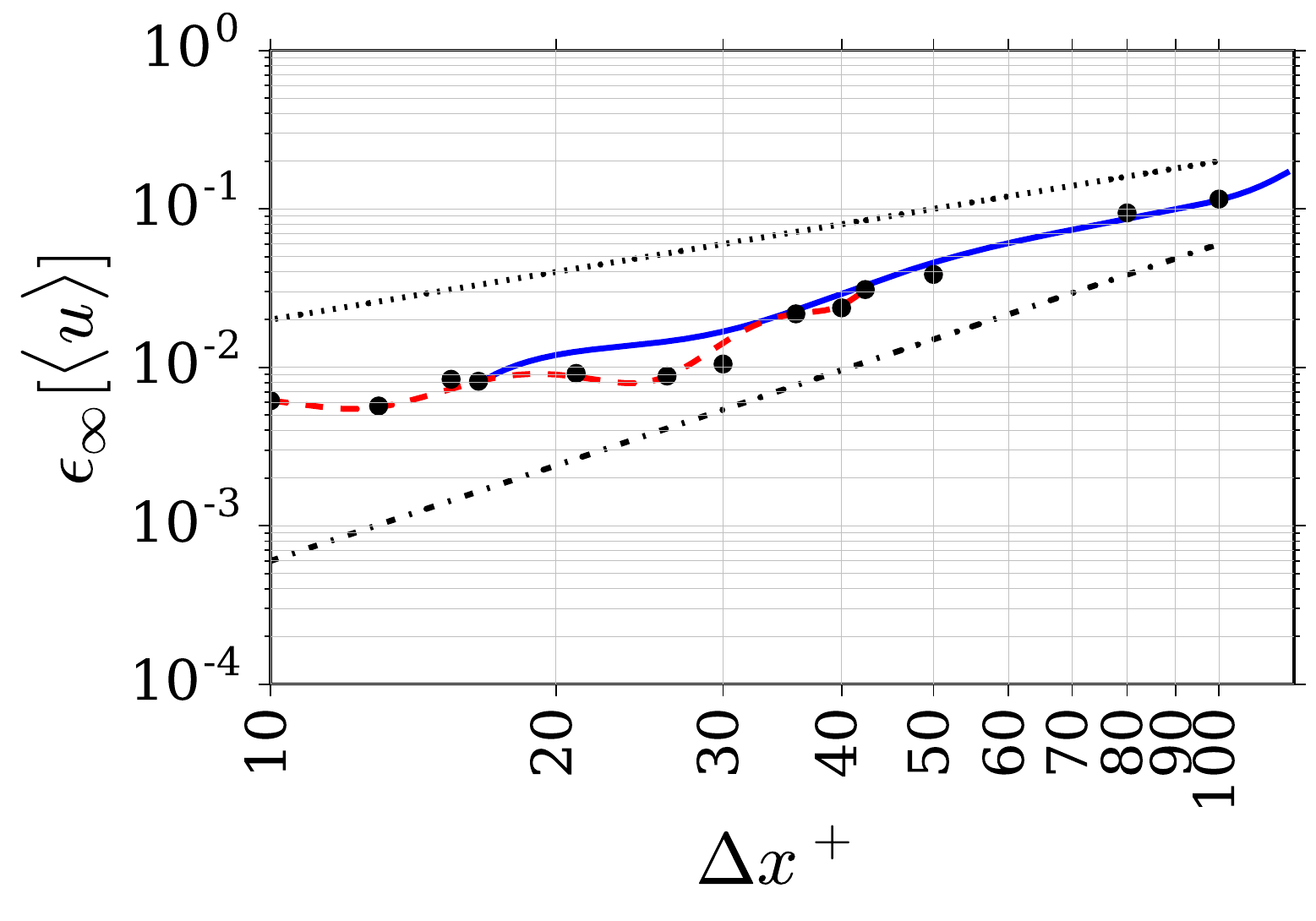} \\      
   {\small (a)} &    {\small (b)} &    {\small (c)} \\      
   \includegraphics[scale=0.38]{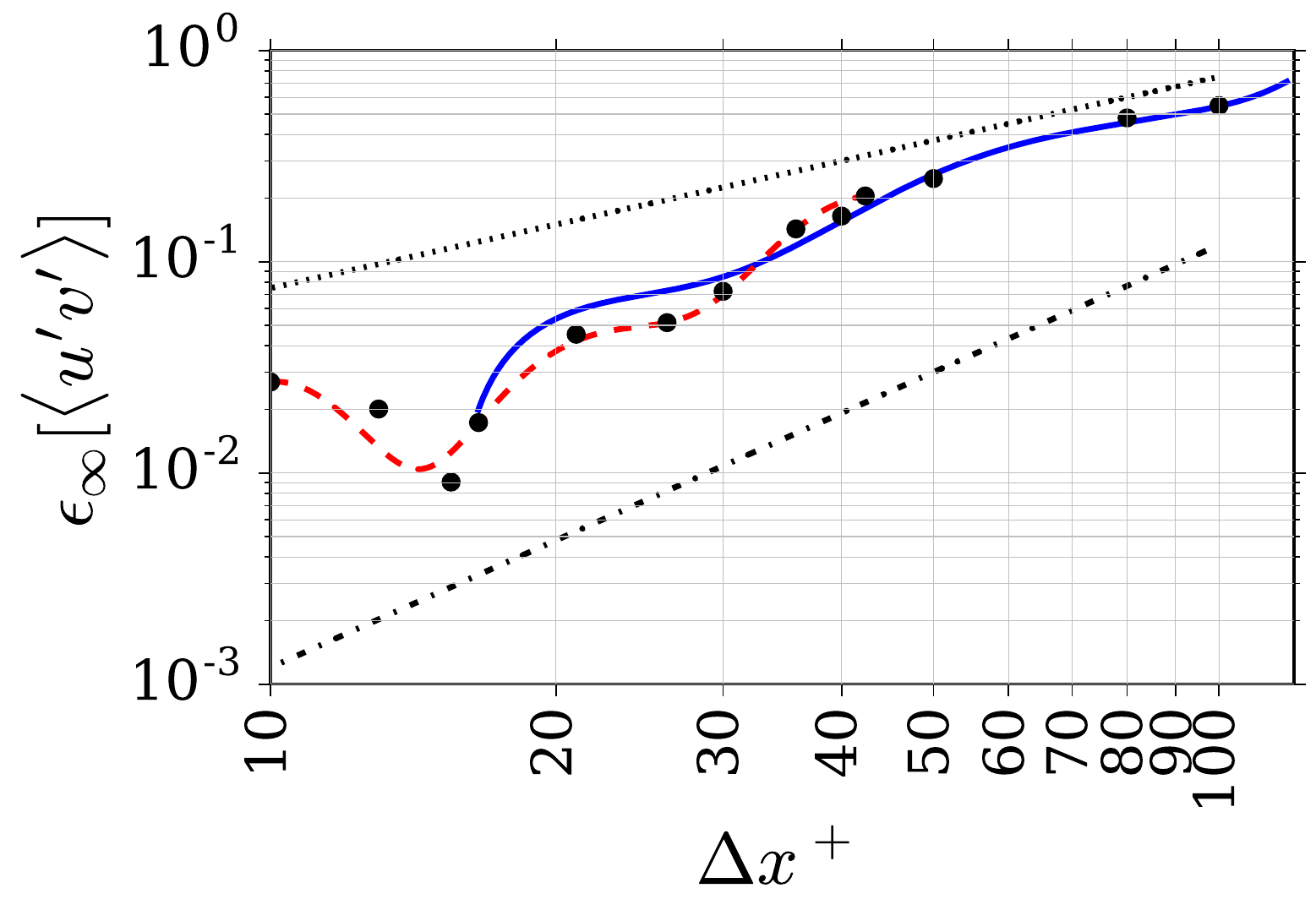} &
   \includegraphics[scale=0.38]{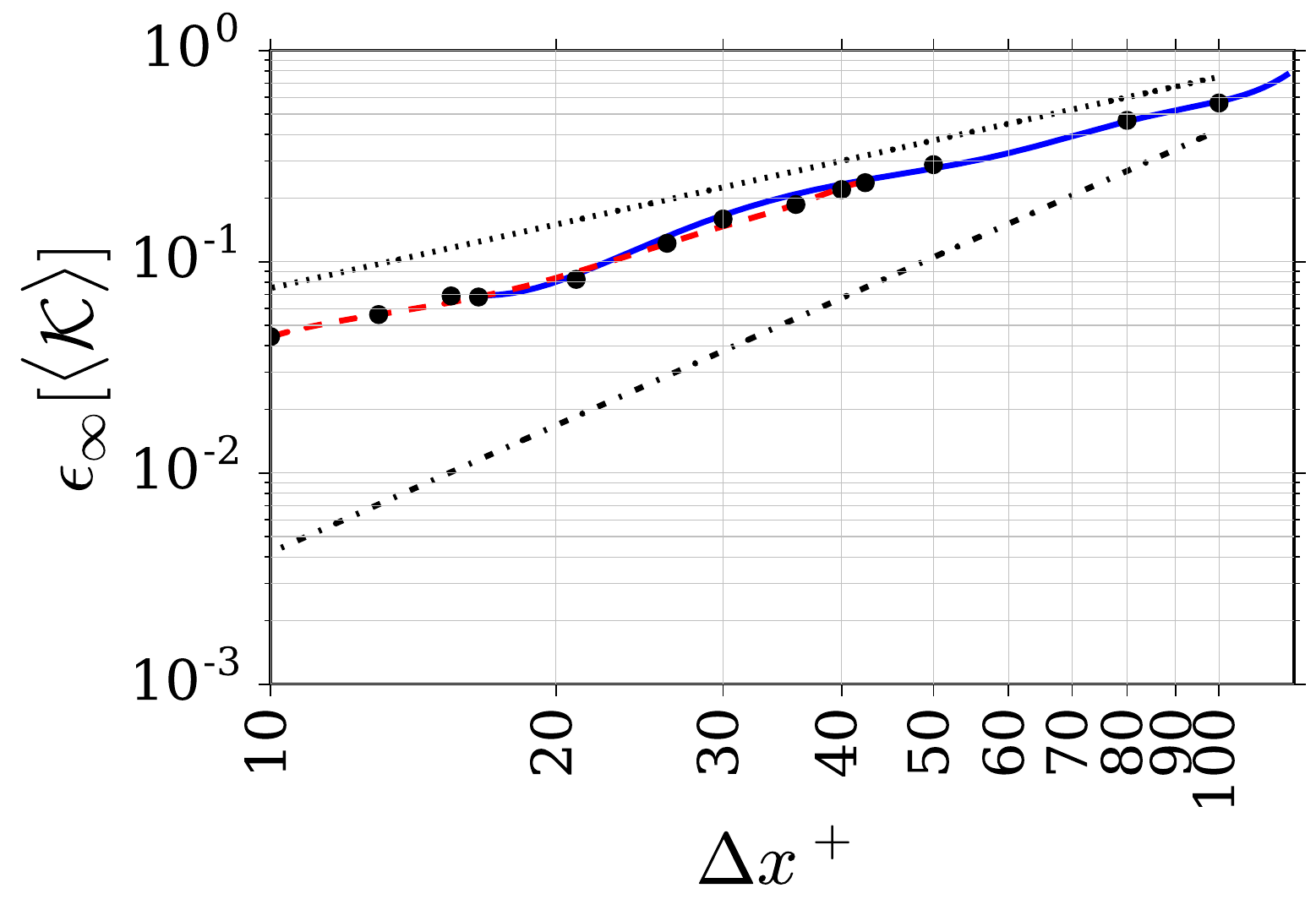} &
   \includegraphics[scale=0.38]{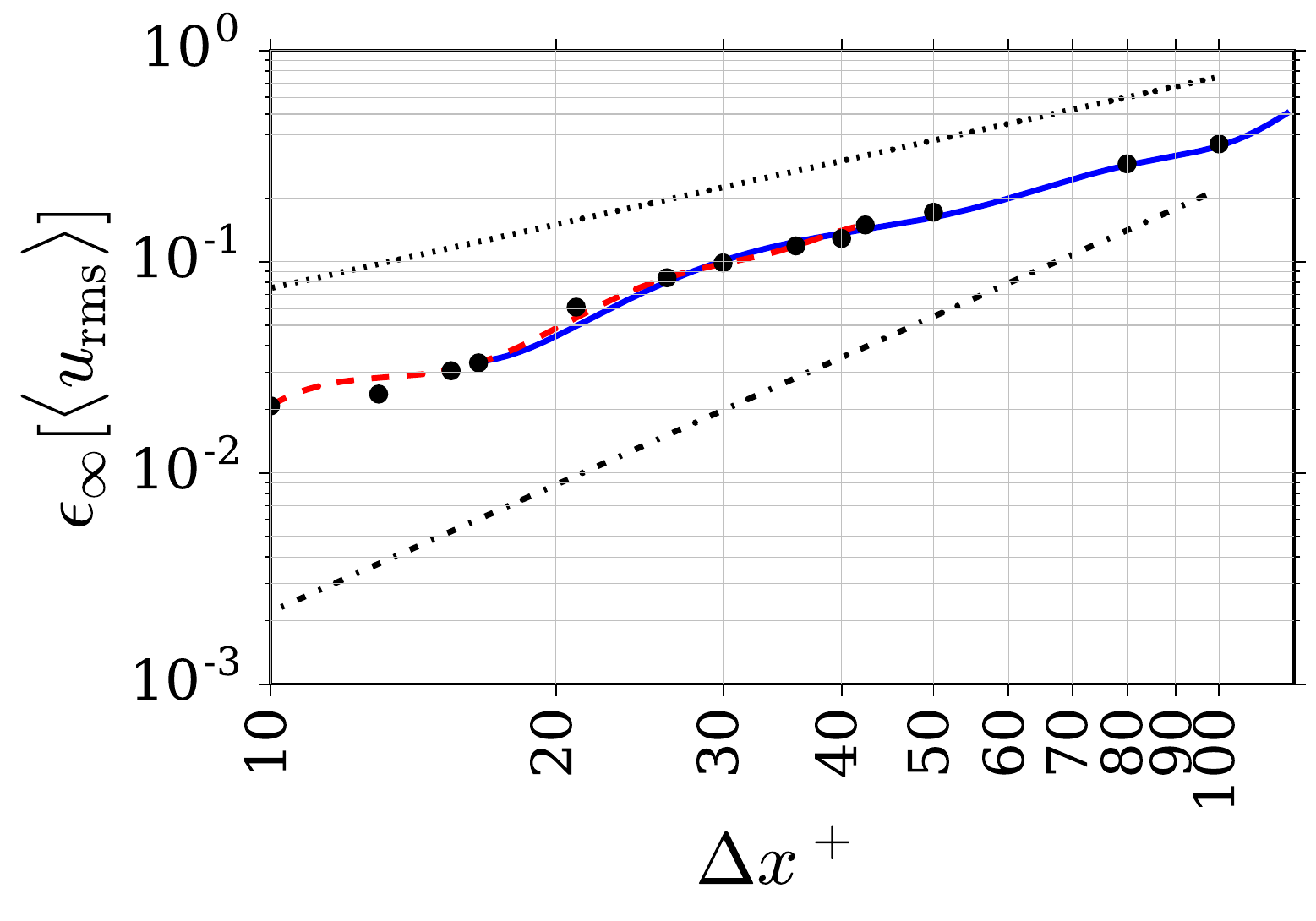} \\    
   {\small (d)} &    {\small (e)} &    {\small (f)} \\   
   \end{tabular}
\caption{Variation of the errors (not in $\%$) in different turbulent channel flow quantities with $\dxp$. For fixed target $\reyt=300$ and $\dyp_w=1.105$, the grid resolution are considered to vary according to $\dzp\propto 0.66 \dxp$. For these settings, the solid and dashed lines respectively represent the predictions of meta model (\ref{eq:pce}) constructed by Set-B and Set-Bf, and the symbols illustrate the errors in a-posteriori channel flow simulations. For reference, the dotted and dash-dotted lines, respectively corresponding to $\epsilon \propto \dxp$ and $\epsilon \propto \Delta x^{+^2}$, are also plotted.}
\label{fig:convergenceRate}
\end{figure}

When refining the grid, $\dyp_w$ is kept constant and the grid cell sizes in the stream- and spanwise directions change according to, $\dzp=\dzp_{\rm m}+a (\dxp-\dxp_{\rm m})$, with $a$ being a constant and $(\dxp_{\rm m},\dzp_{\rm m})$ specifying some particular resolution in the admissible space $\BQ_{\dxp}\times \BQ_{\dzp}$.
The error at different combinations of $\dxp$ and $\dzp$ along the described line is predicted by meta model (\ref{eq:pce}).
In particular for $a=0.66$ and $(\dxp_{\rm m},\dzp_{\rm m})=(10,5.6)$, \fig ~\ref{fig:convergenceRate} illustrates how different errors in Set-B vary with the cell size.
In order to show the validity of the predictions made by meta model (\ref{eq:pce}), the error in the results of a few a-posteriori channel flow simulations \rev{satisfying the same resolution conditions,} are also represented.

For further validation of the predictions of the meta model tuned by the simulations of Set-B, and also for looking at the errors at finer resolutions \rev{than what is considered in the admissible ranges $\BQ_{\dxp}\times \BQ_{\dzp}$}, another set of simulations, hereafter Set-Bf, is considered.
This set covers the admissible space $[8.32,44.04]$ and $[4.49,28.18]$ for $\dxp$ and $\dzp$, respectively, employing $5\times 5$ Gauss quadrature points. 
All other simulation conditions of Set-Bf are the same as Set-B.

According to \fig~\ref{fig:convergenceRate}, when $\dxp$ changes one order of magnitude, the slope of the graphs, representing the rate of change of the errors, may be not fixed, and may even exhibit an oscillatory behavior.
The oscillations are more recognizable, for instance for $\einf[\uv]$, and less observable for $\einf[\tke]$.
They also become more visible when $\dxp$ is finer than $\approx 40$. 
Since the channel flow QoIs whose errors are investigated do not directly appear in governing equations (\ref{eq:LEScont}), providing an explanation for the oscillations is not an easy task.

\rev{At the first look, the oscillations may, at least partially, originate from the use of expansion (\ref{eq:pce}) to predict the error responses.} 
\rev{In fact, a potential drawback of the quadrature-based expansions including the gPCE is that, expansion (\ref{eq:pce}) is by construction, see \sect~\ref{sec:uq}, only constrained to predict the exact value of the responses at the collocation (quadrature) points of the parameters admissible space.
Therefore, no other constraint exists at other interior and boundary points of the parameters space.}
However, as shown in the plots in \fig~\ref{fig:convergenceRate}, the error in the quantities of a few a-posteriori channel flow simulations agrees acceptably well with the predictions of expansion (\ref{eq:pce}). 
Consequently, the oscillations in $\einf[\uv]$, associated with the plateau-like region in the error portrait of $\uv$ in the $\dxp\dash\dzp$ plane discussed in \sect~\ref{sec:yDepQoIs}, seem to really exist rather than being an artefact of the polynomial expansions.

\rev{It is also notified that on the common range of $\dxp$, the deviation between the errors predicted by the meta models of Set-B and Set-Bf is small.
However, in some cases, e.g. $\einf[\U]$, the meta model of Set-Bf predicts slightly more accurate values than that of Set-B, comparing the predicted errors with those found in the a-posteriori simulations.
}

A set of error analysis and quality assessment methods are developed for LES, see \cite{celik:05,klein:05,celik:06} and the references therein, which rely on Richardson extrapolation technique. 
In the framework of these methods, as inspired by the numerical analysis of PDEs (partial differential equations), the error between the numerical solution of the discretized PDE, $\varphi_h$, and the corresponding exact solution, $\varphi$, is assumed to be proportional to some power of a characteristic cell size, $h$, i.e. $\varphi_h=\varphi +c h^r$ \rev{with $c$ and $r$ being constant}.
This fundamental interpretation is further extended by Klein \cite{klein:05} to assess the error induced by SGS modeling in addition to the numerical errors, in the framework of implicitly-filtered LES. 
\rev{However, due to the oscillations observed in \fig~\ref{fig:convergenceRate}, the starting expansion for deriving this type of error assessment methods does not seem to be valid, at least for the averaged QoIs of channel flow and in the settings of the present study.
In other words, the oscillatory reduction of the errors with the grid cell size cannot be described by a constant-value $r$. 
Despite this, if the described error-estimation techniques are to be used for the QoIs with less oscillatory error reduction, such as TKE, an appropriate constant value of $r$ is observed to be between $1$ and $2$, which differs from $r=2$ assumed in \cite{klein:05} for a second-order numerical scheme.
In any case, a fundamental issue to consider in the discussion is that} the characteristic cell size, $h$, is not in the asymptotic range.
This means the grid spacing is not necessarily small enough to let the leading-order term of the error be characterized by $c h^r$.

\subsection{Global sensitivity analysis}\label{sec:GSA}
As a complement to the uncertainty propagation problem, global sensitivity analysis (GSA) can be performed in order to specify how influential \rev{a specific} uncertain factor is on the model response, when all the uncertain factors simultaneously vary over their own admissible spaces. 
\rev{In the context of the present study,} the awareness of this type of sensitivities along with the loci of zero or low errors \rev{in different flow quantities} can be helpful, when refining the grid for obtaining better simulation results.

Here, a variance-based GSA is carried out with the results reported in terms of the total Sobol indices.
Such an index, specifying the sensitivity of response $\cR$ with respect to parameter $q_i$, for $i=1,2,\cdots,p$, is defined by \cite{sobol}, 
$$
S_{T_i}=1-\frac{\var(\BE(\cR|q_{\sim i}))}{\var(\cR)} \,,
$$
where, $q_{\sim i}=\{q_1,\cdots,q_{i-1},q_{i+1},\cdots,q_p\}$, \rev{and $\BE$ and $\var$, respectively, specify expectation and variance}.
When stochastic collocation methods such as gPCE are employed to construct meta models, analytical expressions for obtaining the sensitivity indices can be derived, see \cite{SobolAnova}.

In particular, the focus is on the channel flow simulations with combinations of grid spacings $\dxp\in[10,45]\subset \BQ_{\dxp}$, $\dzp\in[7,30]\subset \BQ_{\dzp}$, and $\dyp_w \in \BQ_{\dyp_w}$ at target $\reyt=300$ and $550$. 
The formed space by these new resolution ranges is a subset of the original admissible space $\BQ_{\dxp} \times \BQ_{\dzp} \times \BQ_{\dyp_w}$ considered to produce the simulation sets in \tab~\ref{tab:caseSummary}. 
\rev{The new admissible ranges for $\dxp$ and $\dzp$ are chosen in accordance with what is practically used in WRLES of wall-bounded flows, to make the resulting conclusions of the GSA more applicable.}
To determine the error responses at the samples taken from the new parameter space, the meta model (\ref{eq:pce}) with known coefficients determined from different sets in \tab~\ref{tab:caseSummary}, is used. 
Especially, $5\times5\times 3$ Gauss points are considered as the deterministic samples to cover the new admissible space of $\dxp\times\dzp\times\dyp_w$.

\begin{figure}[!htbp]
\centering
   \begin{tabular}{cc}
   {\small (a)} &
    \includegraphics[scale=0.5]{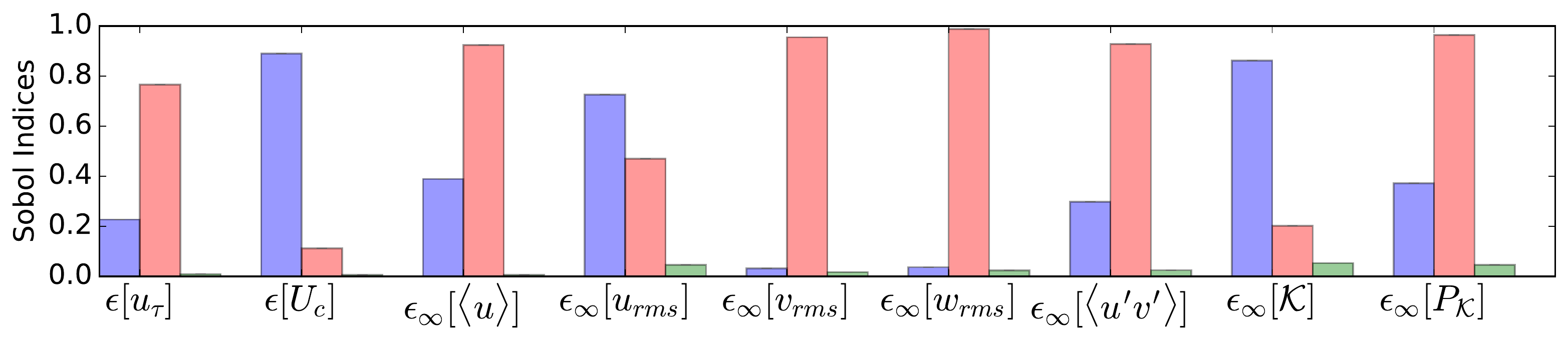} \\
    {\small (b)} &
    \includegraphics[scale=0.5]{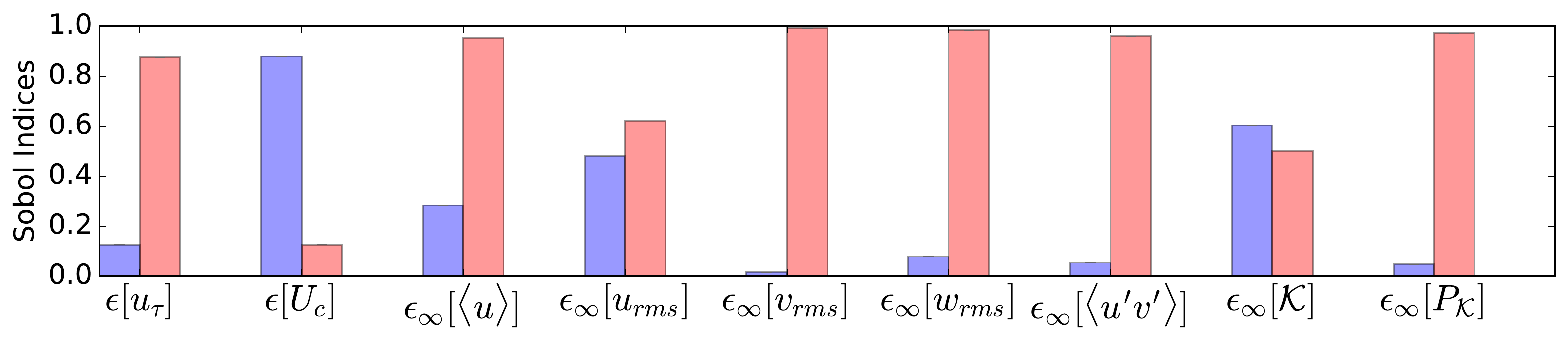} \\
    {\small (c)} &
    \includegraphics[scale=0.5]{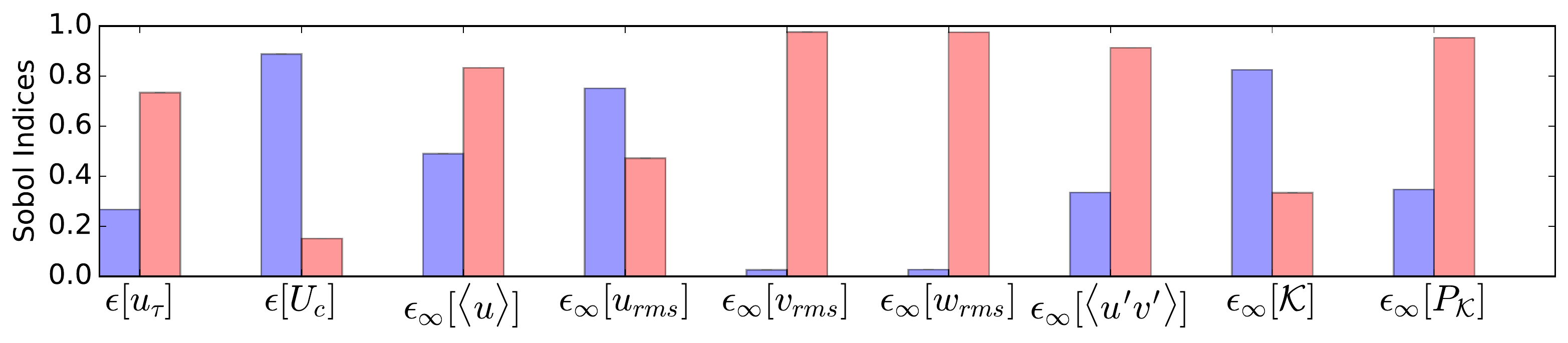} \\      
   \end{tabular}
   \caption{Total sensitivity indices of $\epsilon[{\ut}]$, $\epsilon[\U_c]$, and $\einf$ of the mean streamwise velocity, rms velocity fluctuations, Reynolds stress, turbulent kinetic energy (TKE), and the production rate of TKE with respect to $\dxp$ (blue bars), $\dzp$ (red bars), and $\dyp_w$ (green bars), for Set-ABC (a), Set-Bw (b), and Set-D (c). To produce these results, it is considered that $\dxp\in [10,45]$, $\dzp\in[7,30]$, and $\dyp_w\in [0.25,1.96]$.}\label{fig:sobolIndices}
\end{figure}

According to \fig~\ref{fig:sobolIndices}(a), for target $\reyt=300$, changing $\dyp_w$ over $[0.25,1.96]$ has negligible effect on the errors compared to the variations in $\dxp$ and $\dzp$, as pointed out earlier in \sect~\ref{sec:ResComp3Resolut}.
This may originate from the fact that $\dyp_w$ varies over a comparatively smaller range which is chosen to ensure the fundamental requirement of the wall-resolving LES is satisfied.

By comparing the Sobol indices with respect to the grid spacings in the streamwise and spanwise directions, it turns out that $\epsilon[\ut]$ along with $\einf[\U]$, $\einf[\uv]$, and $\einf[\vrms]$ are most sensitive to the variations in $\dzp$ rather than in $\dxp$, while $\einf[\tke]$, $\einf[\urms]$, and $\einf[\wrms]$ are largely affected by $\dxp$. 
These observations are also valid when an explicit SGS model is included in the simulations at $\reyt=300$, and also when $\reyt$ increases to $550$, as illustrated in \fig ~\ref{fig:sobolIndices} (b) and (c), respectively.

For the three simulation sets shown in \fig~\ref{fig:sobolIndices}, the sensitivity indices of a specific error response with respect to $\dxp$ and $\dzp$ may have different magnitudes, however, their relative importance is maintained. 
This justifies the generalization of the conclusion that, improving the simulation results of the channel flow simulation is hard to achieve by refining the grid solely in either the spanwise or streamwise direction.


\subsection{Suggestions for grid resolution}\label{sec:gridSuggest}
\begin{figure}[!htbp]
\centering
   \begin{tabular}{ccc}
   \includegraphics[scale=0.45]{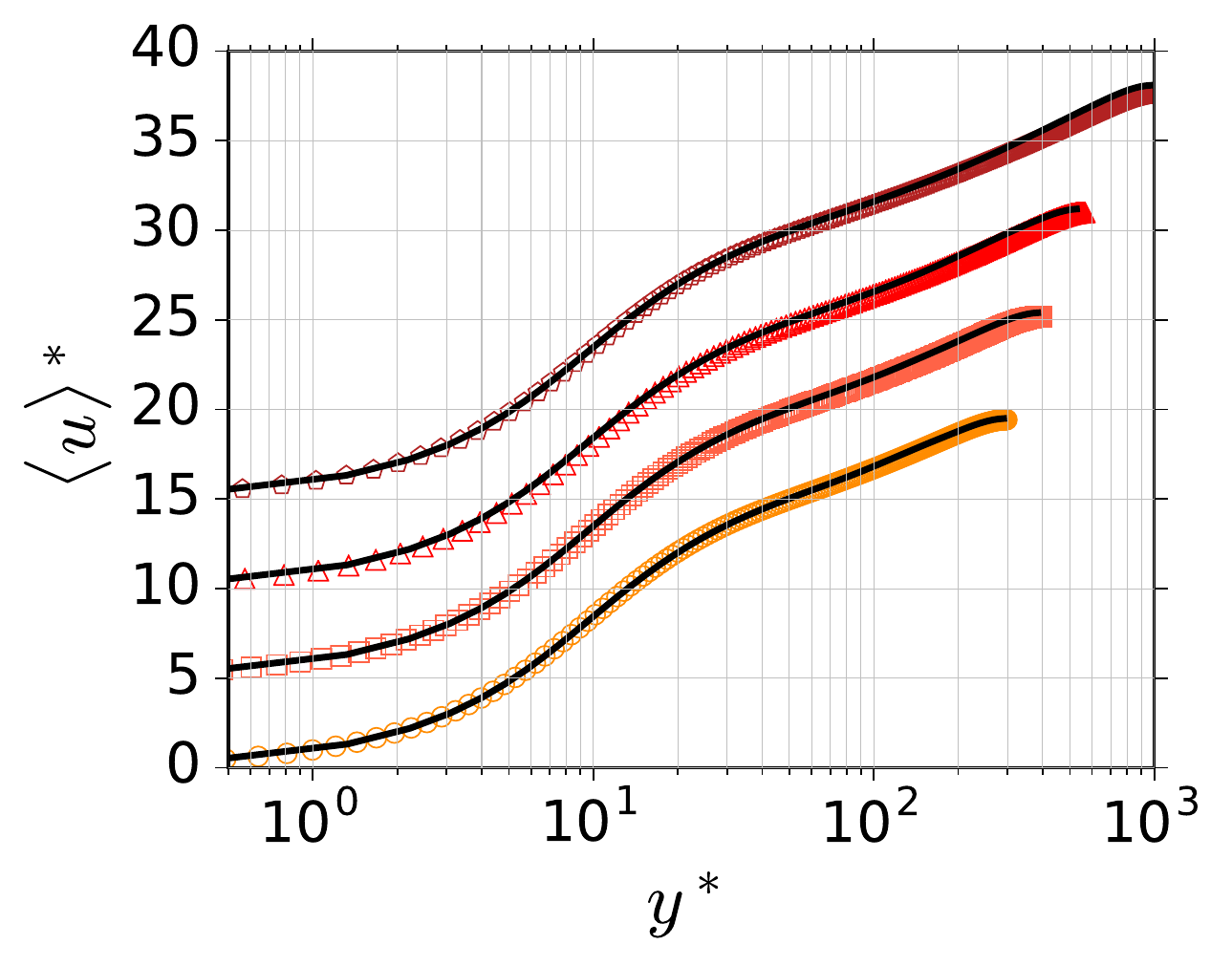} &
   \includegraphics[scale=0.45]{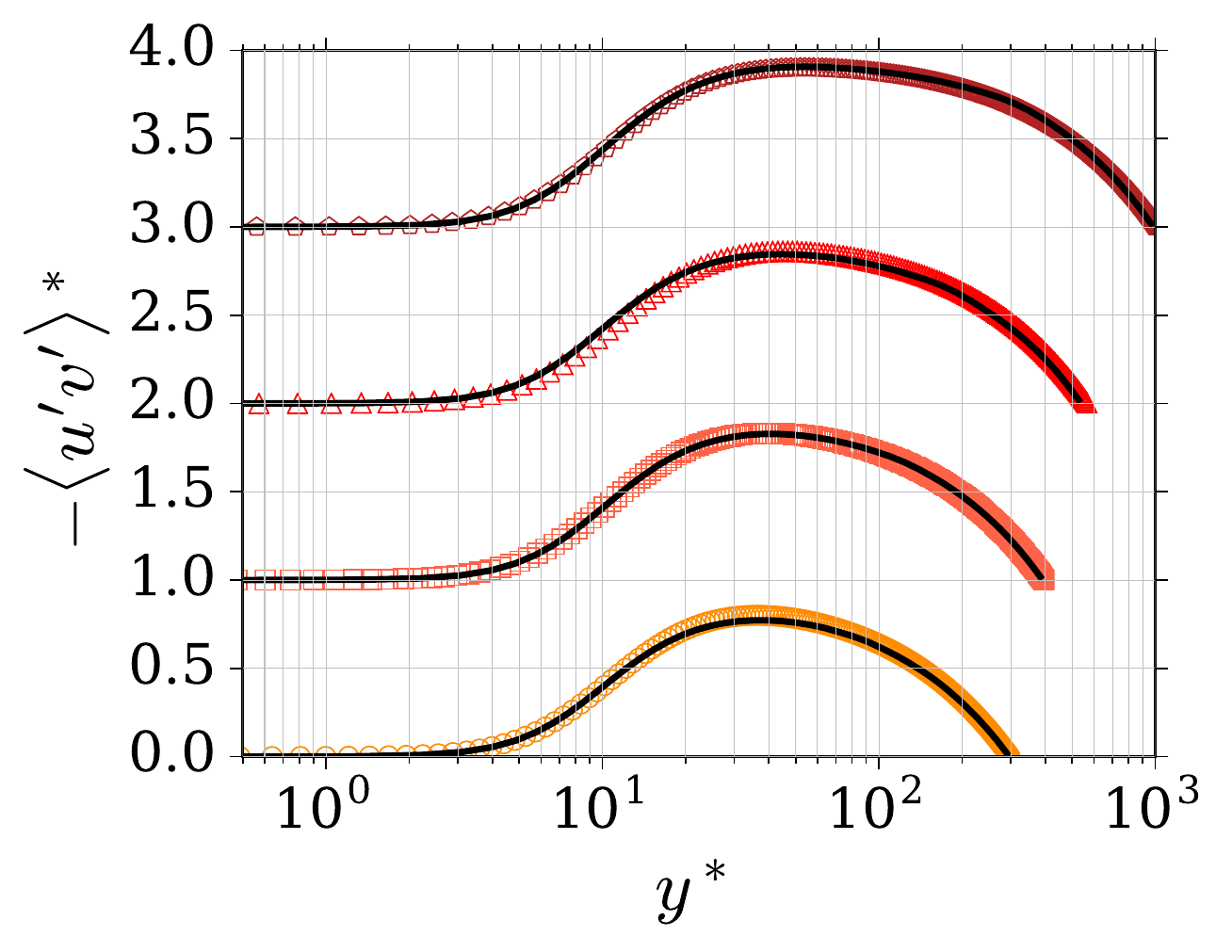} &
   \includegraphics[scale=0.45]{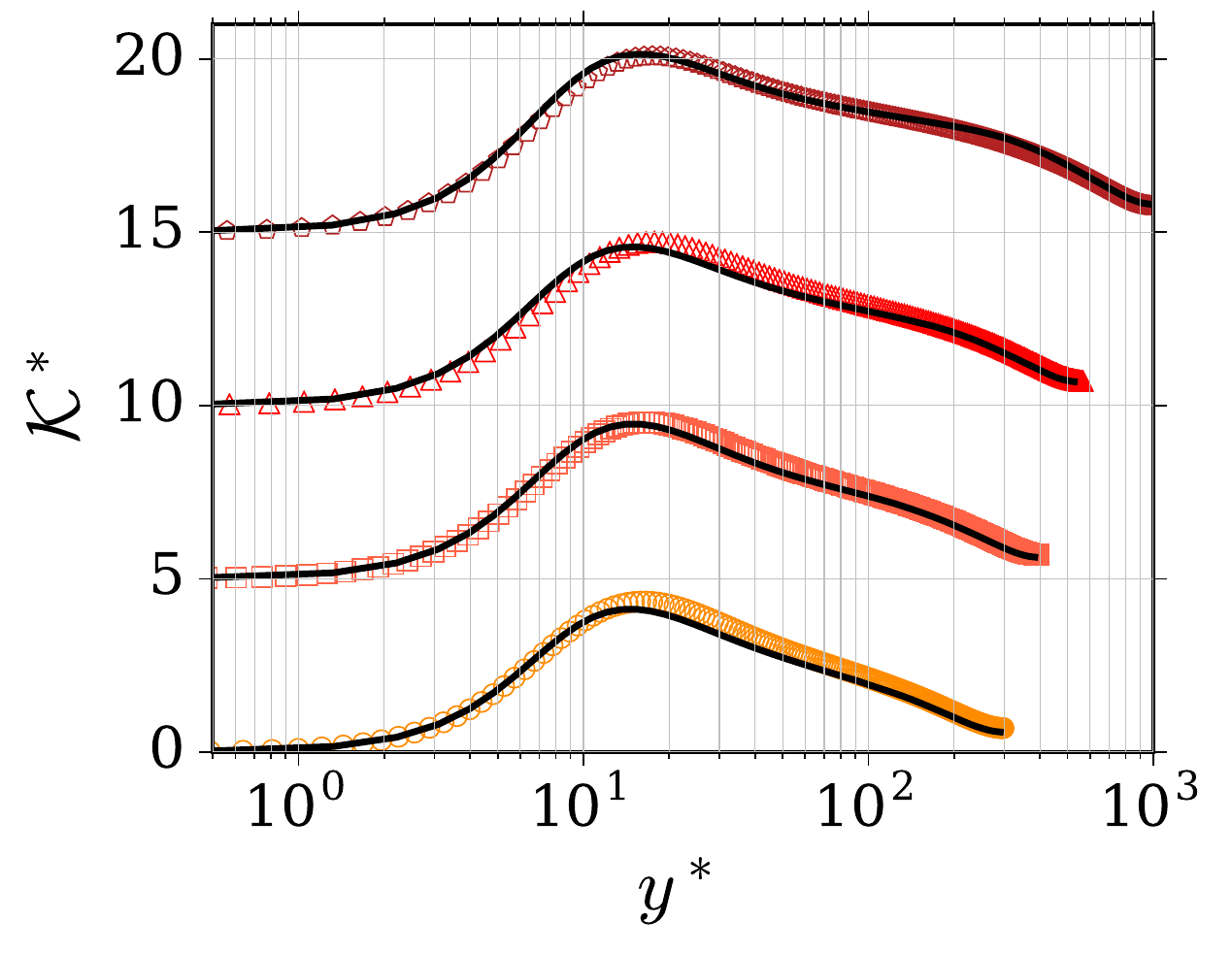} \\
   {\small (a)} &    {\small (b)} &    {\small (c)} \\      
   \end{tabular}
\caption{Inner-scaled profiles of mean velocity, $\U^*$, (a), Reynolds stress, $-\uv^*$, (b), and turbulent kinetic energy, $\tke^*$ (c), versus inner-scaled wall-normal distance $y^*$. The results of the simulations listed in \tab~\ref{tab:fineSimsSummary} (solid lines) are compared to the DNS data of Iwamoto \et~\cite{iwamoto02} for $\reyt=300, \, 400$, and Lee and Moser \cite{lee15} for $\reyt=550,\, 1000$, which are shown by symbols. The graphs are shifted vertically by $5$ wall-units in (a) and (c), and by 1 wall-unit in (b).}\label{fig:plsPlots_finest}
\end{figure}

\begin{figure}[!htbp]
\centering
   \begin{tabular}{cc}
   \includegraphics[scale=0.4]{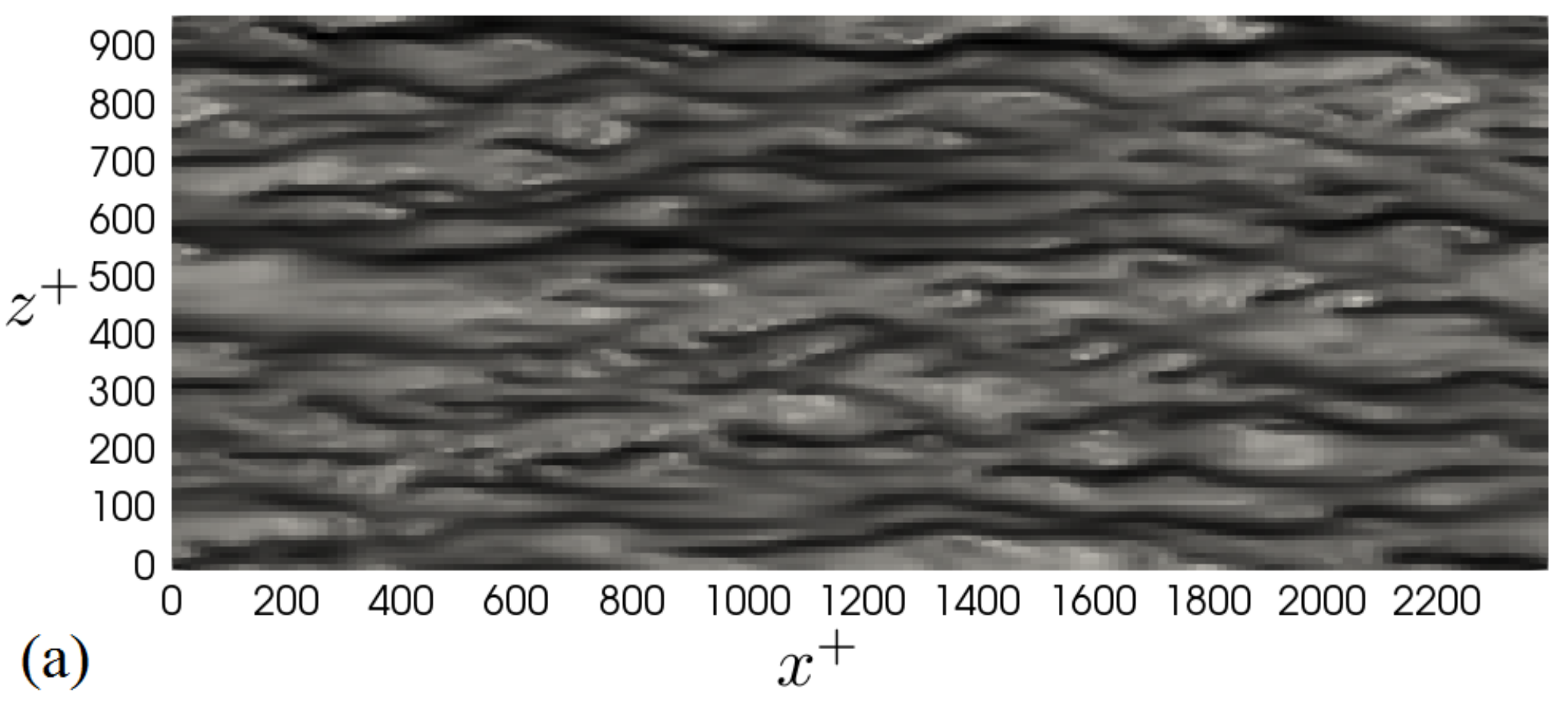} &
   \includegraphics[scale=0.4]{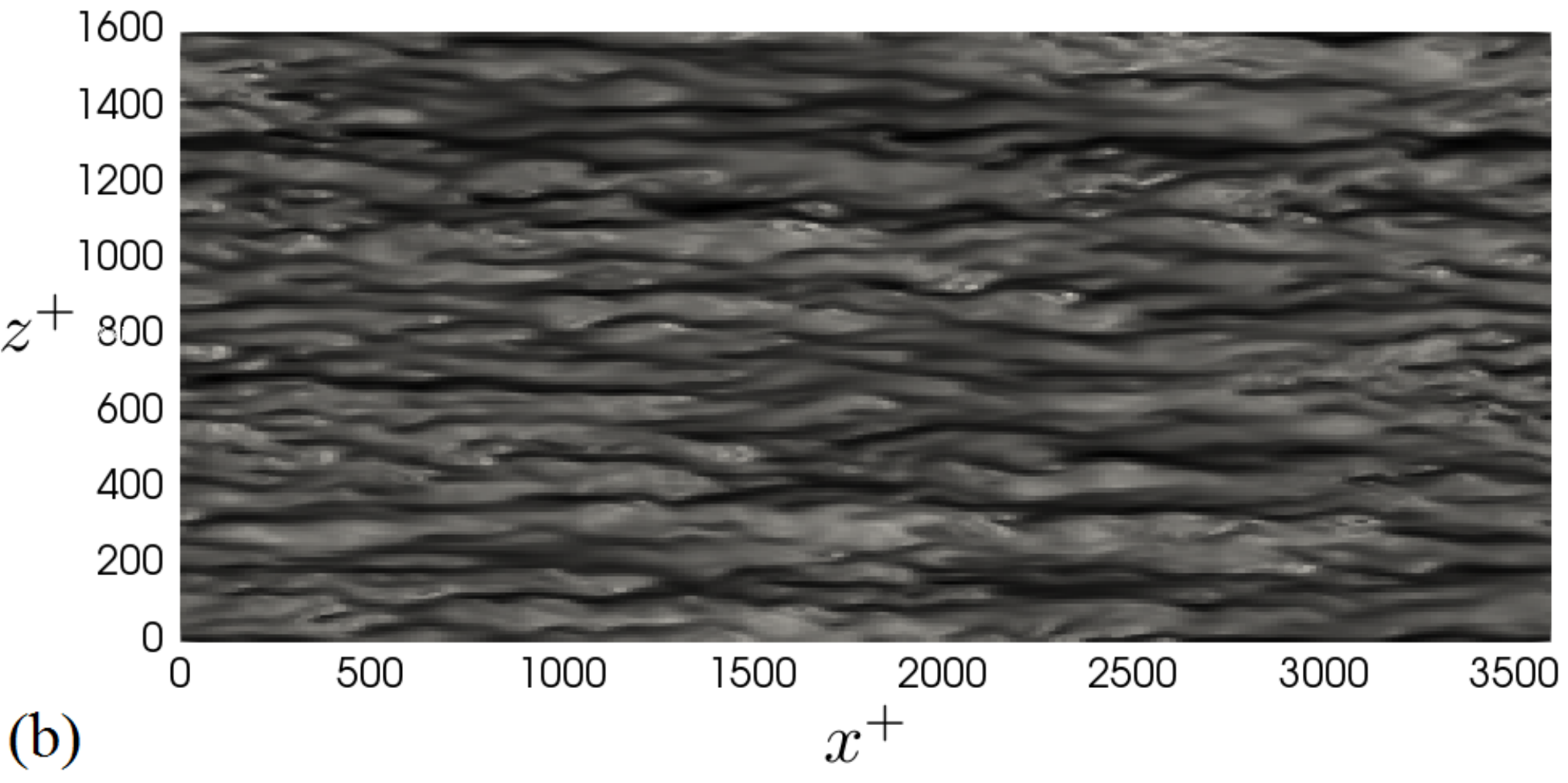} \\
   \includegraphics[scale=0.4]{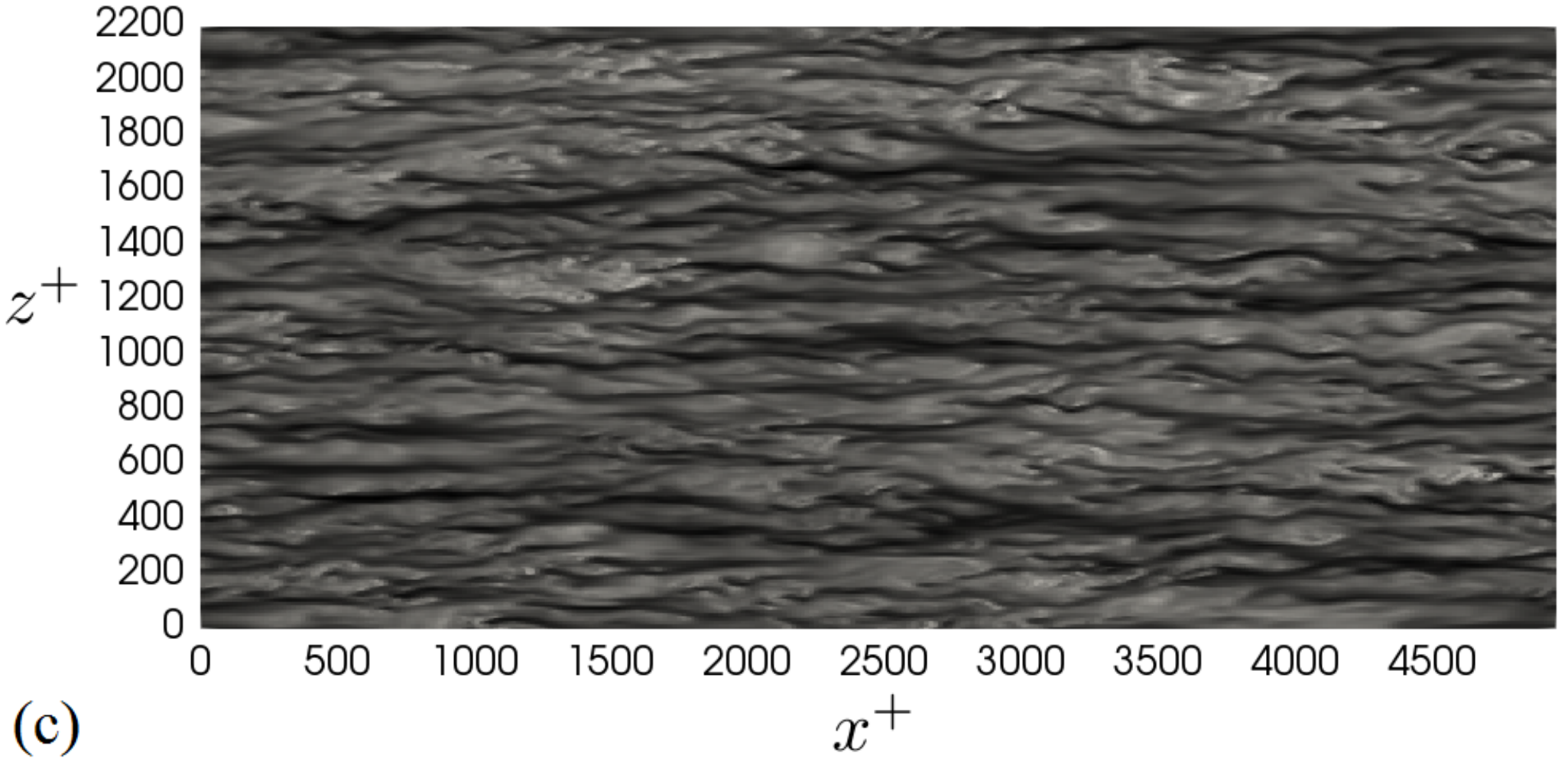} &
   \includegraphics[scale=0.4]{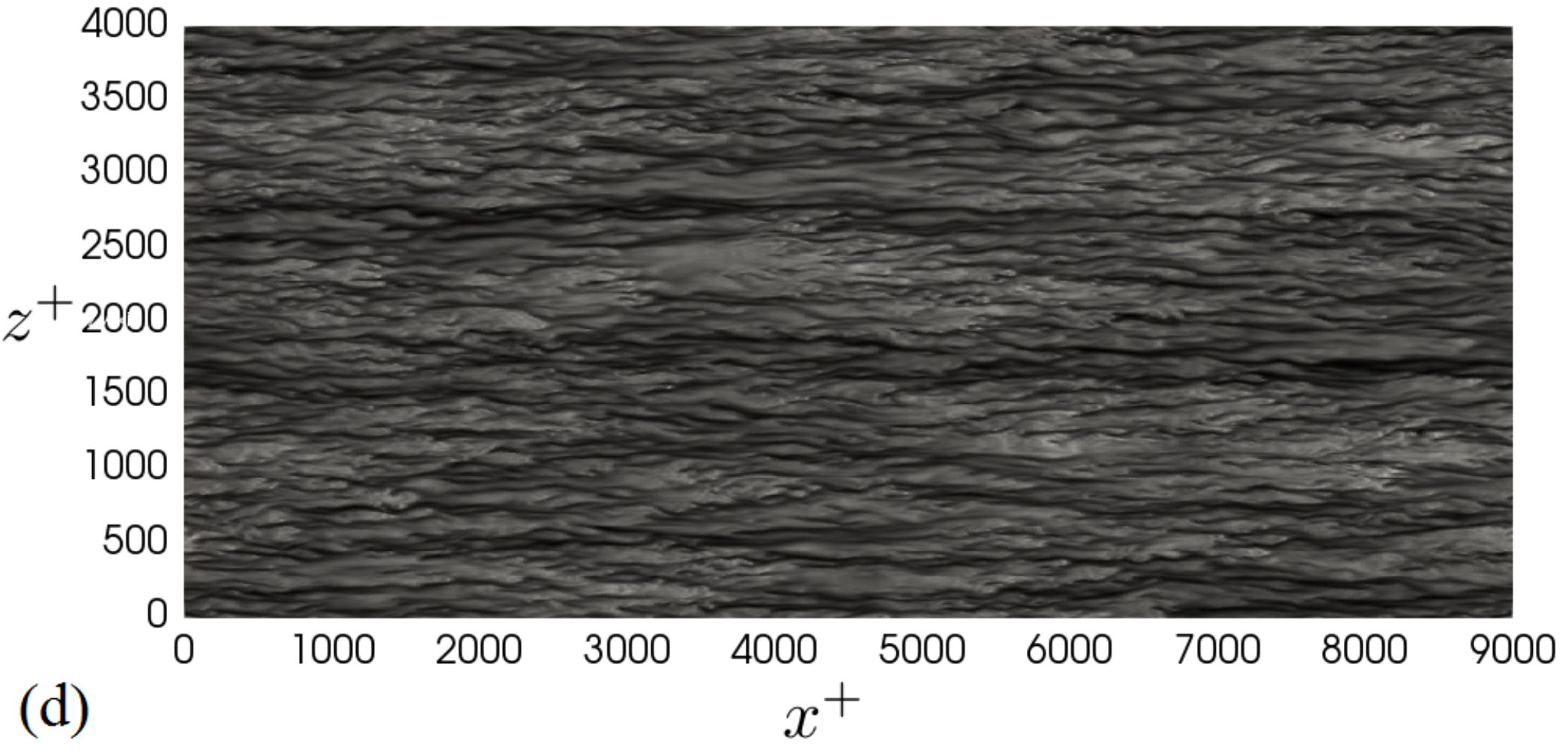} \\   
   \end{tabular}
   \caption{\revCom{Instantaneous streamwise velocity at $y^+=15$ and $\reyt=300$ (a), $400$ (b), $550$ (c), and $1000$ (d) corresponding to the simulations in \tab~\ref{tab:fineSimsSummary}. Darker color corresponds lower magnitude.}}\label{fig:uSnapshot}
\end{figure}

Based on what thoroughly discussed in the previous sections, it is clear that, given a numerical method and a \rev{specific strategy for grid construction} to simulate turbulent channel flow at a target Reynolds number, the errors in different quantities would react differently to the variations in the grid spacings.
Nonetheless, for the particular flow solver described in \sect~\ref{sec:LESCFD} and the distribution of the grid cells in the wall-normal direction by (\ref{eq:delY}), it seems for $\dxp \lesssim 18$, $\dzp\lesssim 12$, and $0.5\lesssim \dyp_w\lesssim 2$, low-error wall-resolving LES of channel flow are obtained.
\rev{In particular, based on the simulations listed in \tab~\ref{tab:caseSummary}, the combination of} $\dxp\approx 16.56$ and $\dzp\approx 9.96$ with $\dyp_w=0.445$ is found to be \rev{appropriate for high-quality WRLES of channel flow at $\reyt=300$\revCom{, see also \app~\ref{sec:twoPointAppendix},} and $\reyt=550$. 
The appropriateness of these resolutions is also confirmed at $\reyt=400$ and $1000$, see \fig~\ref{fig:plsPlots_finest} with the simulation details listed in \tab~\ref{tab:fineSimsSummary} \revCom{ and the corresponding instantaneous streamwise velocities shown in \fig~\ref{fig:uSnapshot}}.}
It is emphasized that, although in producing these simulations, $\dyp_w=0.445$ is employed, relatively good results for other $\dyp_w\in [0.25,1.96]$ are also expected.

\begin{table}[!htbp]
\centering
\caption{Summary of the channel flow simulations carried out at different Reynolds numbers with the grid spacings $\dxp= 16.56$, $\dzp=9.96$, and $\dyp_w=0.445$. Here, $\dyp_c$ specifies the wall-normal grid spacing at the center of the channel. Note that no explicit SGS model is employed in these simulations. \revCom{Fixed time steps $\Delta t = 10^{-3} \delta/U_b$, $3\times 10^{-3}\delta/U_b$, $5\times 10^{-4}\delta/U_b$, and $2\times 10^{-3}\delta/U_b$ are chosen for the simulations at $\reyt=300$, $400$, $550$, and $1000$, respectively.}
}\label{tab:fineSimsSummary}
\begin{small}
\begin{tabular}{*{6}c}
\toprule\toprule
 \multicolumn{3}{c}{$\reyt$} & $\dyp_c$ & Domain Size &  Number of cells \\
Target & DNS ($\reyt^\circ$) & LES  & {}& $l_x \times l_y \times l_z$ &  $n_x \times n_y \times n_z$ \\
\hline
 300 & 297.899, \cite{iwamoto02} & 296.476 & 12 & $7.89\delta \times 2\delta \times 3.17 \delta$ & 141 $\times$ \revCom{92} $\times$ 90 \\
  400 & 395.760, \cite{iwamoto02} & 391.430 &  16 & $9.0\delta \times 2\delta \times 4.0 \delta$ & 215 $\times$ 98 $\times$ 159 \\
550 & 543.496, \cite{lee15} & 539.197 & 22 & $9.0\delta \times 2\delta \times 4.0 \delta$ & 296 $\times$ 102 $\times$ 219 \\
1000 & 1000.512, \cite{lee15} & 983.136 & 40 & $9.0\delta \times 2\delta \times 4.0 \delta$ & 543 $\times$ 116 $\times$ 403 \\
\bottomrule
\end{tabular}
\end{small}
\end{table}

The proposed resolutions agree with the suggestion of Kremer and Bogey \cite{kremer:15}, who reported accurate results for LES of channel flow for $\dxp\leq 30$, $\dyp\leq 1$, and $\dzp\leq 10$, and the same range of $\reyt$ as in \tab~\ref{tab:fineSimsSummary}, however, by employing a higher-order numerical scheme and different function for distributing cells in the wall-normal direction.
In comparison to \cite{kremer:15}, function (\ref{eq:delY}) generates a lower number of cells normal to the wall, yet yielding accurate results.  
This \rev{optimum distribution of cells} is mainly achieved by keeping the number of cells in the outer part of the turbulent boundary layer independent of the Reynolds number, as it is pointed out for instance in \cite{chapman79}.
\rev{However, as driven by the inner-layer and near-wall region, the total number of cells generated in the same manner as the present study, increases with $\rey$-number, see \cite{saleh:2}.}

\section{Summary and Conclusions} \label{sec:concl}
\rev{
The effect of variation of the grid resolutions in the wall-parallel and -normal directions on the wall-resolving LES of turbulent channel flow at target $\reyt=300$ and $550$, is investigated.
The error responses, defined as the error between the LES results and the DNS reference data \cite{iwamoto02,lee15} for different quantities of interest (QoI), are evaluated employing different measures developed in \sect~\ref{sec:Rdefs}. 
The QoIs include the wall mean friction velocity, and the cross-channel profiles of mean velocity and Reynolds stress components. 
The LES are carried out by the open-source finite-volume based library, \of, with the specific numerical schemes described in \sect~\ref{sec:LESCFD}. 
Equi-spaced cells are used in the streamwise and spanwise directions of the flow domain.
Function (\ref{eq:delY}) is employed to construct the grid cell spacings in the wall-normal direction, having specified the inner-scaled distance of the first off-wall cell center, $\dyp_w$, in accordance with the requirement of wall-resolving LES.
}

\rev{To represent the dependence of the error responses on the grid resolution, a meta model for each QoI is constructed based on a limited number of channel flow simulations.
In particular, non-intrusive generalized polynomial chaos expansion (gPCE) is employed for this purpose.}
\rev{To study their impact,} the inner-scaled grid spacings, $\dxp$, $\dzp$, and $\dyp_w$, are considered as uncertain parameters within the UQ (uncertainty quantification) framework. 
\rev{These parameters are assumed to vary over admissible ranges, $10\leq \dxp\leq 150$, $7\leq \dzp\leq 70$, $0.25 \leq \dyp_w\leq 1.96$.
The resulting simulation sets are listed in \tab~\ref{tab:caseSummary}.
 }

For the employed implicitly-filtered LES approach, the errors in the simulations are found to be mostly driven by the numerical errors. 
This is consistent with what formerly observed by other authors, see \cite{vreman96,geurts05}. 
In particular, the errors induced by WALE SGS model \cite{nicoud:99} is shown to be dominated by the numerical error, as inferred by comparing \fig~\ref{fig:Bw_contours} with \figs~\ref{fig:B_Lnorm_1} and \ref{fig:B_Lnorm_2}. 

\rev{Employing the meta models, error isolines for different QoIs of channel flow are constructed in the parameters admissible space.
Observing different patterns of error isolines for different QoIs in the $\dxp\dash\dzp$ plane, reminds the necessity of monitoring the errors in various quantities when grid refinement is carried out seeking for accurate results.}
In particular, for the numerical schemes employed in the present study, see \sect~\ref{sec:LESCFD}, there are combinations of grid spacings for which, the computed $u_\tau$ is the same as the DNS value, however, the errors in other quantities do not necessarily vanish.

\rev{
To make sure low errors in different channel flow quantities are simultaneously achieved, the right-hand-side of criterion (\ref{eq:xi}), that is independent of the LES numerical method and grid construction strategy, must be close to zero. 
For this purpose, the error in $u_\tau$, the error in $|\uv|$ \rev{averaged over the channel half-hight}, and the error in the mean centerline velocity should simultaneously reduce to low values, see \sect~\ref{sec:uTauConv}.
}

The reduction of the errors achieved by reducing the grid spacings is non-monotonous, as discussed in detail in \sect~\ref{sec:errRates}. 
This non-monotonicity may cast doubt on the basic assumption of the quality assessment and error estimation techniques developed based on Richardson extrapolation, see \cite{celik:05,klein:05,celik:06}.

As studied in \sects~\ref{sec:yDepQoIs} and \ref{sec:ReEffects}, respectively, for $\reyt=300$ and $550$, reducing the errors in the cross-channel profiles of rms velocity fluctuations, \rev{and hence in turbulent kinetic energy,} requires finer grid resolutions compared to \rev{what is needed to achieve corresponding accuracy in} the mean velocity and Reynolds stress, $\uv$, profiles. 
As shown in \fig~\ref{fig:someB_profs}, for $\dyp_w=1.105$ and $\reyt=300$, a small improvement in the rms velocities, and turbulent kinetic energy profiles is observed when $\dxp$ and $\dzp$ are refined from $16.56$ and $9.96$ to $10$ and $5.6$, respectively.

Comparing different error contours, correlation between specific errors can be recognized. 
In particular, the error in the peak value of $\uv$ profile, the error in $|\uv|$ \rev{averaged over the channel half-height}, and the error in $\ut$ seem to be correlated, as illustrated in \fig~\ref{fig:B_duv}. 
It is also remarkable that, given $\dyp_w$, for those combinations of $\dxp$ and $\dzp$ which result in zero error in $\ut$, the inner-scaled profiles of $\uv$ approximately collapse on the DNS data, see \fig~\ref{fig:Q_profs}(b).

\rev{In order to quantify the sensitivity of the errors with respect to grid spacings, a variance-based global sensitivity analysis is conducted, as detailed in \sect~\ref{sec:GSA}.}
\rev{To comply with the common resolutions for WRLES of channel flow,} grid spacings are allowed to take values according to $\dxp\in[10,45]$, $\dzp\in[7,30]$, and $\dyp_w\in[0.25,1.96]$, for target $\reyt=300$, and $550$.
The error in $\ut$ along with $\einf[\U]$, $\einf[\uv]$, $\einf[\vrms]$ are found to be most sensitive to the variations in $\dzp$ rather than in $\dxp$. 
In contrast, $\dxp$ is observed to be more influential than $\dzp$, on $\einf[\tke]$, $\einf[\urms]$, and $\einf[\wrms]$. 
Compared to the resolutions in the wall-parallel directions, the influence of $\dyp_w$ on the errors is found to be negligible.

Looking at the errors of different quantities, it is revealed that for $\dxp\leq 18$, $\dzp\leq 12$, and $\dyp_w\lesssim 2$, acceptable low errors in the channel flow quantities can be simultaneously achieved. 
In particular, accurate inner-scaled profiles of mean velocity, Reynolds stress and turbulent kinetic energy at $\reyt=300$, $400$, $550$, and $1000$ are obtained for $\dxp=16.56$, $\dzp=9.96$, and $\dyp_w = 0.445$, see \fig~\ref{fig:plsPlots_finest}.
\rev{These grid resolution guidelines should be of interest for the community of \of~users.}

\rev{The choice of numerical methods, grid construction, and SGS model is expected to influence some of the results.
For instance, see the difference between the error isolines of the mean friction velocity $\ut$ in \revCom{\fig~\ref{fig:ABC_duTau}, \fig~\ref{fig:AL_errorInProfs}} and those in \cite{meyers07}.
However, many of the conclusions are beneficial despite the mentioned potential bias.
In particular, the resulting detailed ``error portraits" for various QoIs represent the challenges that theoretical error estimation techniques have to deal with.
Above all,} the non-intrusive approach described in \sect~\ref{sec:uqcfdLink} to link \rev{the gPCE-based meta model} to a CFD solver,  can be used for the systematic study of various numerical and physical parameters influencing the LES responses.

\section*{Acknowledgements}
The authors would like to thank Timofey Mukha and Gunilla Kreiss, at Uppsala university, for
valuable discussions concerning several aspects of the present work.
All channel flow simulations were performed on resources provided by the Swedish National Infrastructure for Computing (SNIC) at PDC Centre for High Performance Computing (PDC-HPC).
The work was supported by Grant No 621-2012-3721 from the Swedish Research Council.


\appendix

\section{On the uncertainty of the approach} \label{sec:appendix}
A few  potential sources of uncertainty are involved in evaluating the response surfaces, following the procedure discussed in \sect~\ref{sec:results}. 
One uncertainty is related to the ability of expansion (\ref{eq:pce}) in approximate construction of the responses. 
The other factor is initiated from the fact that the quantities of turbulent channel flow whose error response surfaces are constructed by the gPCE, are averaged in both time and space and, hence, are potentially prone to be affected by insufficient averaging. 
In the following two sections, these influential factors are separately discussed.

\subsection{Convergence of the gPCE}\label{sec:pceConv}
As shortly pointed out in \sect~\ref{sec:uq}, one way of showing the accuracy of the predictions made by the meta model (\ref{eq:pce}), is to assess the deviation between $\tilde{f}(\chi,q)$ and the exact $f(\chi,q)$ for some $q^*\in \BQ$.
This is reflected in \sect~\ref{sec:errRates} by the cross-validations made with a limited number of a-posteriori simulations.
However, it is obvious that following this procedure for many samples $q^*$, is computationally expensive. 
An alternative strategy can be built on the fact that the accuracy of (\ref{eq:pce}) is directly dependent on  the maximum polynomial order in the expansion.  
For the particular settings in this study, see \sect~\ref{sec:simCase}, the magnitude of different terms in expansion (\ref{eq:pce}) for different responses is evaluated.
In particular, the norm of the $k$-th term in the expansion (\ref{eq:pce}) normalized by the zero-order term in the expansion, i.e. $\vartheta_k:=\| \hat{f}_{k} \Psi_{k}(q)  \| / |\hat{f}_0|$, is used for this purpose.

\begin{figure}[!htbp]
\centering
   \begin{tabular}{ccc}
      \includegraphics[scale=0.35]{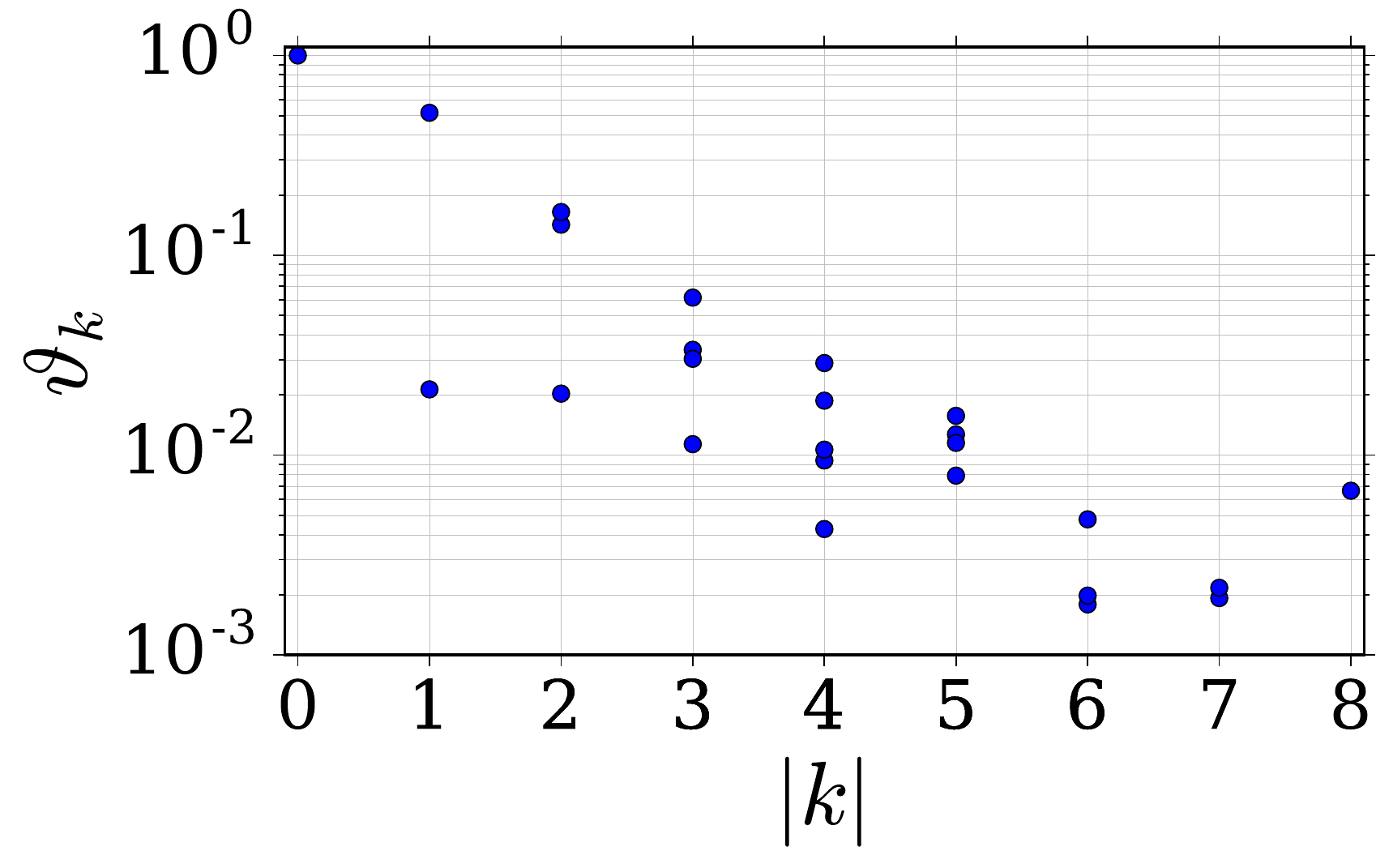} &
      \includegraphics[scale=0.35]{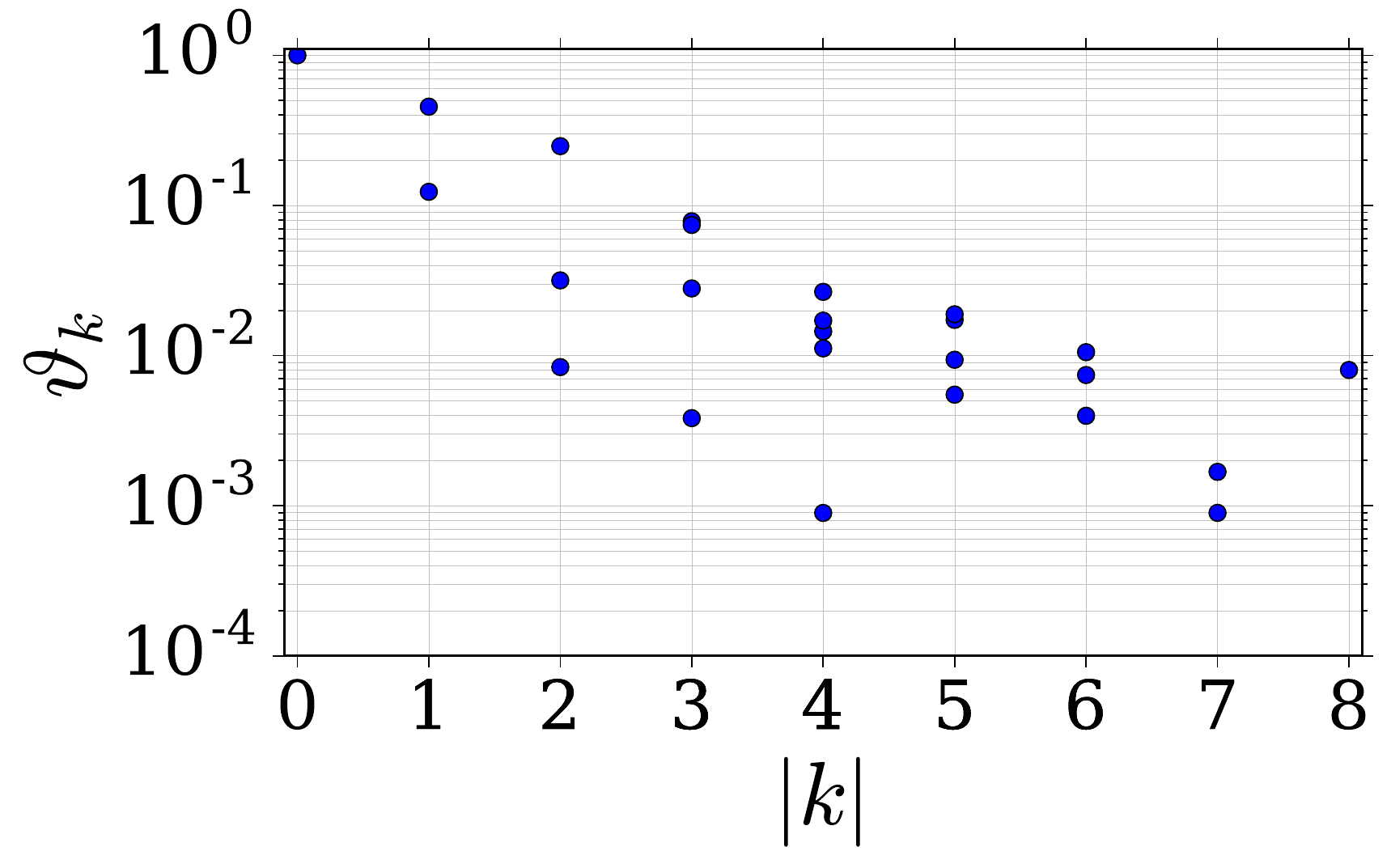} &   
      \includegraphics[scale=0.35]{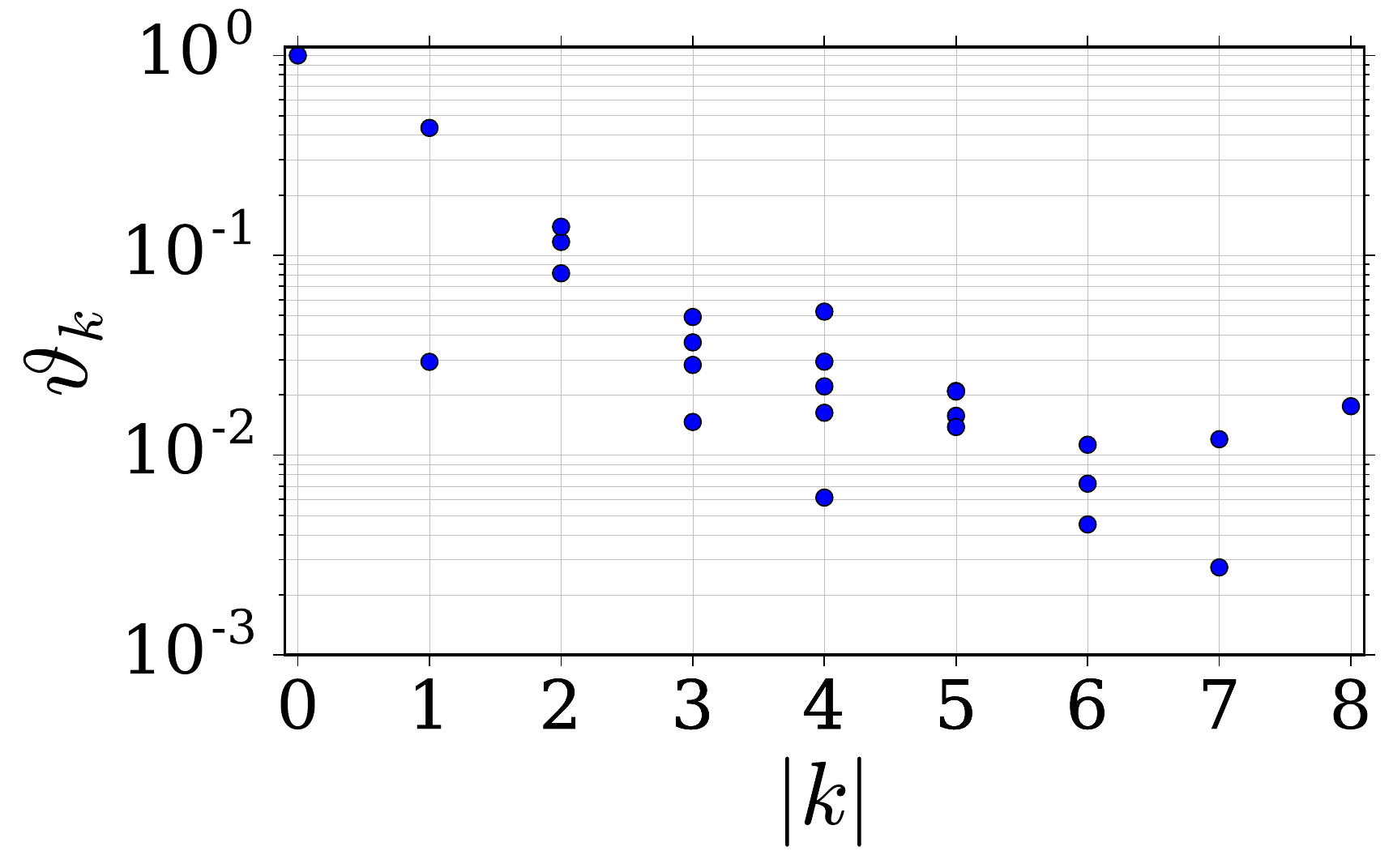} \\  
       {\small (a)} &    {\small (b)} &    {\small (c)} \\            
      \includegraphics[scale=0.35]{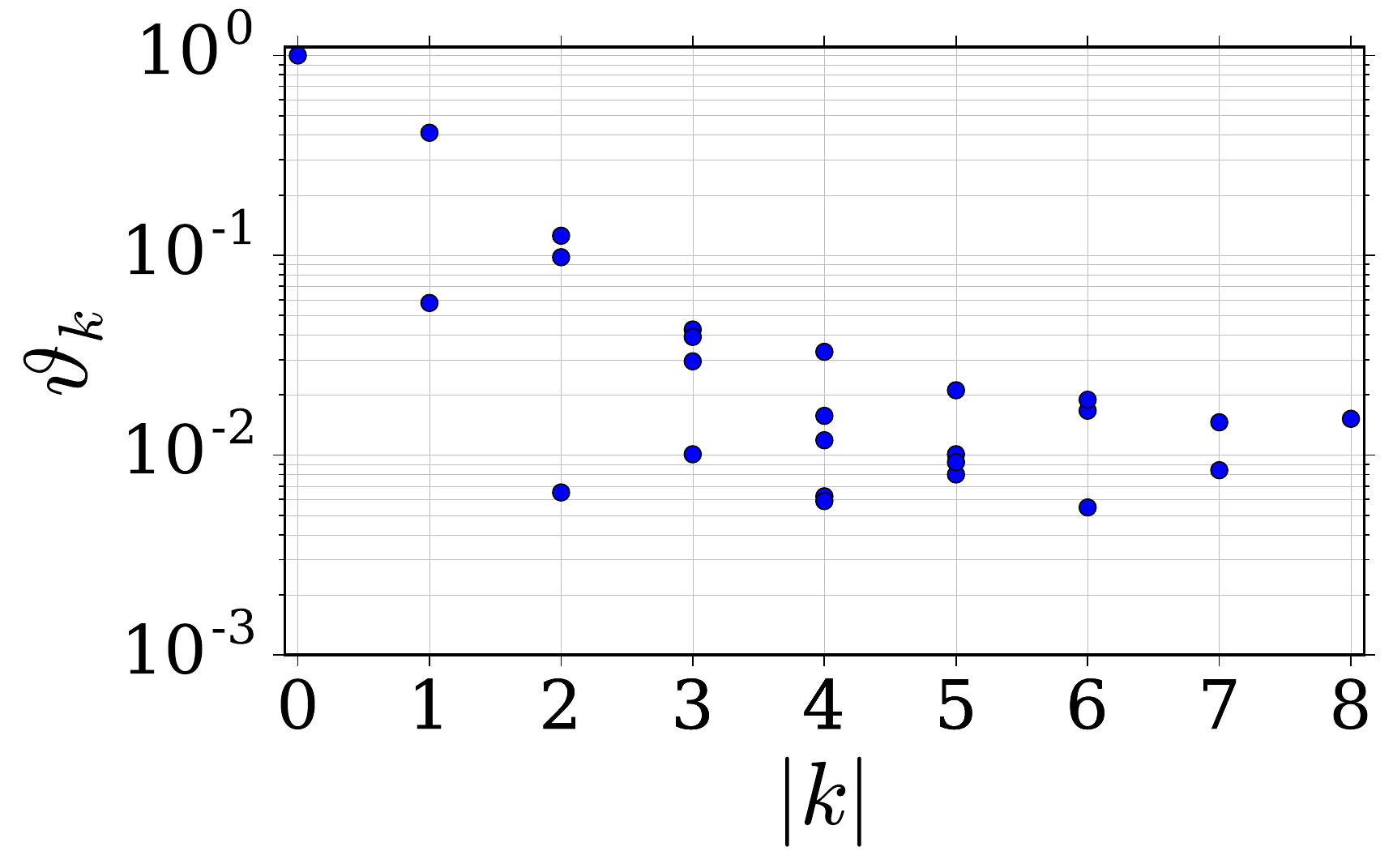} &
      \includegraphics[scale=0.35]{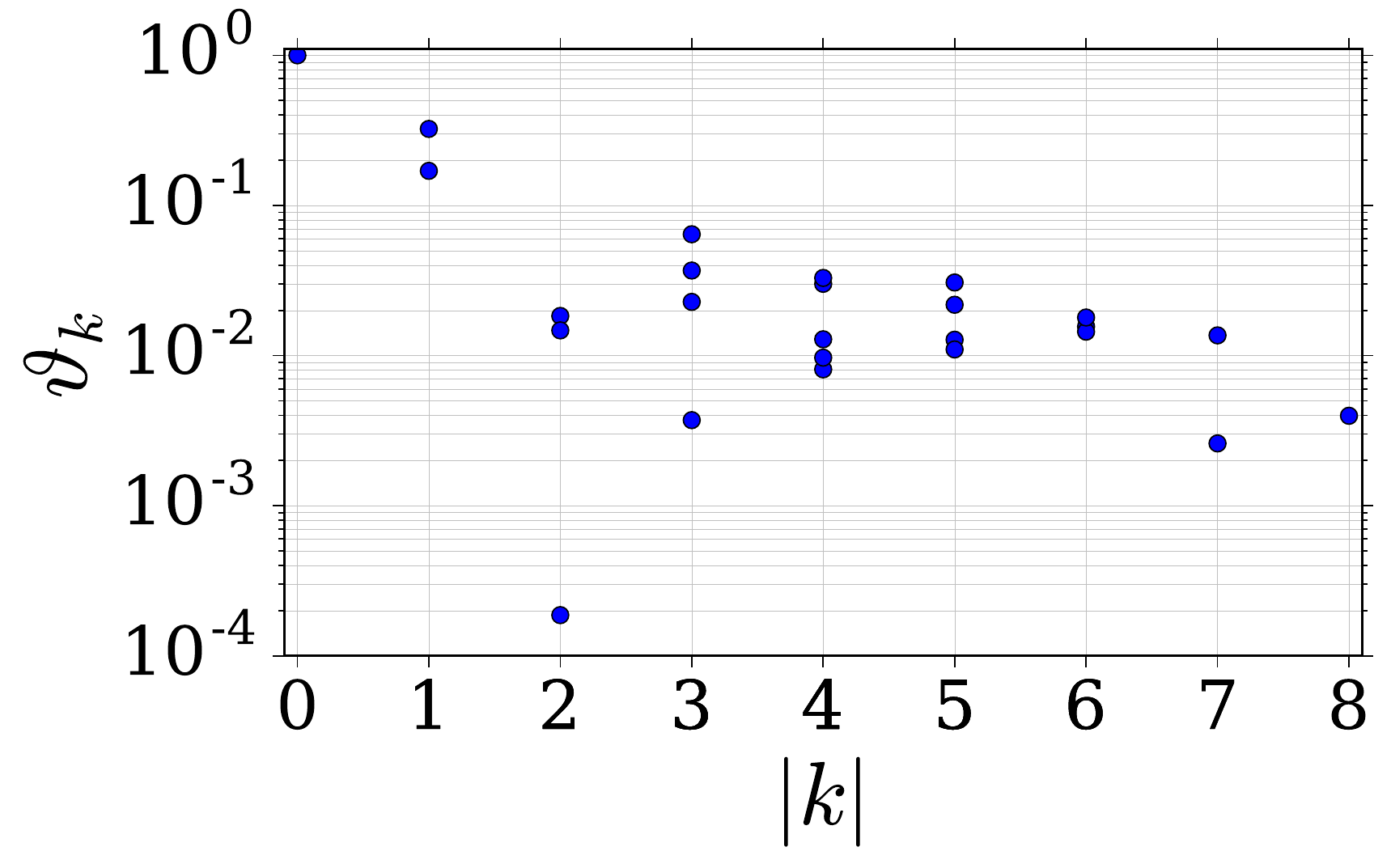} &
      \includegraphics[scale=0.35]{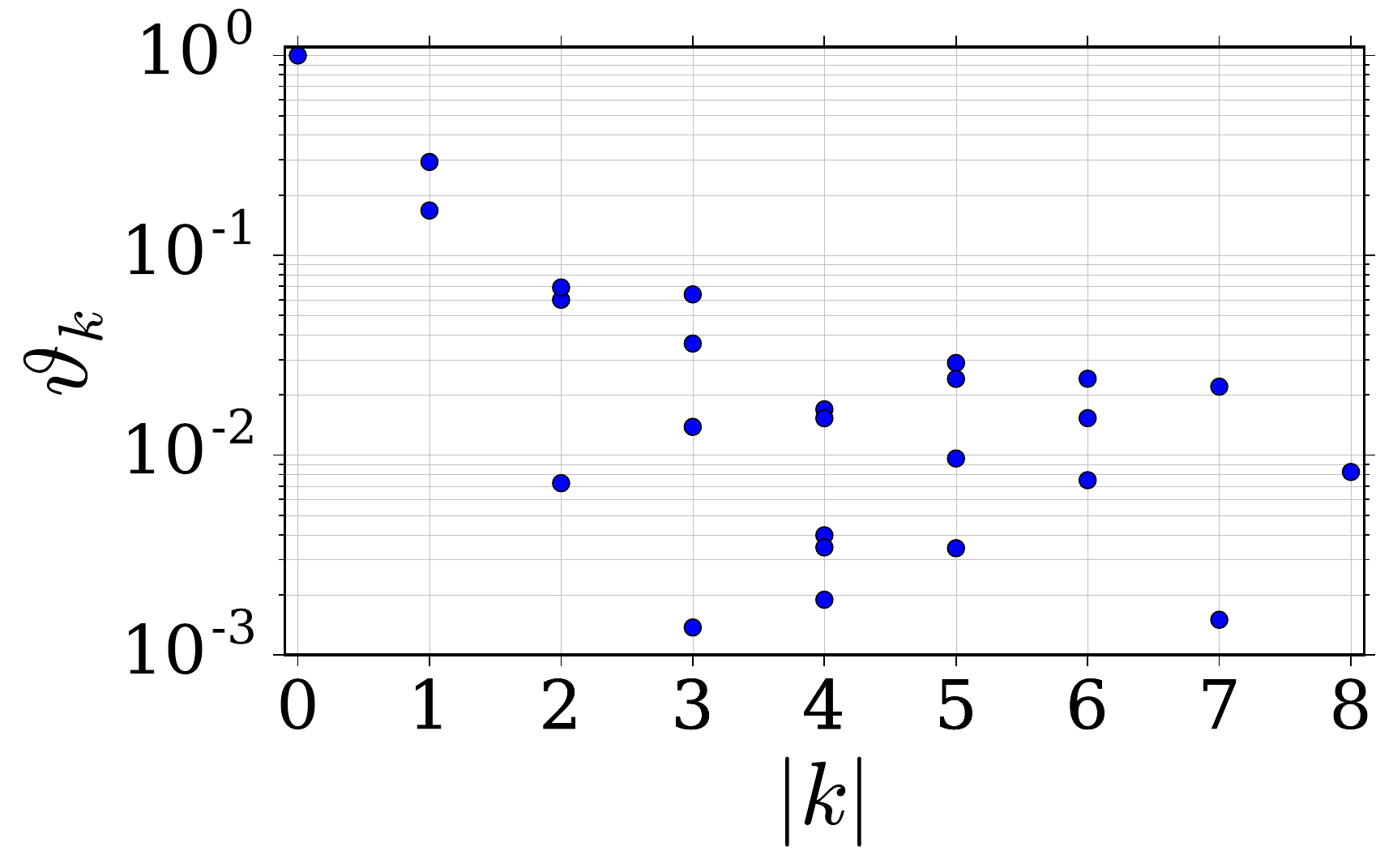} \\      
      {\small (d)} &    {\small (e)} &    {\small (f)} \\            
   \end{tabular}
\caption{Variation of $\vartheta_k=\| \hat{f}_{k} \Psi_{k}(q)  \|_2 / |\hat{f}_0|$ with $|k|=k_1+k_2$
for $k_1, \, k_2 =1,2,3,4$ and $|k|\leq 8$. 
Coefficients $\hat{f}_k$ belong to meta model (\ref{eq:pce}) constructed based on Set-B for $\epsilon[\ut]$ (a), $\epsilon[\U_c]$ (b), $\epsilon_{\infty}[\U]$ (c), $\epsilon_{\infty}[\uv]$ (d), $\epsilon_{\infty}[\tke]$ (e), and $\epsilon_{\infty}[\urms]$ (f).}\label{fig:pceConv}
\end{figure}

\fig~\ref{fig:pceConv} shows how $\vartheta_k$ (with $l_2$-norm used to evaluate the numerator) for different responses of Set-B varies with $|k|:=\sum_{i=1}^p k_i$, where $k_i$ denotes the order of the polynomial bases employed to span the space of the $i$-th parameter. 
For sake of brevity, only these particular plots are illustrated here, noticing that similar pattern is observed for other responses, and also for other simulation sets in \tab~\ref{tab:caseSummary}. 
According to these plots, $\vartheta_k \leq 10^{-1}$ for $|k|\geq 3$, which indicates more relative importance of the lower order terms in expansion (\ref{eq:pce}).

\subsection{Convergence of the flow statistics}\label{sec:timeAvg}
As discussed in \sects~\ref{sec:simCase} and \ref{sec:results}, different averaged quantities of turbulent channel flow can be used in the errors defined in \sect~\ref{sec:Rdefs}. 
These generic errors for QoI $g$ can be written as, 
$$
\epsilon[g]=c\|g-g^\circ \|
$$
with $c$ being a normalization constant and $\|\cdot\|$ denoting an appropriate norm.
The quantity $g$ is the averaged value of $\varphi$ in both time and homogeneous directions, i.e.
$g:=\langle \varphi \rangle$.
The estimator for the averaged value of $\varphi$ is defined by,
\begin{equation}\label{eq:averageOperator}
\langle \varphi \rangle_{nm} = (nm)^{-1} \sum_{i=1}^n \sum_{i=1}^m \varphi_{ij}
\end{equation}
where $n$ represents the number of time samples included in the averaging, and $m$ is the number of cells in the wall-parallel plane, i.e. $m=n_x n_z$.
The basic question is that, given $m$, how much averaging in time is required so that the estimates of the errors made throughout \sect~\ref{sec:results} are valid \rev{(i.e. not significantly contaminated by insufficient time-averaging)}. 
To seek an answer to this question, the estimator of error $\epsilon[g]$ as a function of $n$ is defined as, 
$$
\hat{\epsilon}[g] (n;m)= \left| \epsilon[g] (n;m) - \epsilon[g] (N;m) \right|
$$
in which, $m$ is a fixed parameter and $N$ specifies a large $n$ at which the averaging ends.
In a more practical setting, the elapsed averaging time, $n\Delta t$, is used to define the number of flow-throughs as $T=(U_b (n\Delta t))/l_x$.

\begin{figure}[!htbp]
\centering
   \begin{tabular}{cccc}
   \includegraphics[scale=0.24]{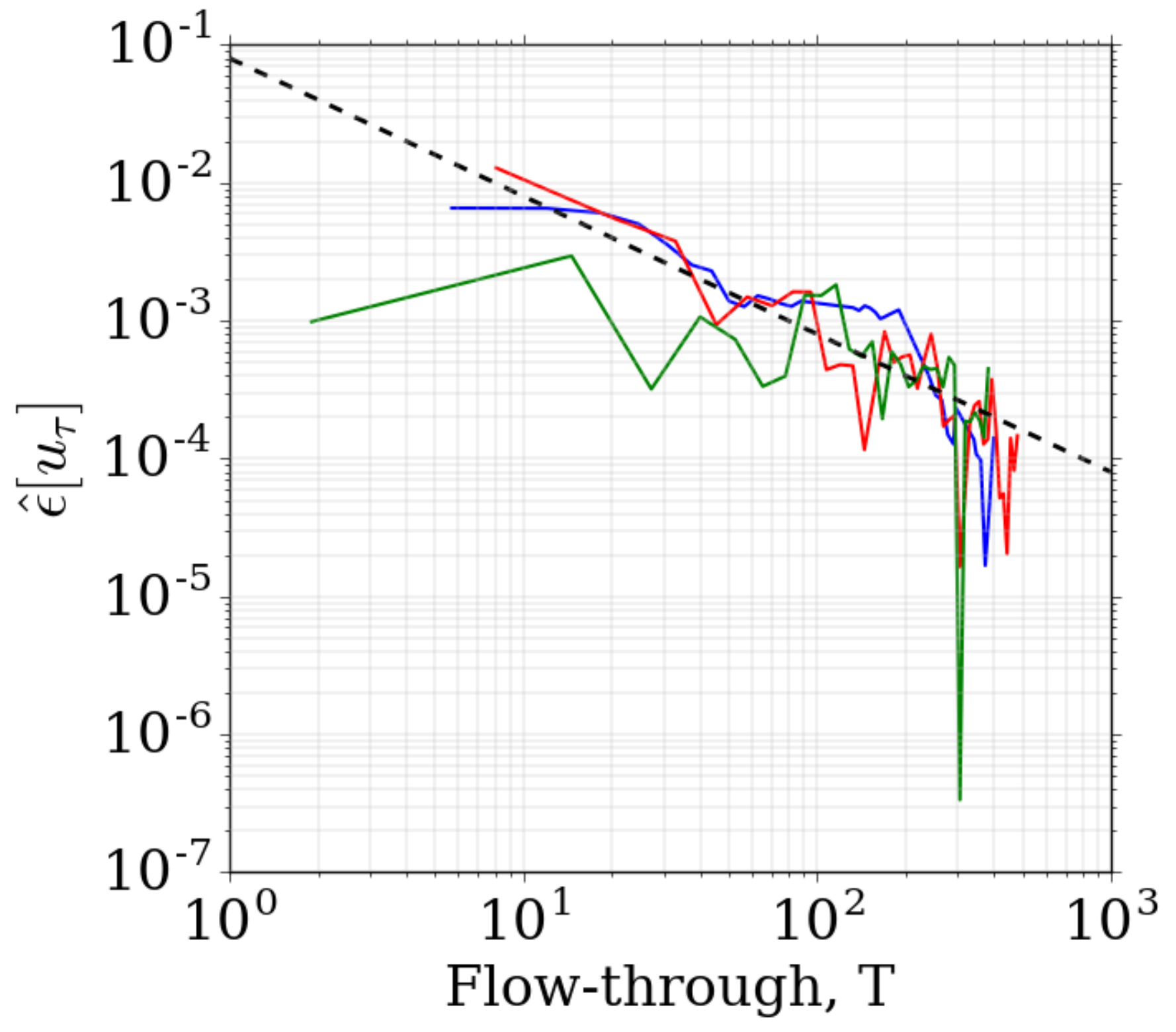} &
   \includegraphics[scale=0.24]{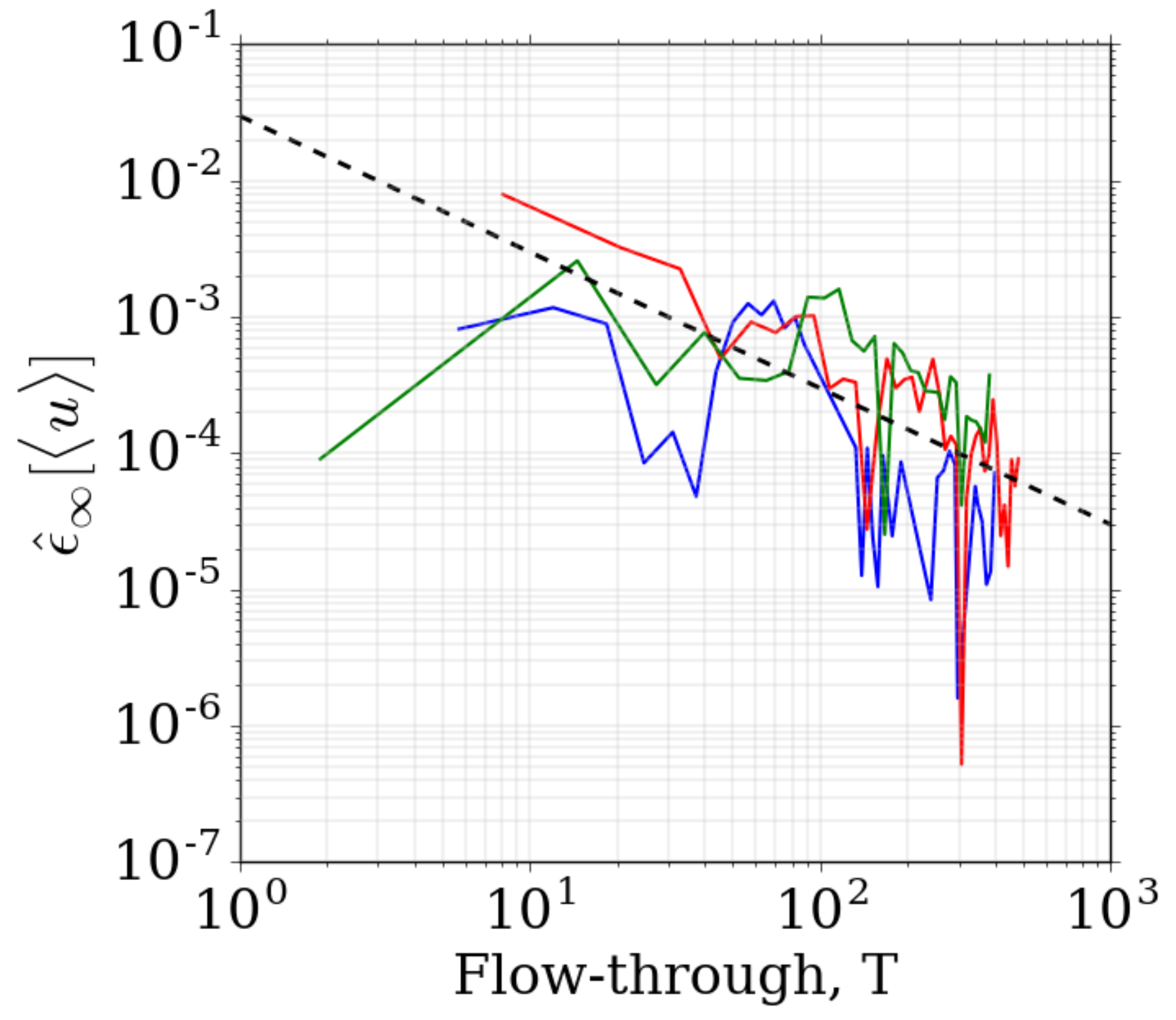} &  
   \includegraphics[scale=0.24]{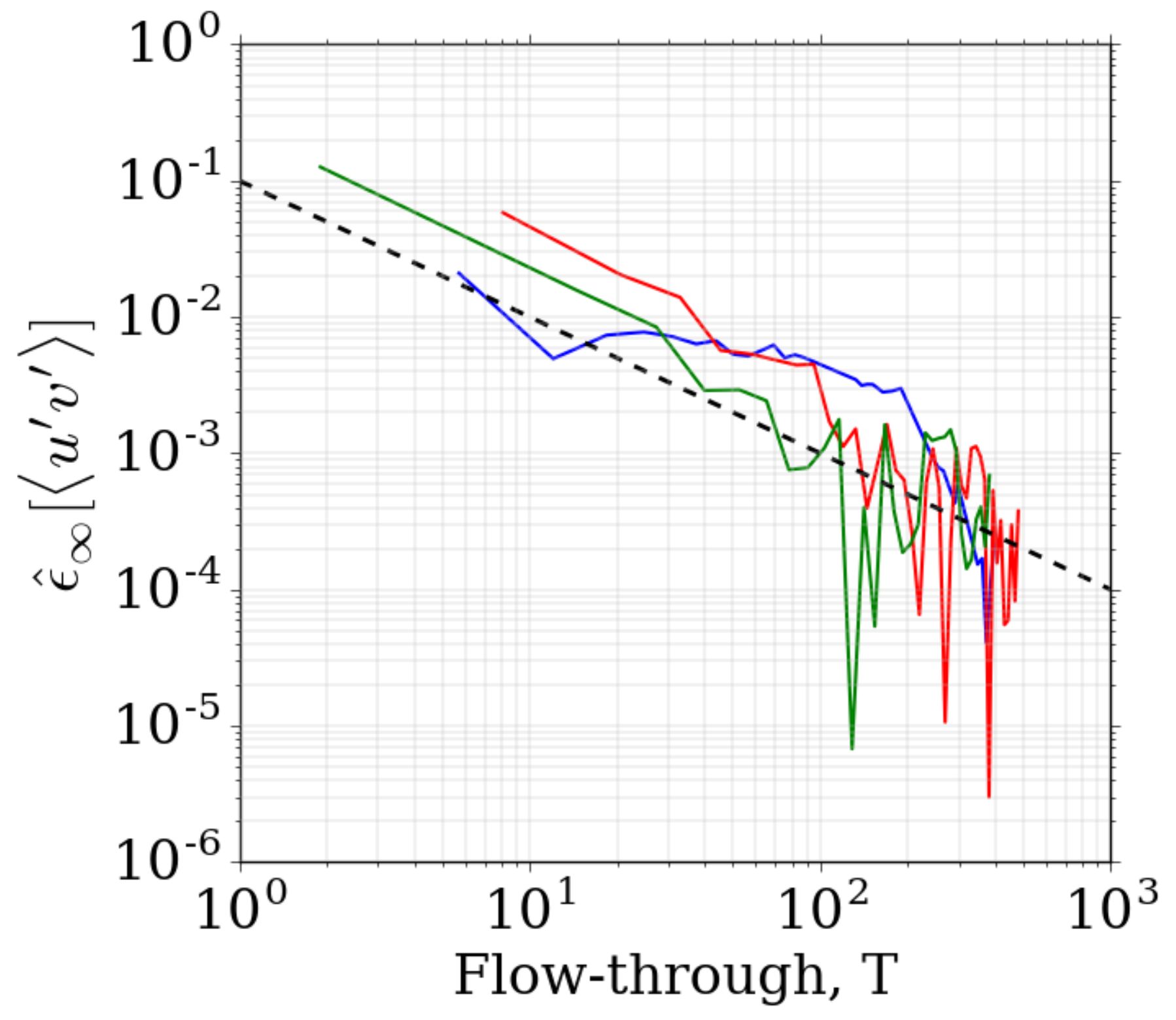} &
   \includegraphics[scale=0.24]{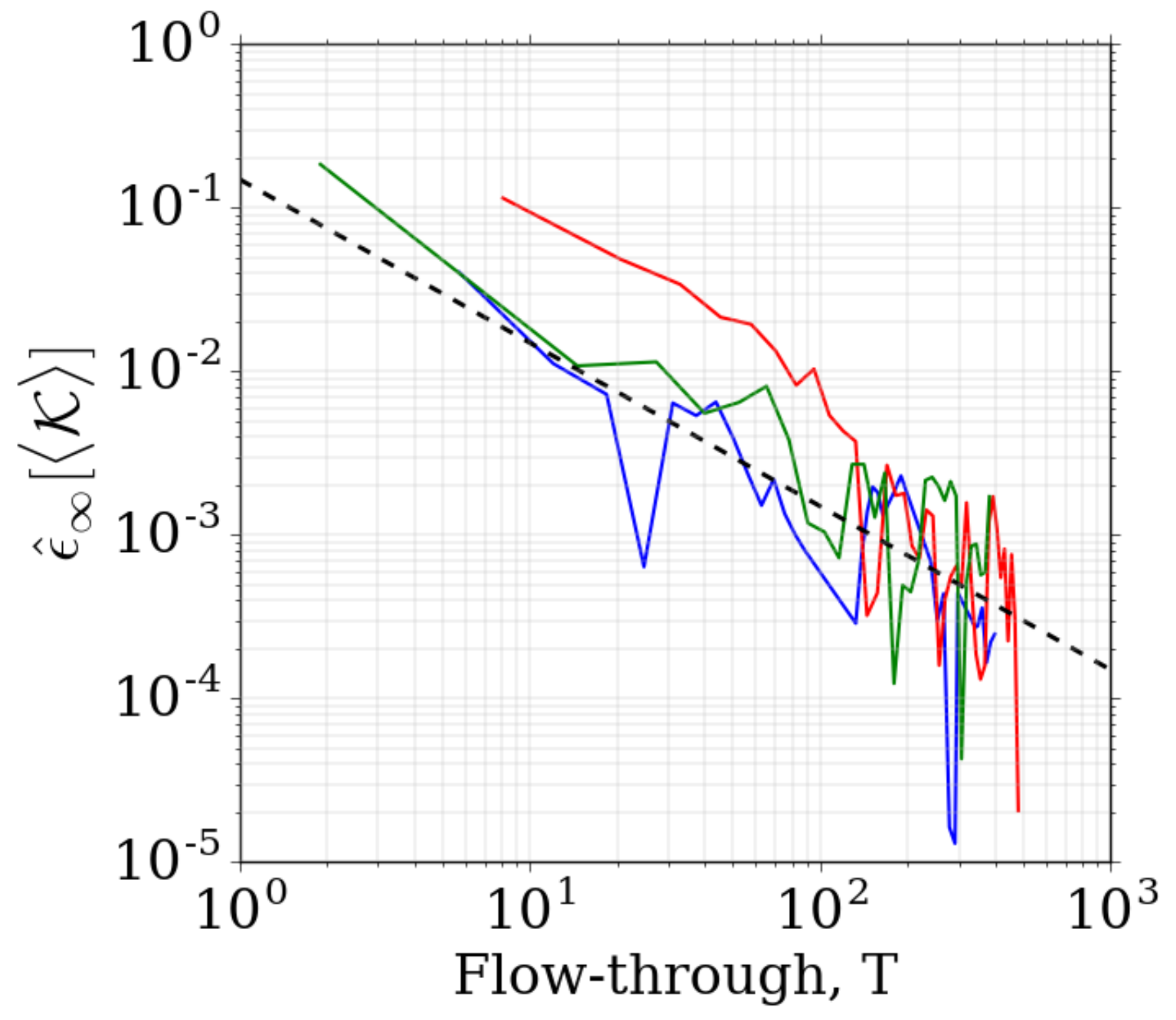} \\   
      {\small (a)} &    {\small (b)} & {\small (c)} &    {\small (d)} \\
   \end{tabular}
\caption{Variation of error estimators (not in $\%$) $\hat{\epsilon}[\ut]$ (a), $\hat{\epsilon}_{\infty}[\U]$ (b),  $\hat{\epsilon}_{\infty}[\uv]$ (c), and $\hat{\epsilon}_{\infty}[\tke]$ (d), with the number of flow-throughs, $T=(U_b (n\Delta t))/l_x$, for three simulations in Set-C with resolutions $\dxp=16.56$, $\dzp=9.96$ (blue line), $\dxp=80.0$, $\dzp=38.5$ (red line), and $\dxp=16.56$, $\dzp=67.0$ (green line). The dashed line represents $\hat{\epsilon} \propto T^{-1}$.}\label{fig:timeConvError}
\end{figure}
 
Figure \ref{fig:timeConvError}, illustrates the error estimators $\hat{\epsilon}[\ut]$, $\hat{\epsilon}_{\infty}[\U]$,  $\hat{\epsilon}_{\infty}[\uv]$, and $\hat{\epsilon}_{\infty}[\tke]$ for three simulations of Set-C in \tab~\ref{tab:caseSummary}. 
These particular combinations are chosen for sake of brevity, taking the fact into account that, a similar trend is expected for other simulations and QoIs. 
The first observation is that if the statistics are gathered at least for $T\approx 115$, as it is the case for the simulations in \sect~\ref{sec:results}, the estimated error in different quantities would be deviated by less than $\approx 10^{-3}$ \rev{from the value that would be achieved by continuing averaging for several hundred more $T$.}
Beside this, it seems the overall rate of reduction of the estimated errors, measured by different norms for different quantities and for different $n_x$ and $n_z$, is proportional to $T^{-1}$.
This observation that could be also expected from (\ref{eq:averageOperator}), clearly shows the relatively low convergence rate of the channel flow statistics.

\section{\revCom{Influence of the numerical scheme and $\dy$ distribution}}\label{sec:SetALAm}
\revCom{
Throughout \sect~\ref{sec:results}, all discussions were made based on the simulations listed in \tab~\ref{tab:caseSummary} which were carried out using the same numerical scheme, see \sect~\ref{sec:LESCFD}, and the same function (\ref{eq:delY}) for distributing grid points in the wall-normal direction. 
In order to address how the conclusions are sensitive to these two factors, two new sets of simulations are considered, both at target $\reyt=300$.}

\revCom{
To show the influence of the numerical scheme, Set-AL is considered which has the same conditions as Set-A in \tab~\ref{tab:caseSummary}, but LUST (linear-upwind stabilized transport) scheme~\cite{weller:12} is used instead of the linear interpolation to obtain face-center values of the fields from the values at the finite volume cell centers. 
In particular, the LUST scheme constructed as a fixed blend of $25\%$ linear upwind and $75\%$ linear central schemes is employed.
The conditions of the other new set of simulations, Set-AM, are kept the same as those of Set-A, except that the grid points in the wall-normal direction are distributed differently as explained below.
Starting from the wall, the $j$-th grid spacing in the wall-normal direction is, 
\begin{equation}\label{eq:delY2}
\dy_j=\gamma^{j-1} \dy_w \,,\quad j=1,2,\cdots,n_{y,2} \,,
\end{equation}
where, $n_{y,2}$ is the number of grid cells between the wall and channel half-height, and $\dy_w=(\delta /\reyt^\circ)  \dyp_w$, see \cite{saleh:2}.
The grid increment ratio, $\gamma$, is equal to ${\rm W}/({\rm W}-1)$, where ${\rm W}=\delta/\dy_c$, and $\dy_c$ is the grid cell height at the channel half-height. 
By choosing $\dyp_w=0.445$ and ${\rm W}=23$, the resulting number of cells to cover $\delta$ is $n_{y,2}=64$ which is equal to what used by Meyers and Sagaut \cite{meyers07}, and is higher than $46$ resulted from (\ref{eq:delY}) as employed in Set-A. }

\begin{figure}[!htbp]
\centering
\begin{tabular}{ccc}
   \includegraphics[scale=0.26]{A_DuTau.pdf} &
   \includegraphics[scale=0.26]{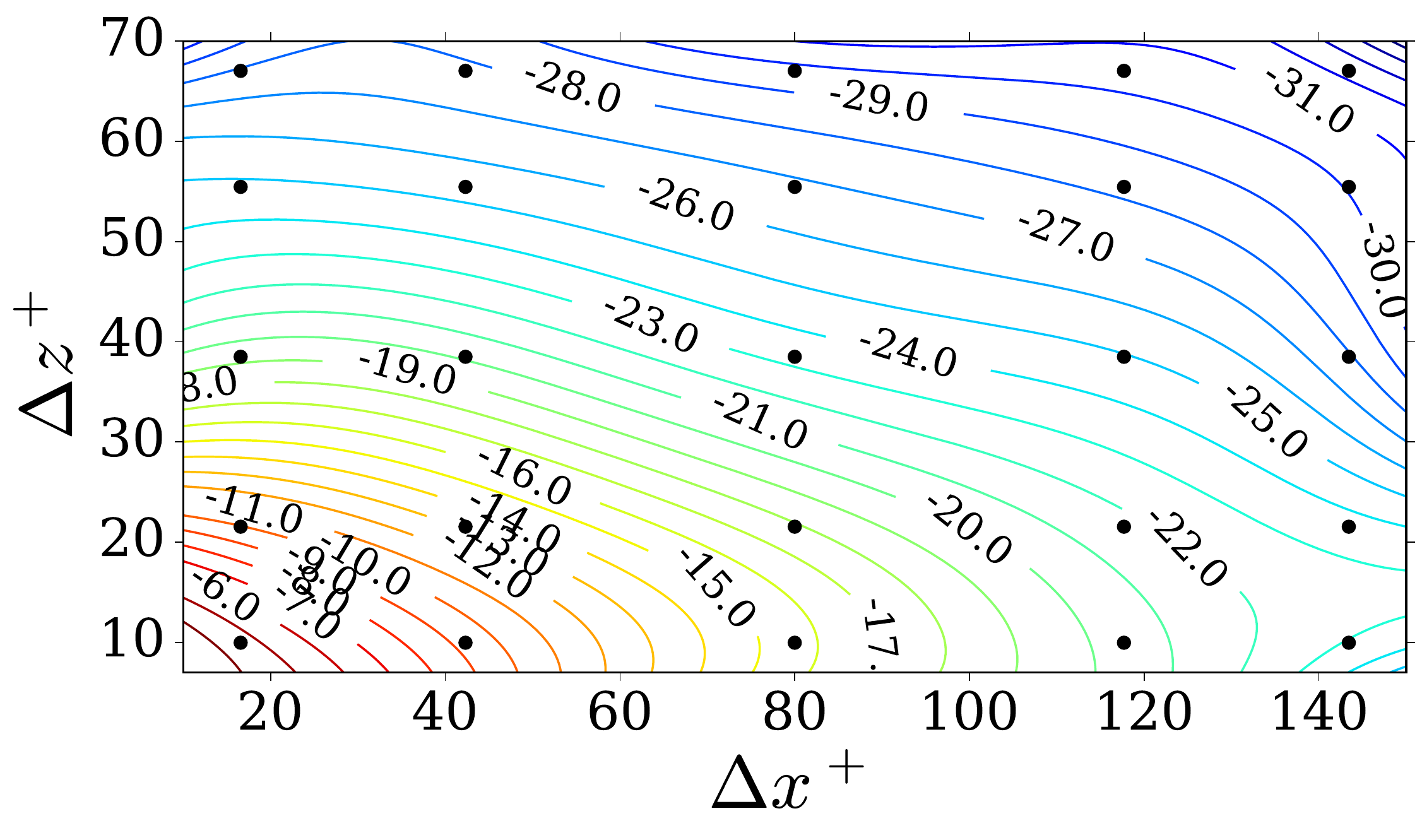} &
   \includegraphics[scale=0.26]{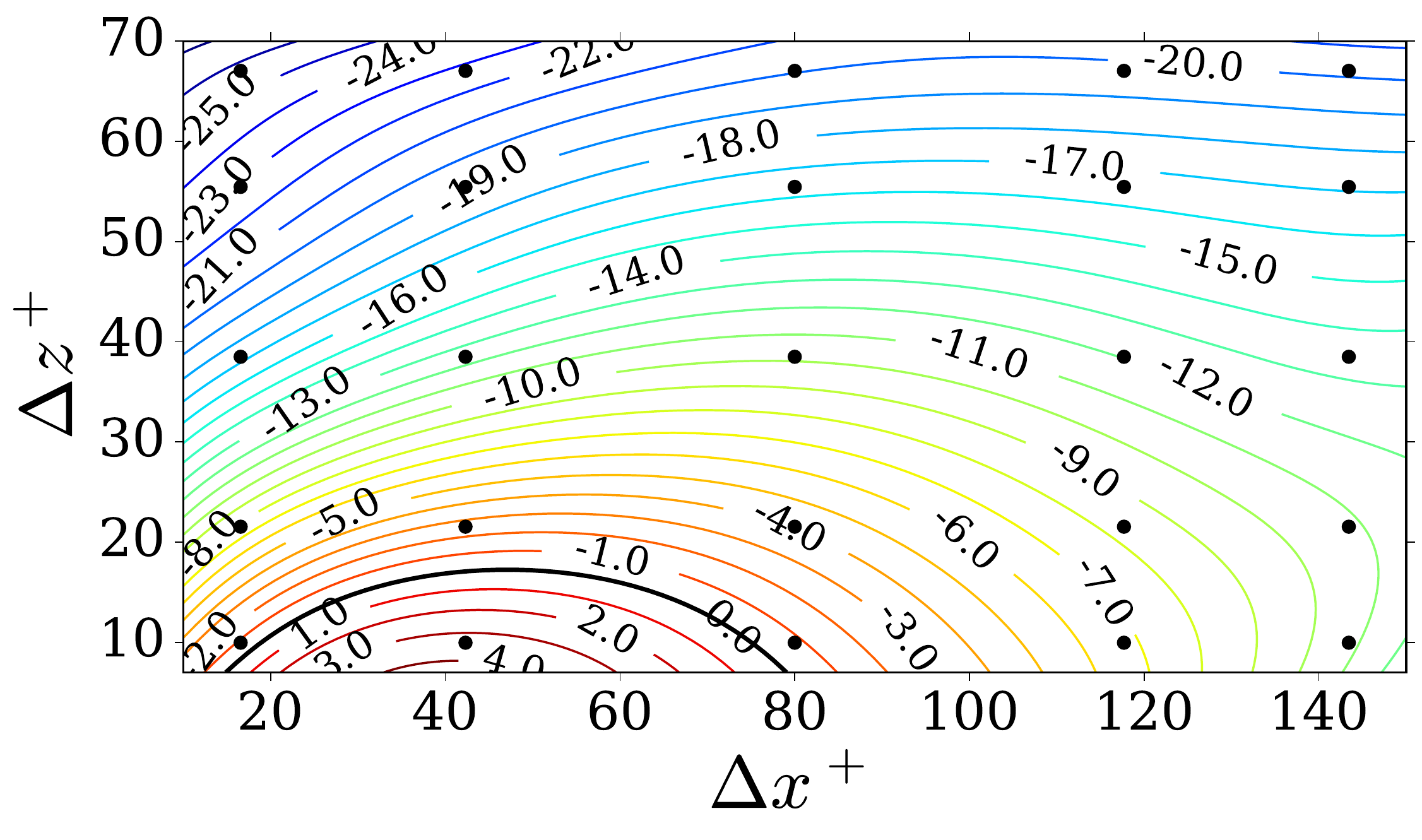} \\
   {\small (a)} & {\small (b)} & {\small (c)} \\
\end{tabular}
\caption{\revCom{Isolines of $\epsilon[{\ut}]$ in the $\dxp\dash\dzp$ plane at target $\reyt=300$ and $\dyp_w=0.445$ resulted from, the linear scheme with (\ref{eq:delY}) (Set-A) (a), the LUST scheme with (\ref{eq:delY}) (Set-AL) (b), and the linear scheme with (\ref{eq:delY2}) (Set-AM) (c).}}\label{fig:AALAM_duTau}
\end{figure}

\begin{figure}[!htbp]
\centering
   \begin{tabular}{ccc}
   \includegraphics[scale=0.26]{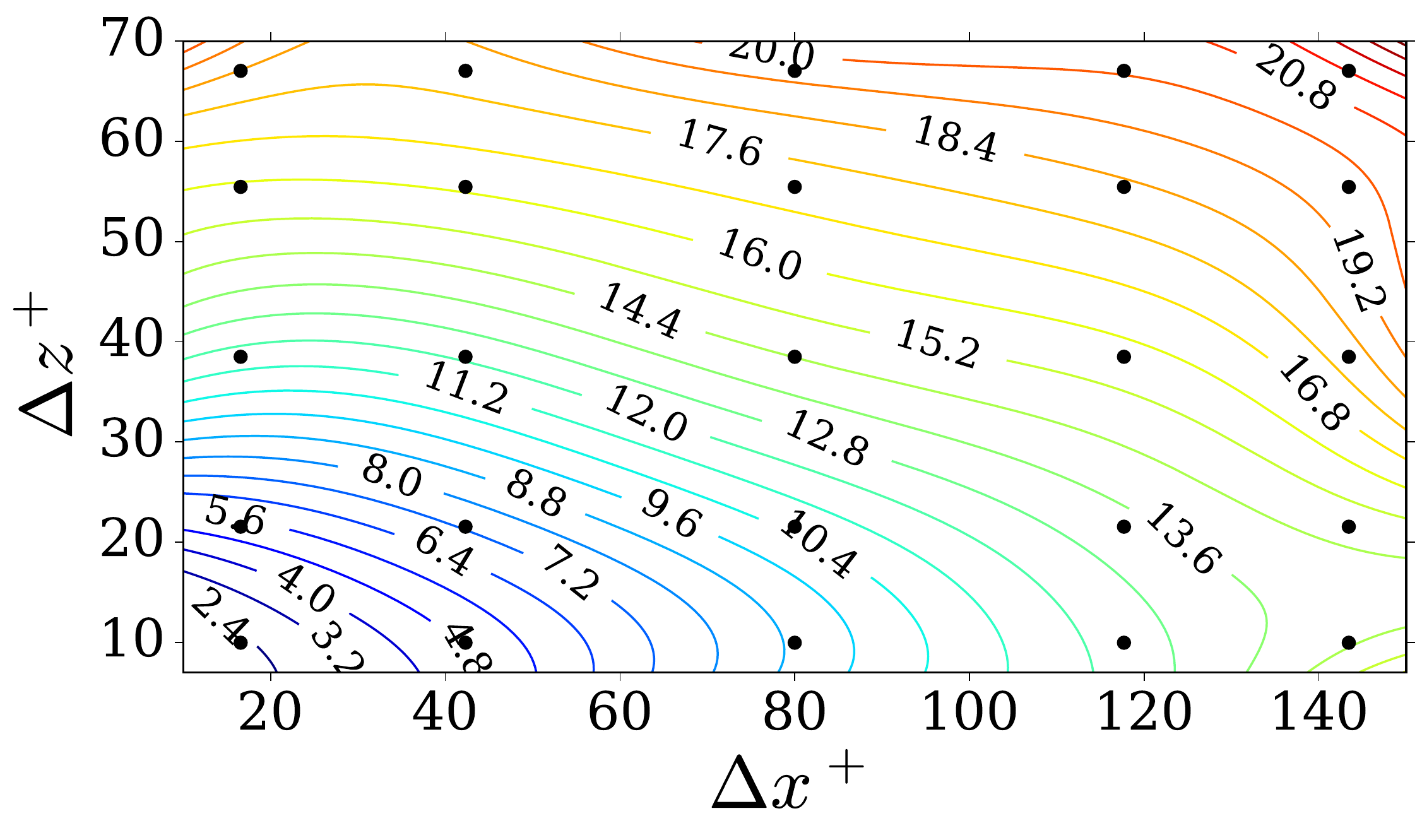} &
   \includegraphics[scale=0.26]{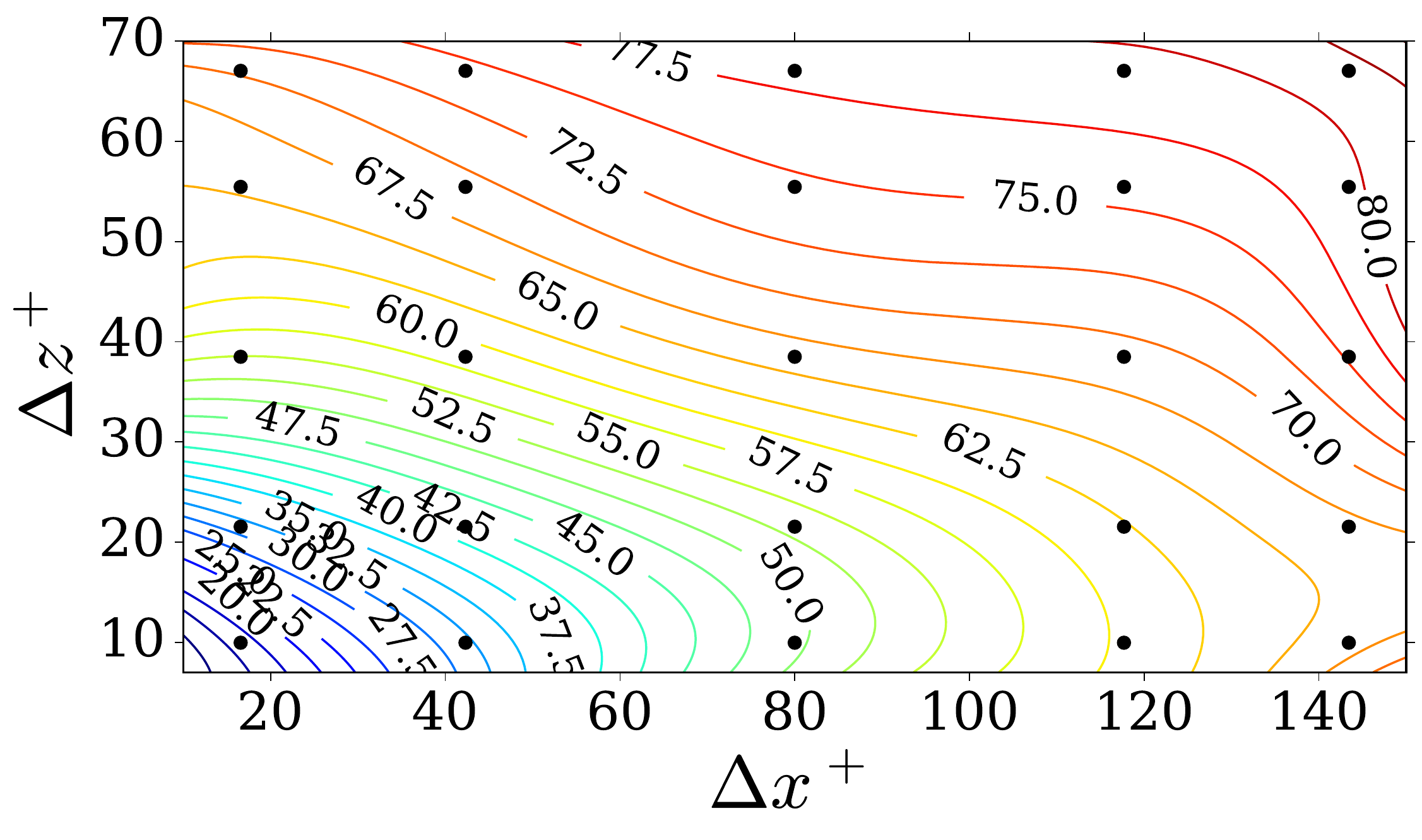} &
   \includegraphics[scale=0.26]{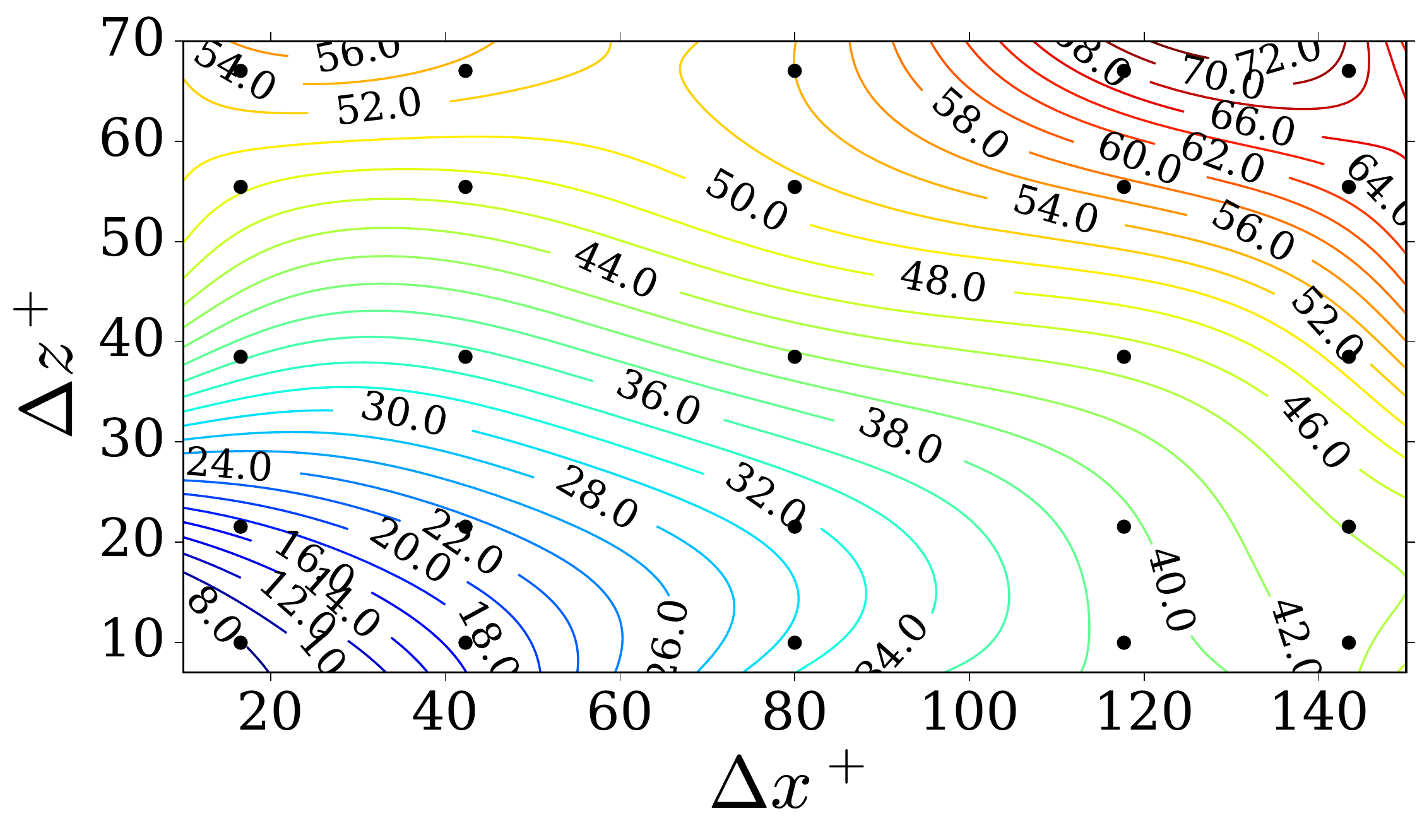} \\   
   {\small (a)} & {\small (b)} & {\small (c)} \\   
   \end{tabular}
\caption{\revCom{Isolines of $\einf[\U]$ (a), $\einf[\uv]$ (b), and $\einf[\tke]$ (c), plotted in the $\dxp\dash\dzp$ plane for Set-AL.}}\label{fig:AL_errorInProfs}
\end{figure}

\revCom{
The isolines of $\epsilon[{\ut}]$ of Set-AL and Set-AM are plotted in the $\dxp\dash\dzp$ plane in \fig~\ref{fig:AALAM_duTau} (b) and (c), respectively.
For ease of comparison, the associated plot of Set-A is also added. 
The clearly different pattern of the $\epsilon[{\ut}]$ isolines of Set-AL compared to that of Set-A, reveals the strong influence of the numerical scheme. 
It is observed that not only the loci of $\epsilon[{\ut}]=0$ do not exist for Set-AL, but also the associated values of $\epsilon[{\ut}]$ are larger than those of Set-A at the same resolution.
Moreover, even at the finest resolution, $\ut$ of Set-AL does not converge to the DNS value. 
The other main characteristic of the errors in QoIs resulted from the LUST scheme is that the error reduction in  the $\dxp\dash\dzp$ plane is relatively monotonous without any fictitious convergence for some QoIs, for instance see \fig~\ref{fig:AL_errorInProfs}.
This is an evidence that the loci of $\epsilon[\ut]=0$ in the $\dxp\dash\dzp$ plane achieved by the linear scheme in Set-A is due to the numerical effects.
}

\revCom{
In contrast to the large influence of the numerical scheme, the similarity of the errors in different QoIs resulted from Set-AM and Set-A, suggests very low impact of the grid cell distribution in the wall-normal direction. 
This is clearly visible for $\epsilon[u_\tau]$ in \fig~\ref{fig:AALAM_duTau}, and also for other errors which are not illustrated here for sake of brevity. 
}

\section{\revCom{Two-point velocity correlations and integral length scales}}\label{sec:twoPointAppendix}
\begin{figure}[!htbp]
\centering
   \begin{tabular}{cc}
   \includegraphics[scale=0.35]{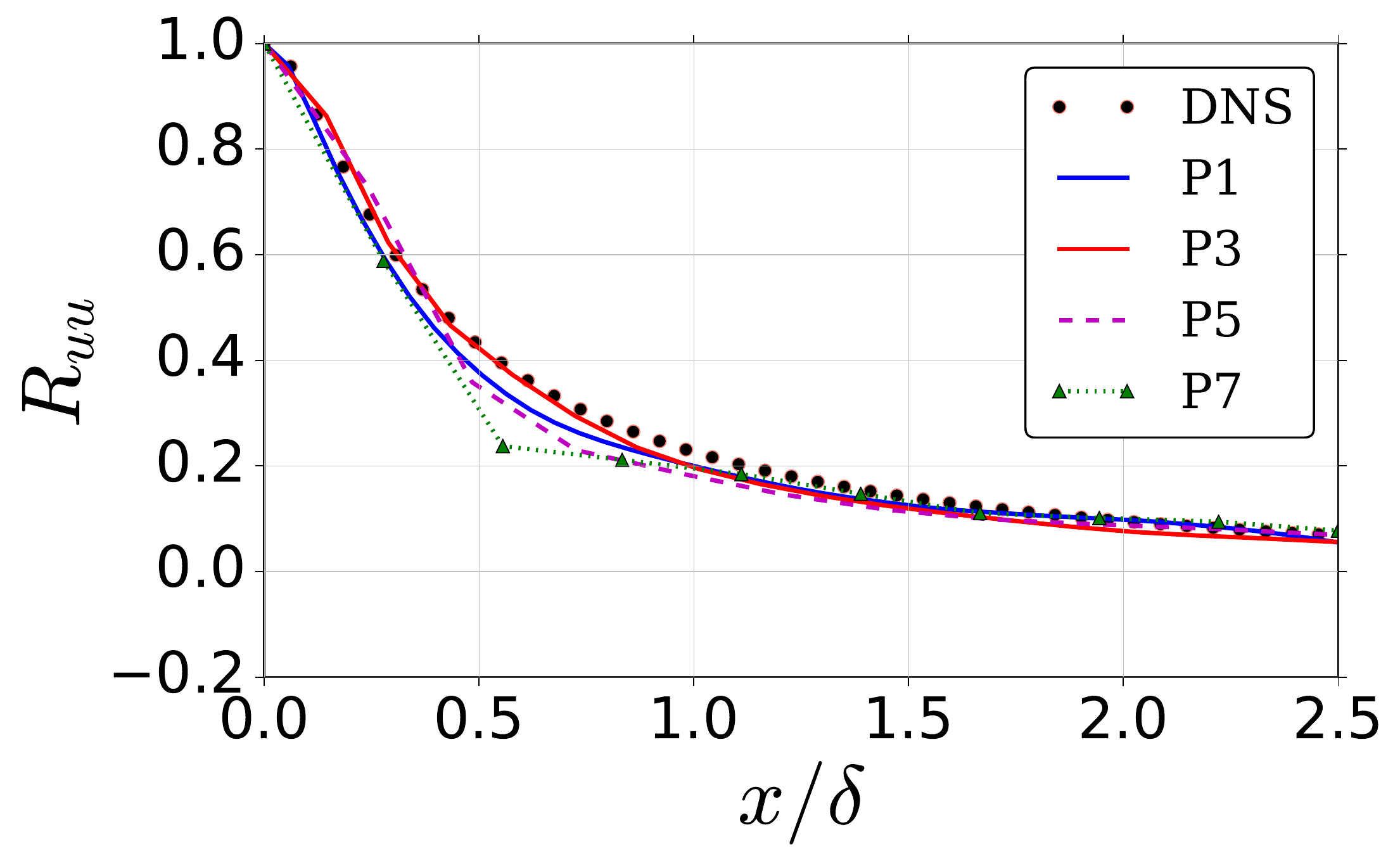} &
   \includegraphics[scale=0.35]{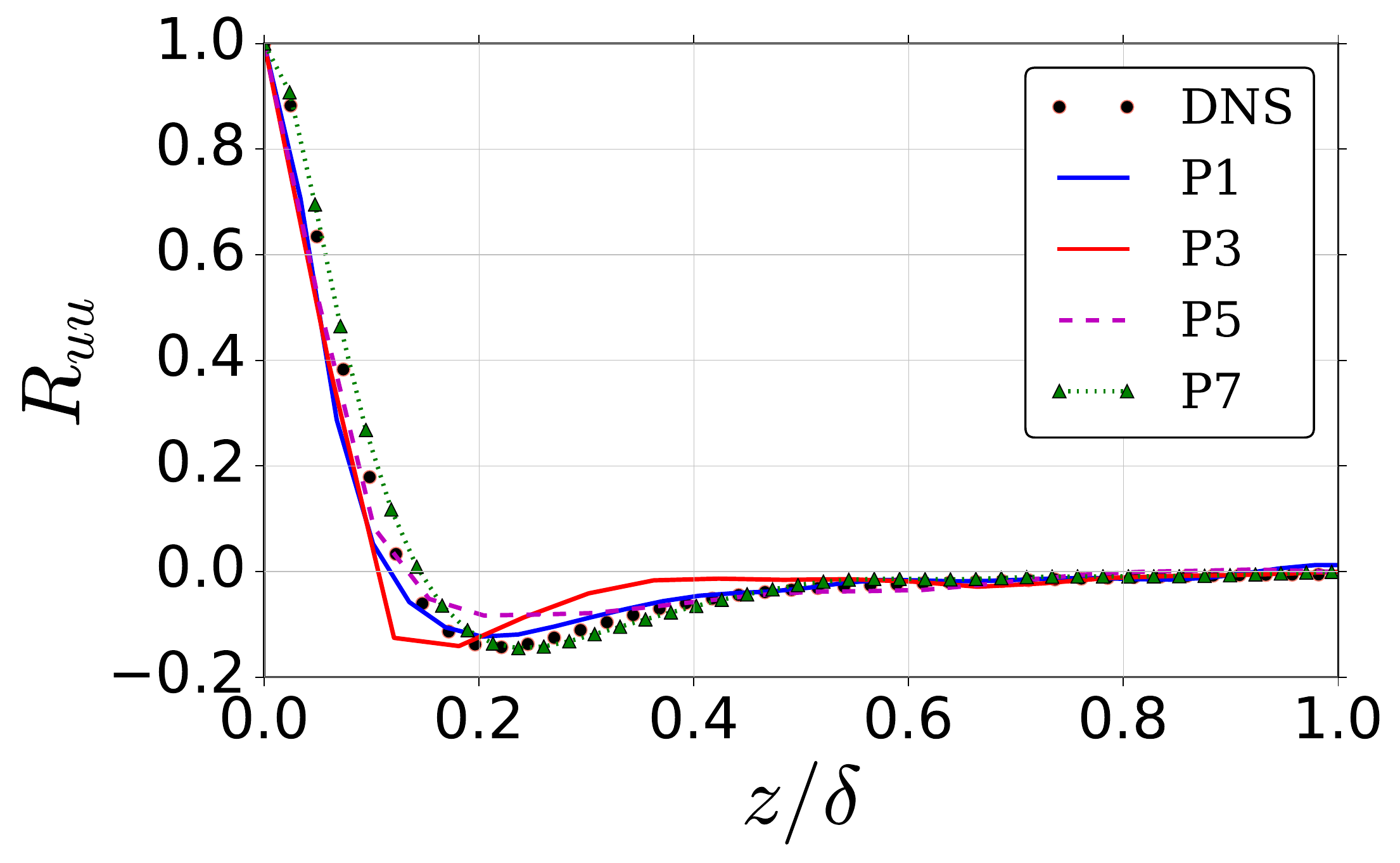} \\
   {\small (a)} &    {\small (b)} \\      
   \includegraphics[scale=0.35]{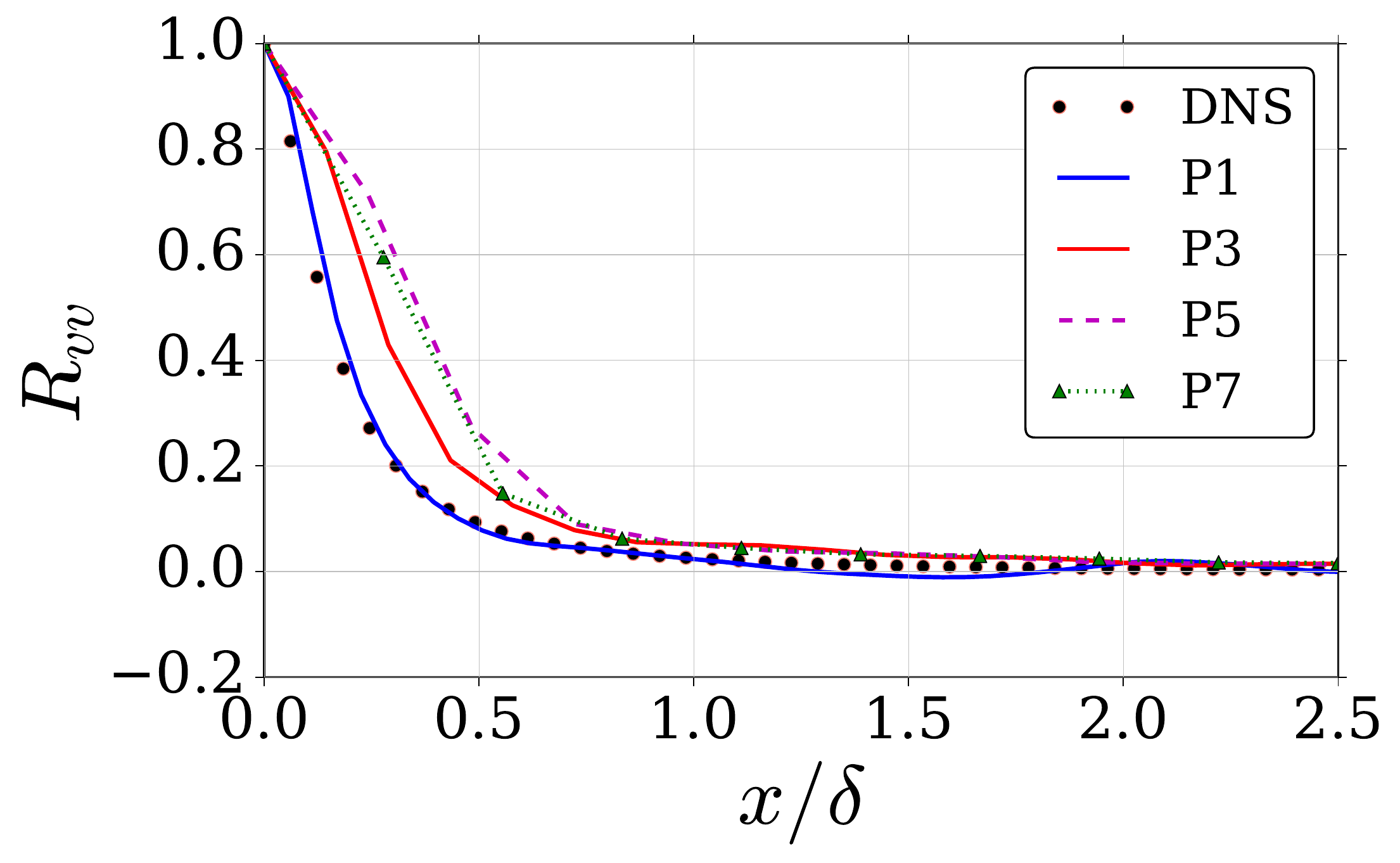} &
   \includegraphics[scale=0.35]{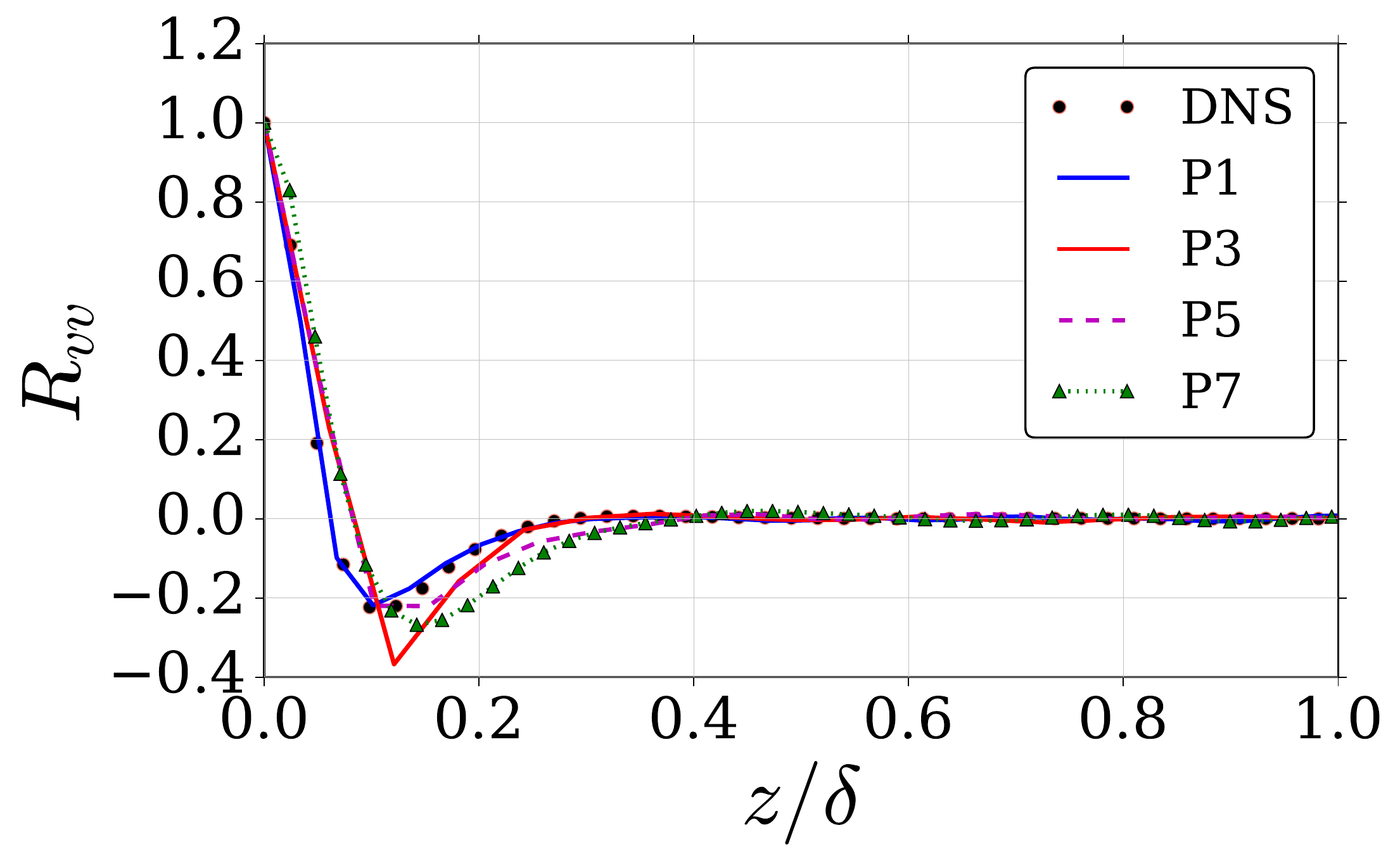} \\  
   {\small (c)} &    {\small (d)} \\      
   \end{tabular}
\caption{\revCom{Two-point correlations $R_{uu}$ (top) and $R_{vv}$ (bottom) in the streamwise and spanwise directions for a few resolutions with $\epsilon[u_\tau]=0$ based on the simulation condisions of Set-A, see \fig~\ref{fig:AuTauLoci} and \tab~\ref{tab:integLengths}. $y^+_0 = 15$, $x_0=l_x/2$, and $z_0=l_z/2$. }}\label{fig:twoPtCorrs}
\end{figure}

\revCom{
\fig~\ref{fig:twoPtCorrs} shows the two-point velocity correlations $R_{uu}$ and $R_{vv}$ in $x$ and $z$ directions for a selected number of simulations associated with the resolutions on the $\epsilon[\ut]=0$ curve in \fig~\ref{fig:AuTauLoci}.
The two-point velocity correlations in the streamwise and spanwise directions are defined as, see \eg~\cite{pope}, 
$$
R^{(x)}_{{u_i}{u_j}} = \langle u'_i(x_0,y_0,z_0) u'_j(x_0+x,y_0,z_0) \rangle_{x_0},\quad   
R^{(z)}_{{u_i}{u_j}} = \langle u'_i(x_0,y_0,z_0)u'_j(x_0,y_0,z_0+z)\rangle_{z_0} \,,
$$
where, $\langle \cdot \rangle_{x_0}$ and $\langle \cdot \rangle_{z_0}$ denote averaging over all locations $x_0$ and $z_0$, respectively, in addition to averaging over time. 
According to \fig~\ref{fig:twoPtCorrs}, no relation between the plots of different cases can be recognized.
The $R_{uu}$ and $R_{vv}$ of the simulation with the finest resolution, P1, have good agreement with the DNS data of \cite{iwamoto02}.
However, other cases may have accurate correlations as well, for instance see the $R_{uu}$ versus $x/\delta$ of P3.
Compared to $R_{uu}$, correlation $R_{vv}$ seems to be more reliable in assessment of the resolutions: $R_{vv}$ is only well predicted by the highest considered resolution. 
}

\revCom{
For a quantitative assessment, integral length scales corresponding to the two-point correlations can be employed, which are defined as, $L^{(x)}_{{u_i}{u_j}}=R^{{(x)}^{-1}}_{{u_i}{u_j}}(0) \int R^{(x)}_{{u_i}{u_j}}(x) \dd x$ and $L^{(z)}_{{u_i}{u_j}}=R^{{(z)}^{-1}}_{{u_i}{u_j}}(0) \int R^{(z)}_{{u_i}{u_j}}(z) \dd z$.
The integrals of $L^{(x)}_{uu}$ and $L^{(x)}_{vv}$ are taken from $0$ to $l_x/2$.
But, due to the sign change of $R^{(z)}_{uu}$ and $R^{(z)}_{vv}$, the integrals of $L^{(z)}_{uu}$ and  $L^{(z)}_{vv}$ are calculated respectively from $0$ to $z^+\approx 36.6$ and from $0$ to $z^+\approx 19.5$ for target $\reyt=300$. 
Using the DNS data of Iwamoto \et~\cite{iwamoto02} at $y_0^+\approx 15$, it is resulted that 
$L^{{(x)}^+}_{uu} \approx 226.6$, $L^{{(z)}^+}_{uu} \approx 19.1$, $L^{{(x)}^+}_{vv} \approx 68.3$, $L^{{(z)}^+}_{vv} \approx 10.4$.
Based on these, the number of cells per unit integral length scales for the channel flow simulations associated to the resolutions specified in \fig~\ref{fig:AuTauLoci} is obtained and listed in \tab~\ref{tab:integLengths}.
}

\begin{table}[!htbp]
\centering
\caption{\revCom{Number of cells per integral length scales in the streamwise and spanwise directions for the simulations denoted by P1, ..., P7 in \fig~\ref{fig:AuTauLoci}. Integral length scales are computed by the DNS data of~\cite{iwamoto02} at $y^+=15$.}}\label{tab:integLengths}
\begin{small}
\begin{tabular}{ccc|cccc}
\toprule\toprule
Simulation & $\dxp$ & $\dzp$ & $L^{(x)}_{uu}/\dx$ & $L^{(z)}_{uu}/\dz$ & $L^{(x)}_{vv}/\dx$ & $L^{(z)}_{vv}/\dz$\\
\hline
P1 & 16.56 & 9.96 & 13.6 & 1.9 & 4.1 & 1.0\\
P2 & 30 & 15.25 & 7.5 & 1.3 & 2.3 & 0.7\\
P3 & 42 & 17.67 & 5.4 & 1.1 & 1.6 & 0.6\\
P4 & 50 & 18.10 & 4.5 & 1.1 & 1.4 & 0.6\\
P5 & 70 & 15 & 3.2 & 1.3 & 1.0 & 0.7\\
P6 & 78 & 9.96 & 2.9 & 1.9 & 0.9 & 1.0\\
P7 & 80 & 7 & 2.8 & 2.7 & 0.85 & 1.5\\
\bottomrule
\end{tabular}
\end{small}
\end{table}

\revCom{
The $\dxp$ of P1 is the smallest, so the resulting $L^{(x)}_{uu}/\dx$ and $L^{(x)}_{vv}/\dx$ of this case can be used as the guidelines for the resolution in the streamwise direction.
Specially, $L^{(x)}_{vv}/\dx$ is uniquely the indicator of the best simulation, according to  \fig~\ref{fig:twoPtCorrs}(c).
To resolve $L^{(z)}_{uu}$ and $L^{(z)}_{vv}$ in the spanwise direction, respectively, two and one cell(s) are/is at least required. 
It is noted that, although the resolutions listed in \tab~\ref{tab:integLengths} lead to accurate predictions of $\ut$ at $\reyt=300$, most of them are clearly insufficient to resolve the near wall structure as it is essential for WRLES.
It is recalled that, see~\cite{robinson91}, in the viscous sublayer of the TBL there are low- and high-velocity streaks with length $\approx 400\dash 500$ wall units which are spaced about $100$ wall units apart. 
There are also quasi-streamwise vortices in the buffer layer with length $200\dash400$ and diameter $50$ wall units.
}


\bibliographystyle{plain}
\bibliography{bib_uqWRChanGrid}

\begin{thebibliography}{10}

\bibitem{dakotaMan}
B.~M. Adams, L.~E. Bauman, W.~J. Bohnhoff, K.~R. Dalbey, M.~S. Ebeida, J.~P.
  Eddy, M.~S. Eldred, P.~D. Hough, K.~T. Hu, J.~D. Jakeman, J.~A. Stephens,
  L.~P. Swiler, D.~M. Vigil, and T.~M. Wildey.
\newblock Dakota, {A} {M}ultilevel {P}arallel {O}bject-{O}riented {F}ramework
  for {D}esign {O}ptimization, {P}arameter {E}stimation, {U}ncertainty
  {Q}uantification, and {S}ensitivity {A}nalysis: {V}ersion 6.3 {U}ser{'}s
  {M}anual.
\newblock {\em Sandia Technical Report}, SAND2014-4633, 2015.

\bibitem{bae:17}
H.~J. Bae, A.~Lozano-Dur\'an, S.~T. Bose, and P.~Moin.
\newblock Turbulence intensities in large-eddy simulation of wall-bounded
  flows.
\newblock {\em Phys. Rev. Fluids}, 3:014610, 2018.

\bibitem{celik:06}
I.~Celik, M.~Klein, M.~Freitag, and J.~Janicka.
\newblock Assessment measures for {URANS/DES/LES}: an overview with
  applications.
\newblock {\em Journal of Turbulence}, 7:N48, 2006.

\bibitem{celik:05}
I.~B. Celik, Z.~N. Cehreli, and I.~I. Yavuz.
\newblock Index of resolution quality for large eddy simulations.
\newblock {\em ASME. J. Fluids Eng.}, 127(5):949--958, 2005.

\bibitem{chapman79}
D.~R. Chapman.
\newblock Computational aerodynamics development and outlook.
\newblock {\em AIAA journal}, 17(12):1293--1313, 1979.

\bibitem{choi12}
H.~Choi and P.~Moin.
\newblock Grid-point requirements for large eddy simulation: Chapman's
  estimates revisited.
\newblock {\em Physics of Fluids}, 24(1):--, 2012.

\bibitem{chow03}
F.~K. Chow and P.~Moin.
\newblock A further study of numerical errors in large-eddy simulations.
\newblock {\em Journal of Computational Physics}, 184(2):366 -- 380, 2003.

\bibitem{FP:1}
J.~H. Ferziger and M.~Peri\'{c}.
\newblock {\em Computational Methods for Fluid Dynamics}.
\newblock Springer, 1996.

\bibitem{geurts:06}
B.~J. Geurts.
\newblock Interacting errors in large-eddy simulation: a review of recent
  developments.
\newblock {\em Journal of Turbulence}, 7:N55, 2006.

\bibitem{geurts:02}
B.~J. Geurts and J.~Fr{\"o}hlich.
\newblock A framework for predicting accuracy limitations in large-eddy
  simulation.
\newblock {\em Physics of Fluids}, 14(6):L41--L44, 2002.

\bibitem{geurts05}
B.~J. Geurts and F.~van~der Bos.
\newblock Numerically induced high-pass dynamics in large-eddy simulation.
\newblock {\em Physics of Fluids}, 17(12):125103, 2005.

\bibitem{ghanem:91}
R.~G. Ghanem and P.~D. Spanos.
\newblock {\em Stochastic Finite Elements: A Spectral Approach}.
\newblock Springer-Verlag, New York, NY, USA, 1991.

\bibitem{GMR:1}
F.~F. Grinstein, L.~G. Margolin, and W.~J. Rider, editors.
\newblock {\em {Implicit Large Eddy Simulation, Computing Turbulent Fluid
  Dynamics}}.
\newblock Cambridge University Press, 2007.

\bibitem{gullbrand:02}
J.~Gullbrand and F.~K. Chow.
\newblock Investigation of numerical errors, subfilter-scale models, and
  subgrid-scale models in turbulent channel flow simulations.
\newblock {\em Center for Turbulence Research, Annual Research Brief}, pages
  87--104, 2002.

\bibitem{issa86}
R.~I. Issa.
\newblock {Solution of the implicitly discretised fluid flow equations by
  operator-splitting}.
\newblock {\em Journal of Computational Physics}, 62(1):40--65, 1986.

\bibitem{iwamoto02}
K.~Iwamoto, Y.~Suzuki, and N.~Kasagi.
\newblock Reynolds number effect on wall turbulence: toward effective feedback
  control.
\newblock {\em International Journal of Heat and Fluid Flow}, 23(5):678--689,
  2002.

\bibitem{jimenez:pof13}
J.~Jim{\'e}nez.
\newblock Near-wall turbulence.
\newblock {\em Physics of Fluids}, 25(10):101302, 2013.

\bibitem{klein:05}
M.~Klein.
\newblock An attempt to assess the quality of large eddy simulations in the
  context of implicit filtering.
\newblock {\em Flow, Turbulence and Combustion}, 75(1):131--147, 2005.

\bibitem{kremer:15}
F.~Kremer and C.~Bogey.
\newblock Large-eddy simulation of turbulent channel flow using relaxation
  filtering: Resolution requirement and {R}eynolds number effects.
\newblock {\em Computers \& Fluids}, 116:17 -- 28, 2015.

\bibitem{kreys}
E.~Kreyszig.
\newblock {\em Introductory Functional Analysis with Applications}.
\newblock Wiley classics library. Wiley India Pvt. Limited, 2007.

\bibitem{lee15}
M.~Lee and R.~D. Moser.
\newblock Direct numerical simulation of turbulent channel flow up to {$Re_\tau
  = 5200$}.
\newblock {\em J. Fluid Mech.}, 774:395--415, 2015.

\bibitem{LF:3}
M.~Liefvendahl and C.~Fureby.
\newblock Grid requirments for {LES} of ship hydrodynamics in model and full
  scale.
\newblock {\em Ocean Engineering}, 143:259--268, 2017.

\bibitem{MaHM:1}
T.~Mari\'{c}, J.~H\"{o}pken, and K.~Mooney.
\newblock {\em {The OpenFOAM Technology Primer}}.
\newblock sourceFlux, 2014.

\bibitem{mariotti17}
A.~Mariotti, L.~Siconolfi, and M.V. Salvetti.
\newblock Stochastic sensitivity analysis of large-eddy simulation predictions
  of the flow around a 5:1 rectangular cylinder.
\newblock {\em European Journal of Mechanics - B/Fluids}, 62:149 -- 165, 2017.

\bibitem{lesReli1}
J.~Meyers, B.~Geurts, and P.~Sagaut, editors.
\newblock {\em Quality and Reliability of Large-Eddy Simulations}.
\newblock Springer, Netherlands, 2010.

\bibitem{meyers07}
J.~Meyers and P.~Sagaut.
\newblock Is plane-channel flow a friendly case for the testing of large-eddy
  simulation subgrid-scale models?
\newblock {\em Physics of Fluids}, 19(4):048105, 2007.

\bibitem{montomoli}
F.~Montomoli, M.~Carnevale, A.~D'Ammaro, M.~Massini, and S.~Salvadori.
\newblock {\em Uncertainty Quantification in Computational Fluid Dynamics and
  Aircraft Engines}.
\newblock Springer International Publishing, Springer Science+Business Media
  Dordrecht, 1st edition, 2015.
\newblock SpringerBriefs in Applied Sciences and Technology.

\bibitem{najm}
H.~N. Najm.
\newblock Uncertainty quantification and polynomial chaos techniques in
  computational fluid dynamics.
\newblock {\em Annual Review of Fluid Mechanics}, 41(1):35--52, 2009.

\bibitem{nicoud:99}
F.~Nicoud and F.~Ducros.
\newblock Subgrid-scale stress modelling based on the square of the velocity
  gradient tensor.
\newblock {\em Flow, Turbulence and Combustion}, 62(3):183--200, 1999.

\bibitem{oberkampf}
W.~L. Oberkampf and T.~G. Trucano.
\newblock Verification and validation in computational fluid dynamics.
\newblock {\em Progress in Aerospace Sciences}, 38(3):209 -- 272, 2002.

\bibitem{Patankar:1}
S.~V. Patankar.
\newblock {\em {Numerical Heat Transfer and Fluid Flow}}.
\newblock Taylor \& Francis, 1980.

\bibitem{pope}
S.~B. Pope.
\newblock {\em Turbulent Flows}.
\newblock Cambridge University Press, 10th printing edition, 2000.

\bibitem{pope04}
S.~B. Pope.
\newblock Ten questions concerning the large eddy simulation of turbulent
  flows.
\newblock {\em New J. Phys.}, 6:1--24, 2004.

\bibitem{saleh:2}
S.~Rezaeiravesh and M.~Liefvendahl.
\newblock Grid construction strategies for wall-resolving large eddy simulation
  and estimates of the resulting number of grid points.
\newblock Technical Report 2017-005, Department of Information Technology,
  Uppsala University, April 2017.

\bibitem{robinson91}
S.~K. Robinson.
\newblock Coherent motions in the turbulent boundary layer.
\newblock {\em Annu. Rev. Fluid Mech.}, 23:601--639, 1991.

\bibitem{sagaut}
P.~Sagaut.
\newblock {\em Large Eddy Simulation for Incompressible Flows, An
  Introduction}.
\newblock Springer, 3rd edition, 2006.

\bibitem{lesReli2}
M.~V. Salvetti, B.~G., J.~Meyers, and P.~Sagaut, editors.
\newblock {\em Quality and Reliability of Large-Eddy Simulations {II}}.
\newblock Springer, 2011.

\bibitem{smith}
R.~C. Smith.
\newblock {\em Uncertainty Quantification Theory, Implementation, and
  Applications}.
\newblock SIAM, 1st edition, 2014.

\bibitem{sobol}
I.~M. Sobo{l'}.
\newblock Global sensitivity indices for nonlinear mathematical models and
  their {M}onte {C}arlo estimates.
\newblock {\em Mathematics and Computers in Simulation}, 55:271--280, 2001.

\bibitem{SobolAnova}
G.~Tang, G.~Iaccarino, and M.~S. Eldred.
\newblock Global sensitivity analysis for stochastic collocation.
\newblock In {\em 51st AIAA/ASME/ASCE/AHS/ASC Structures, Structural Dynamics,
  and Materials Conference}, Orlando, Florida, 2010.

\bibitem{tyacke13}
J.~Tyacke, P.~Tucker, R.~Jefferson-Loveday, N.~R. Vadlamani, R.~Watson,
  I.~Naqavi, and X.~Yang.
\newblock Large eddy simulation for turbines: Methodologies, cost and future
  outlooks.
\newblock {\em Journal of Turbomachinery}, 136(6):061009, 2013.

\bibitem{deVilliers}
E.~De Villiers.
\newblock The potential of large eddy simulation for the modeling of wall
  bounded flows.
\newblock {\em PhD Thesis, Imperial College of Science, Technology and
  Medicine, London, UK}, 2006.

\bibitem{vreman96}
B.~Vreman, B.~Geurts, and H.~Kuerten.
\newblock Comparision of numerical schemes in large-eddy simulation of the
  temporal mixing layer.
\newblock {\em International Journal for Numerical Methods in Fluids},
  22(4):297--311, 1996.

\bibitem{weller:12}
H.~Weller.
\newblock Controlling the computational modes of the arbitrarily structured {C}
  grid.
\newblock {\em Monthly Weather Review}, 140(10):3220--3234, 2012.

\bibitem{xiu_gPCE}
D.~Xiu and G.~E. Karniadakis.
\newblock The {W}iener--{A}skey polynomial chaos for stochastic differential
  equations.
\newblock {\em SIAM Journal on Scientific Computing}, 24(2):619--644, 2002.

\end{thebibliography}

\end{document}